\documentclass[12pt]{article}

\usepackage{graphicx}

\newcommand{\blackslug}{\hbox{\hskip 1pt
        \vrule width 4pt height 8pt depth 1.5pt\hskip 1pt}}
\newcommand{\myQED}{\hfill \blackslug}
\newcommand{\la}{\langle}
\newcommand{\ra}{\rangle}

\newenvironment{proof}
    {\pagebreak[1]{\narrower\noindent {\bf Proof:\nopagebreak}}}%
    {\myQED}

    {\myQED}

\newtheorem{lemma}{Lemma}

\newtheorem{definition}{Definition}

\begin{document}

\begin{center}
{\large \bf
Environmental Sensing Options for Robot Teams: \\
A Computational Complexity Perspective}

\vspace*{0.2in}

Todd Wareham \\
Department of Computer Science \\
Memorial University of Newfoundland \\
St.\ John's, NL Canada \\
(Email: {\tt harold@mun.ca}) \\

\vspace*{0.1in}

Andrew Vardy \\
Department of Computer Engineering \\
Department of Computer Science \\
Memorial University of Newfoundland \\
St.\ John's, NL Canada \\
(Email: {\tt av@mun.ca}) \\

\vspace*{0.1in}

\today
\end{center}

\begin{quote}
{\bf Abstract}:
Visual and scalar-field (e.g., chemical) sensing are two of the options robot teams can 
use to perceive their environments when performing tasks. We give the first comparison of
the computational characteristic of visual and scalar-field sensing, phrased in
terms of the computational complexities of verifying and designing teams of
robots to efficiently and robustly perform distributed construction tasks.  This is
done relative a basic model in which teams of robots with deterministic finite-state
controllers operate in a synchronous error-free manner in 2D
grid-based environments. Our results show that for both types of sensing, all of our
problems are polynomial-time intractable in general and remain intractable under
a variety of restrictions on parameters characterizing robot controllers, teams, and
environments. That being said, these results also include
restricted situations for each of our problems in which those problems are effectively
polynomial-time tractable. Though there are some differences, our results suggest that 
(at least in this stage of our investigation) verification and design problems relative 
to visual and scalar-field sensing have roughly the same patterns and types of 
tractability and intractability results.
\end{quote}

\newpage

\section{Introduction}

\label{SectIntro}

\begin{quote}
``As if I had been totally color-blind before, and suddenly found myself in a
world full of color \ldots I had dreamt I was a dog --- it was an olfactory
dream --- and now I awoke to an infinitely redolent world --- a world in
which all other senses, enhanced as they were, paled before smell.'' \\
--- Oliver Sacks, ``The Dog Beneath the Skin'' \cite{Sac85}
\end{quote}

We human beings perceive much of our world by sight, using visible light and binocular 
vision to determine both the presence and nature of objects in our environment and their 
distances and orientations relative to ourselves. Visual senses exist that
exploit other forms of radiated energy (e.g., infrared light, 
ultrasound (see \cite[Chapters 8 and 9]{Dus92}, \cite[Sections 2.4 and 2,7]{Ste13},
and references). There are also senses such as smell based on the proximal detection of 
scalar fields.
Not greatly developed in modern human beings (but revivable
to stunning effect, as shown by the medical student Stephen D.\ quoted by Sacks), smell is
critical in the activities of many creatures (e.g., bacteria following chemical
gradients to approach food or flee toxins, moths finding mates by pheromones). 
Senses also exist that exploit other types of scalar fields
(e.g., blue-green bacteria approaching sunlight to optimize photosynthesis, fish using 
electrical fields to find prey in muddy water, birds navigating using Earth's magnetic 
field (see \cite[Chapters 6, 7, and 10]{Dus92}, 
\cite[Sections 2.2, 2.3, 2.5, and 2.6]{Ste13}, and references). 

It is perhaps not surprising that when we have built artificial systems, they have 
first been endowed with visual senses \cite{Cor17,SNS11} based not only on those seen in
nature (e.g., binocular cameras, sonar) but on other forms of radiation as well (e.g., 
lidar, radar). Much work has also been done on endowing artificial systems with various
types of scalar-field-based senses (e.g., the detection and localization of noxious 
chemical leaks) \cite{BC+17,IWM12} as well as integrating both types of sensory 
information to aid in performing tasks (e.g., lidar and GPS in autonomous vehicles). 

Visual and scalar-field sensing have different characteristics and thus different
advantages and disadvantages. In general, visual sensors are more complex but
yield more information (in terms of distance and perceived object characteristics),
while scalar-field sensors are simpler but yield less information unless
augmented by spatial gradient sensing and the ability to discriminate fine variation in
perceived scalar quantities. A finer-grained description of these
characteristics would be useful, particularly in applications such as micro- and 
nano-robotics \cite{KIK20,Sit17,WK+21} where conventional types of visual
sensing are extremely curtailed or impossible.

Investigations of these issues have previously been done via experiments and simulations
\cite{BC+17,IWM12,RB+03}. Recent theoretical research 
\cite{TW19,War15, War19, WK+11, WV18_SI, WV18_AT,WV21} complements this work by 
exploring the computational characteristics of problems associated with the design of 
robot controllers, teams, and environments for robot teams that perceive using 
visual and scalar-field senses. This work uses computational \cite{GJ79} and 
parameterized \cite{DF99} complexity analysis to characterize those situations in which 
each of the investigated problems is and is not efficiently solvable, where situations
are described in terms of restrictions on sets of one or more aspects characterizing 
individual robot controllers, robot teams, and operating environments.

It would be of great interest to compare the computational characteristics of visual
and scalar-field sensing, as both an aid for robotics researchers in interpreting 
experimental observations and an alternative perspective on the relative 
advantages and disadvantages of both kinds of sensing. 
Characterizations of the computational tractability and intractability of problems are
nothing new --- indeed, they have been a central part of computer science since the
dawn of algorithm design and computational complexity analysis. However, we believe that 
such work becomes more useful if it also incorporates the following:

\begin{enumerate}
\item A focus on determining the mechanisms in problems that interact to produce intractability.
\item A focus on making the results of analyses {\em as well as the techniques by which
       these results were derived} comprehensible to researchers who are
       not theoretical computer scientists.
\end{enumerate}

\noindent
Focus (1) implies that the complexity analyses should be done (at least initially) 
relative to simplified versions of problems that occur in the real world. This is done 
not only to simplify analysis but also (as in the classic thought experiments
of \cite{Bra84}) to allow easier exploration of mechanism interactions in these
problems independent of real-world complications. 
Focus (2) implies that result details (in particular those underlying 
intractability results) should be part of the presentation of results. This is done to
allow robotics researchers to better appreciate the restrictions under which our 
tractability and intractability results hold and hence more easily collaborate in deriving
results for more realistic and useful instances of verification and design
problems.

\subsection{Previous Work}

Various work has been done on the computational complexity of both verifying if a 
given multi-entity system can perform a task and designing such systems for tasks.
The systems so treated include groups of agents \cite{DLW03,Ste03,WD02}, robots 
\cite{DHX14,HSS84}, game-pieces \cite{FH+03}, and tiles \cite{AC+02}. Much of this
work, e.g. \cite{AC+02,DHX14,FH+03,HSS84}, assumes that the entities being moved
cannot sense, plan, or move autonomously. In the work where entities do have these
abilities, e.g. \cite{DLW03,Ste03,WD02}, the formalizations of control mechanisms
and environments are very general and powerful (e.g., arbitrary Turing machines or
Boolean propositional formulae), rendering both the intractability of these 
problems unsurprising and the derived results unenlightening with respect to 
possible restrictions that could yield tractability. 

Eight complexity-theoretic papers to date incorporate both autonomous 
robots and a suitably simple and explicit model of robot architecture and 
environment \cite{TW19,War15,War19,War22,WK+11,WV18_SI, WV18_AT,WV21}. 
Four of these papers consider navigation tasks
performed by robots with deterministic Brooks-style subsumption \cite{War15,WK+11}
and finite-state \cite{War22,WV18_AT} reactive controllers, respectively. The
other four consider construction-related tasks performed by robots with 
deterministic finite-state controllers relative to 
robot controller and environment design in a given environment where robot 
controllers are designed from scratch \cite{WV18_SI,WV21}, robot team design where teams 
are designed by selection from a provided library \cite{War19,WV21}, and robot team / 
environment co-design where robot teams are designed by selection from a provided
library \cite{TW19}. All eight papers use a 2D grid environment model.
Seven of these papers have robots that can visually 
sense the type of any square within a specified Manhattan-distance radius of a robot's 
position, with the robots in \cite{WV18_AT} using
single line-of-sight sensors. Only one complexity-theoretic analysis to date has 
considered environment and robot-team design problems under scalar-field sensing 
\cite{WV21}; however, due to conference page limits, no formal proofs of cited
results were given.

\subsection{Summary of Results}

In this paper, we give the first comparison of the computational characteristics of 
visual and scalar-field sensing. This is phrased in terms of the results of both
computational and parameterized complexity analyses of the following five verification and
design problems for robot teams and environments:

\begin{enumerate}
\item {\bf Team / Environment Verification}: Does a given team perform a
        specified task in a given environment?
\item {\bf Controller Design by Library Selection}: Can a robot controller be
        constructed from a given library of controller components that allows
        a team endowed with that controller to perform a specified task in
        a given environment?
\item {\bf Team Design by Library Selection}: Can the controllers of the
        members of a team be selected from a given library of robot controllers such
        that the resulting team performs a specified task in a given environment?
\item {\bf Environment Design}: Can an environment be designed such that a given
        team can perform a specified task in that environment?
\item {\bf Team / Environment Co-design by Library Selection}: Can the controllers of the
        members of a team be selected from a given library of robot controllers and
        an environment be designed such that the resulting team performs a specified task 
        in that environment?
\end{enumerate}

\noindent
Results for scalar-field sensing are derived relative to a basic model of scalar fields,
\cite{WV21} in which these fields are generated by field-quantities spreading outwards 
from discrete sources located in the environment.\footnote{
More complex models not treated here incorporate fields whose scalar quantities are not 
necessarily associated with discrete environmental sources (e.g., temperature, pressure) 
or are generated by mathematical operations (e.g., distance transforms \cite{FP06}).
\label{FnCSF}
}
We use the robot team operation and task models proposed in \cite{WV18_SI} in which teams
of robots with finite-state controllers operate in a non-continuous, synchronous, and 
error-free manner to perform distributed construction tasks \cite{PN+19}.

Our results show that for both types of sensing, all of our
problems are polynomial-time intractable in general and remain intractable under
a variety of restrictions on aspects characterizing robot controllers, teams, and
environments, both individually and in many combination and often when aspects
are restricted to small constant values.  That being said, our results also include
restricted situations for each of our problems in which those problems are effectively 
polynomial-time tractable. Our results show that, though there are some differences, both
kinds of sensing have (at least in this stage of our investigation) roughly the same types
and patterns of intractability and tractability results.

Before we close out this subsection, clarification is in order regarding the
provenance of results reported here. Results for problems (1--5) have been given
previously in other papers with proof for visual sensing (Results A.ST.1 and D.ST.1--6 
\cite{WV18_SI}; Result B.ST.1 \cite{War22}; Results C.ST.1--3, C.ST.6, and C.ST.7 \cite{War19}; Results E.ST.1--6 
\cite{TW19}) and in other papers without proof for scalar-field sensing (Results 
C.SF.1--7 and D.SF.1--7 \cite{WV21}). All other results (Result A.SF.1 (Problem (1)); 
Results B.ST.2--4 and B.SF.1--5 (Problem (2)); Results C.ST.4 and C.ST.5,(Problem (3)); 
Results E.SF.1--7 (Problem (5)) are new here, as well as the proofs of all results cited
previously in \cite{WV21}. Hence, if one counts the results in \cite{WV21} as new,
there are 19 previous and 32 new results in this paper. Though a number of these new 
results build on proofs of results described previously, many also involve non-trivial 
modification of and additions to those previous proofs and hence are not simply 
incremental extensions of previous work.

\subsection{Organization of Paper}

This paper is organized as follows. In Section \ref{SectForm}, we describe the 
environment, target structure, robot controller, and robot team models used to formalize 
our verification and design problems relative to visual and scalar-field sensing. In 
Section \ref{SectRes}, we consider the viable algorithmic options for these problems 
relative to two popular types of exact and restricted efficient solvability. To allow
focus on this comparison and its implications in the main text, all proofs of previously 
unpublished results are given in an appendix. In Section \ref{SectDisc}, we discuss the 
overall implications of our results for our simplified problems and real-world
robotics, and propose a metaphor for thinking about comparative computational complexity 
analyses. Finally, our conclusions and directions for future work are given in Section 
\ref{SectConc}.

\section{Formalizing Scalar-field Sensing for Distributed Construction}

\label{SectForm}

In this section, we first describe the basic entities in our models (given previously
in \cite{WV18_SI,WV21}) of distributed
construction relative to visual and scalar-field sensing --- namely, environments, 
structures, individual robots and robot teams. We then formalize the computational 
problems associated with robot controller, team, and environment design (based on those 
defined in \cite{TW19,War19,WV18_SI,WV21}) that we will analyze in the remainder of this paper. 

\begin{figure}[t]
\begin{center}
\includegraphics[height=3.5in]{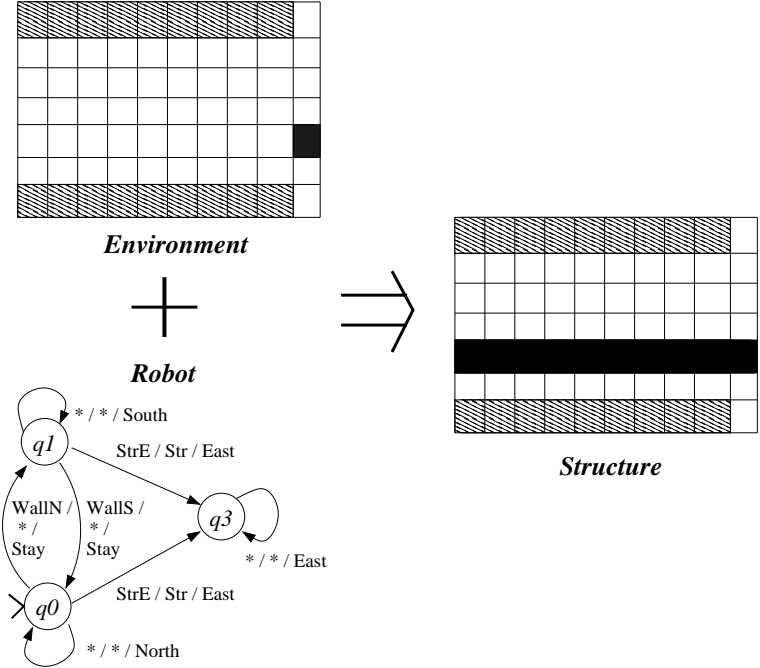}
\end{center}
\caption{An example of distributed construction by a robot team (Adapted from Figures
          1 and 2 in \cite{WV18_SI}). In the initial and final
          environments, wall and structure squares are indicated by hatching and 
          black fill, respectively. The team consists of 9 copies of the pictured
          finite-state robot, which can only sense immediately-adjacent squares.  
          Each transition is 
          labeled with a triple $x/y/z$ where $x$ is the transition-activation formula, 
          $y$ is the square-change (if any), and $z$ is the movement-action
          performed thereafter. See the main text for further details.
          }
\label{FigRobotEx}      
\end{figure}

The basic entities are as follows:

\begin{itemize}
\item{
{\bf Environments}: Our robots operate in a finite 2D square-based environment $E$
based on a 2D grid $G$; for simplicity, we will assume that there are no obstacles
and hence robots can move and fields can propagate freely through any square. Let
$E_{i,j}$ denote the square that is in the $i$th column and $j$th
row of $E$ such that $E_{1,1}$ is the square in the southwest-most corner of $E$. 

We have two types of environments, square-type and scalar-field, which are associated
with visual and scalar-field sensing, respectively. These two types of environments
are necessary as the environmental aspects sensed by visual and scalar-field sensing
behave in different ways.
In a {\bf square-type environment}, each square in $G$ has a square-type, e.g.,
grass, gravel, wall, drawn from a set $E_T$. An example square-type environment is shown in
Figure \ref{FigEnvType}(a). A {\bf scalar-field environment} has 
a set of scalar-field instances drawn from a field-set $S$ that are 
placed within $G$.  A {\bf scalar field}
$s \in S$ is defined by a field-quantity $fq$, a field-type, a source-value
$s_s$ that is a positive real number, and a decay $s_d$ that is a non-negative
real number. There are two types of fields:

\begin{enumerate}
\item Point-fields that radiate outwards radially from a specified grid-square,
       and
\item Edge-fields that radiate outward linearly from a specified grid-edge, 
       which can be North, South, East or West, in the opposite compass
       direction of the source edge.
\end{enumerate}

\noindent
The value of $fq$ for a field-instance $s$ in a square $p$ in scalar-field environment
$E$ is $s_s - (s_d \times z)$, where $z$ is the Manhattan distance between $p$ and the 
field-source in the case of point-fields and the shortest Manhattan distance between $p$ and
the specified grid-edge in the case of edge-fields. As such, our scalar fields are generated
by field-quantities spreading outwards from discrete sources in the
environment (see Footnote \ref{FnCSF}). Example point- and edge-fields are shown in Figures 
\ref{FigEnvType}(b) and \ref{FigEnvType}(c), respectively. Each square in $E$ can have at 
most one point-field, though each grid-edge may have multiple edge-fields. As there may be 
multiple instances of fields based on the same field-quantity $fq$ in $E$, the value of 
$fq$ in a square $p$ in $E$ is the sum of the values of all field-instances
based on $fq$ in $E$ at $p$. An example of such an environment summed relative to
one edge-field and two point-fields based on the same field-quantity is shown in
Figure \ref{FigEnvType}(d).

\begin{figure}[p]
\begin{center}
\includegraphics[width=4.0in]{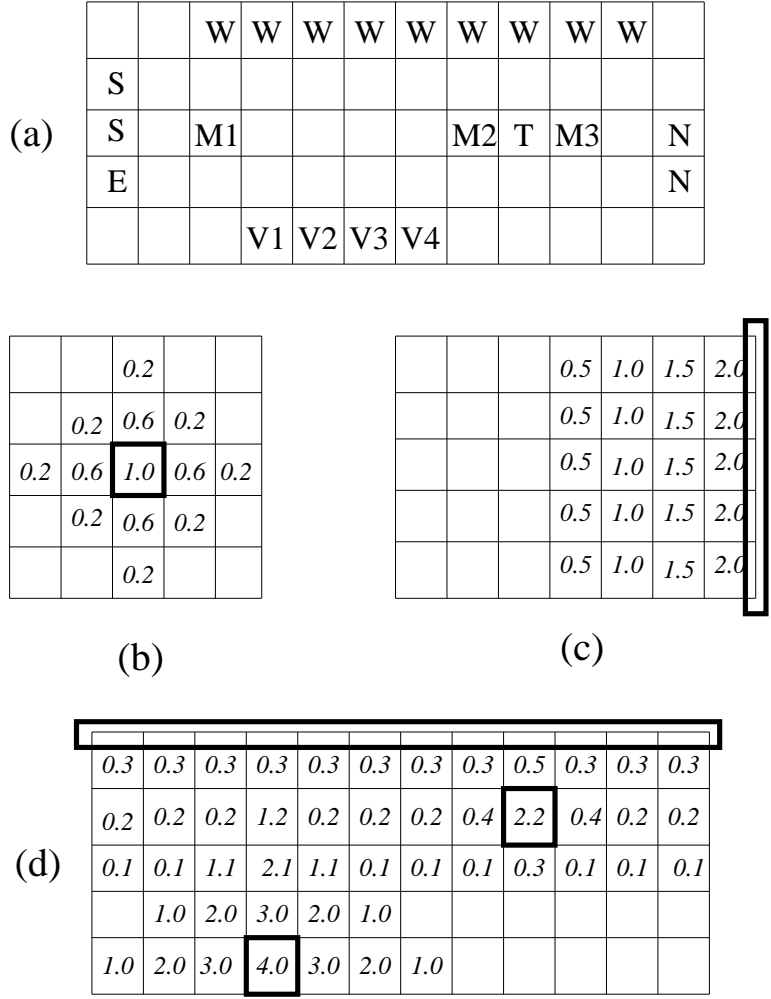}
\end{center}
\vspace*{0.1in}
\caption{Example square-type and scalar-field environments (Modified from Figure 1
          in \cite{WV21}). a) A 
          square-type environment based on the square-type set $E_T = \{e_{V1}, e_{V2},
          e_{V3}, e_{V4}, e_N, e_S, e_W,$ $e_E, e_{M1}, e_{M2}, e_{M3}, e_T, e_B\}$;
          all blank squares have type $e_B$ (adapted from Figure 2 in \cite{War19}).
          b) A scalar-field environment with a point-field based at square $E_{3,3}$ having
          source-value 1.0 and decay 0.4. c) A 
          scalar-field environment with an East-based edge-field with source-value
          2.0 and decay 0.5. d) A scalar-field environment with three
          fields based on the same field-quantity (a North-based edge-field with
          source-value 0.3 and decay 0.1; a point-field based at square $E_{4,1}$
          with source-value 4.0 and decay 1.0; a point-field based at square $E_{9,4}$
          with source-value 2.0 and decay 1.8). The source-region of each 
          field is indicated with a boldfaced box.}
\label{FigEnvType}
\end{figure}

In square-type environments, every square-type set $E_T$ contains the special
square-types $e_X$ (that is used to specify parts of structures) and 
$e_{robot}$ (that is used to indicate the positions of robots). In scalar-field
environments, every field-set $S$ contains special
point-fields $s_X$ (with source-value 1 and decay-value 0.5 that is used to
specify parts of structures) and $s_{robot}$ (with source-value 1 and
decay-value 0.5 that is used to indicate the positions of robots).
}
\item{
{\bf Structures}: A structure $X$ in an environment $E$ is a
two-dimensional pattern of structural elements in an $m \times n$ 
grid whose location in $E$ is specified relative to the position $p_X$ of 
southwest-most corner of the structure-grid in the environment-grid. 
The structural elements are squares of type $e_X$ and point-fields $s_X$
in square-type and scalar-field environments, respectively. An example
linear structure is shown in Figure \ref{FigRobotEx}.
}
\item{
{\bf Robots}: Each robot occupies a square in an environment $E$ and in a 
basic movement-action
can either move exactly one square to the north, south, east or west of its 
current position or elect to stay at its current position. In square-type
environments, a robot can sense the type of any square out to specified
Manhattan distance $r$ and modify the type of any square out to
Manhattan distance one from the robot's current position using the predicates 
$enval()$ and $enmod()$ (see \cite{WV18_SI} for details); note that the square at 
Manhattan distance zero (e.g., $r = 0$) is the robot's current position.
In this type of environment, square-type modifications correspond to real-world
activities such as agents placing construction materials or signs to guide other agents,
e.g., ``turn right'', ``this way to food'', ``do not go beyond this point''.
Each robot has an associated square-type $e_{robot}$ at its current position.

In scalar-field environments, a robot can sense both the absolute value of any 
field-quantity in and the manner in which that quantity changes in any square
immediately adjacent to the square in which that robot is currently positioned. These two 
types of sensing are done using the following predicates:

\begin{enumerate}
\item $fval(fq,rel,val)$, $rel \in \{=, <, >, \leq, \geq\}$, which returns $True$ if 
       $fq ~ rel ~ val$ in the robot's current position and $False$ otherwise; and
\item $fgrd(fq,rel,dir)$, $rel \in \{=, <, >, \leq, \geq\}$ and $dir \in \{North, South,
       East,$ \linebreak $West\}$, which returns $True$ if $fq ~ rel ~ fq_{dir}$, where $fq_{dir}$
       is the value of $fq$ in the square immediately adjacent to the robot's 
       current position in direction $dir$, and $False$ otherwise.
\end{enumerate}

\noindent 
We refer to these two types as value and gradient sensing (the latter so named by
analogy with the ability of various organisms to sense spatial chemical gradients in their
environments).
Figure \ref{FigSFSChar} illustrates these kinds of sensing and shows how both types are
necessary and useful. One of the fundamental problems with value sensing is that it can only
establish that a particular field-source is nearby (Figure \ref{FigSFSChar}(a));
to determine the direction to that source, gradient sensing is required (Figure
\ref{FigSFSChar}(b)). Given appropriate field source- and decay-values and the
ability to distinguish small variations in field-quantity values, value and gradient 
can be used in tandem to implement sensing at a distance (Figure 
\ref{FigSFSChar}(c)) as well as determine appropriate robot actions in complex
environments with multiple field-sources (Figure \ref{FigSFSChar}(d)); both of these
techniques will be exploited frequently in the proofs of our complexity
results in Section \ref{SectRes}.

\begin{figure}[p]
\begin{center}
\includegraphics[width=3.0in]{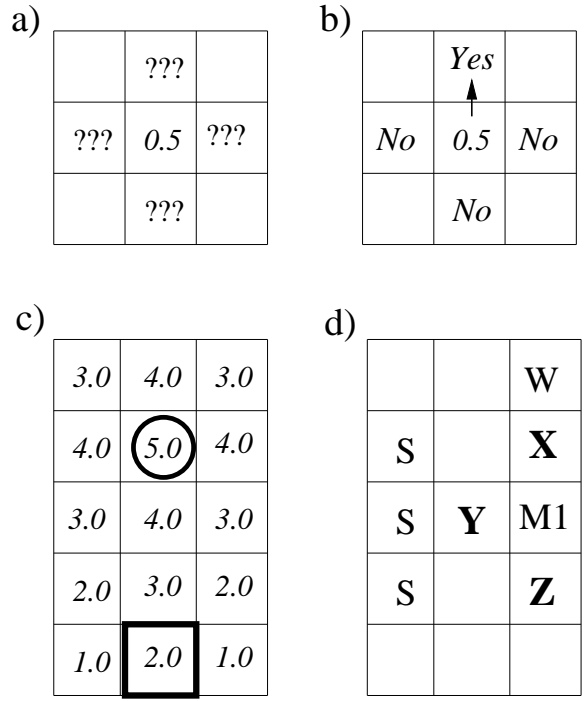}
\end{center}
\vspace*{0.1in}
\caption{Characteristics of value and gradient sensing in scalar-field environments.
          a) The fundamental problem with value sensing. Given a point-field with 
          source-value and decay (in this case, 1 and 0.5, respectively), value sensing 
          alone (in this case, $fval(fq, =, 0.5)$) can determine that the point-field is 
          within a particular Manhattan distance (in this case, 1) of the current 
          position but not the direction to that point-field. b) Appropriate gradient sensing
          (in this case, $fgrd(fq, >, North)$) can
          mitigate the problem noted in (a). c) Sensing at a distance can be 
          implemented using a combination of value and gradient sensing. In this case,
          the point-field of interest has source-value 5 and decay 1; gradient
          sensing gives the direction to that field and value sensing gives the
          distance (namely five minus the sensed value). d) Determining appropriate
          robot actions in a complex environment, in this case the westmost three
          columns of a point-field version of the square-type environment in
          Figure \ref{FigEnvType}(a). At positions {\bf X}, {\bf Y}, and {\bf Z}, the wanted
          robot movements are West, South, and East. Assuming point-fields $s_W$,
          $s_S$, and $s_{M1}$ with source-value 1 and decay 0.5, appropriate movement at
          {\bf X} and {\bf Y} can be 
          triggered with $fval(fq_W, =, 0.5)$ and $fval(fq_S, =, 0.5)$ respectively, but this would
          be contradicted by using $fval(fq_{M1}, =, 0.5)$ to trigger eastward movement at
          {\bf Z} (which would additionally attempt to
          trigger eastward movement at {\bf X} and {\bf Y}). Two correct
          triggers for movement at {\bf Z} are thus $(fval(fq_{M1}, =, 0.5) ~\rm{and}~
          fval(fq_S, =, 0.0) ~\rm{and}~ fval(fq_W, =, 0.0))$ or $(fval(fq_{M1}, =, 0.5)
          ~\rm{and}~ fgrd(fq_{M1}, >, North))$.
}
\label{FigSFSChar}
\end{figure}

A scalar-field sensing robot can either add a point-field to
or modify an existing point-field at any square at any position within
Manhattan distance one of the robot's current position to type $s \in S$
via predicates of the form $fmod(s,pos)$ where $pos$ is specified in terms of 
a pair $(x,y)$ specifying an environment-square $E_{i+x,j+y}$ if the robot is 
currently occupying $E_{i,j}$. Analogous to modifications to square-type environments, 
scalar-field additions and modifications correspond to real-world activities such as agents 
placing construction materials, e.g., termites dropping pheromone-laden mud pellets, or 
signs to guide other agents, e.g., ``turn right'', ``this way to food'',
``do not go beyond this point''.
Each robot has an associated point-field of type $s_{robot}$ at its current position. 
For simplicity, we shall assume that the 
field changes caused by point-field addition or modification or 
robot motion propagate instantaneously to all squares in $E$.

Each robot has a finite-state controller and is hence known as a 
Finite-State Robot (FSR). Each such controller consists of a set $Q$ of states 
linked by transitions, where each transition $(q, f, x, move, q')$ between states 
$q$ and $q'$ has a propositional logic trigger-formula $f$, an environment 
modification specification $x$, and a movement-specification $move \in \{goNorth, goSouth, 
goEast, goWest,$ \linebreak $stay\}$. In robots operating in square-type environments, this 
formula and modification specification are based on predicates $enval()$ and 
$enmod()$, respectively; in robots operating in scalar-field environments , the
corresponding predicates are $fval()$ and $fmod()$, respectively. The transition
trigger-formulas and/or environment modification specifications can also be a special
symbol $*$, which is interpreted as follows: (1) if $f \neq x \neq *$ and
the transition's trigger-formula evaluates to $True$, i.e, the transition is enabled, this
causes the environment-modification specified by $x$ to occur, the robot to move either
one or no square as specified by $move$, and the robot's state to change from $q$ to $q'$;
(2) if $f = *$, the transition enables and executes if no other non-$*$ transition is 
enabled (making this in effect the default transition), and (3) if $x = *$, no 
environment-modification is made. Note that when assessing the length of a 
transition-trigger formula, each predicate counts as a single symbol. 

Let $L$ be a library of transition templates of the form $(q, f, x, move, q')$ above. Such
a library is used in problems ContDesLSSF and ContDesLSST defined below
to construct an FSR controller from a specified set of states by instantiating transition 
templates relative to those states. Note it may be the case that $q = q'$ in such a 
construction, i.e., a transition may loop back on the same state.
}
\item{
{\bf Robot teams}: A team $T$ consists of a set of the robots described above, 
where there may be more than one robot with the same controller on a team. Let 
$T_i$ denote the $i$th robot on the team. Each square in $E$ can hold at most 
one member of $T$; if at any point in the execution of a task two robots in a 
team attempt to occupy or modify the same square or a robot attempts to move to a square 
outside the environment, the execution terminates and is considered unsuccessful. A 
{\bf positioning} of $T$ in $E$ is an assignment of the robots in $T$ to a set of $|T|$ 
squares in $E$. For simplicity, team members move synchronously and do not communicate 
with each other directly (though they may communicate indirectly through
changes they make to the environment, i.e., via stigmergy \cite{BDT99}). Note that once 
movement is triggered, it is atomic in the sense that the specified movement is completed.
}
\end{itemize}

\noindent
We use the notion of deterministic robot and team operation introduced in \cite{WV18_SI}
as extended in \cite{TW19} (i.e., requiring that at any time as the team operates in an 
environment, all transitions enabled in a robot relative to the current state of that
robot perform the same environment modifications and progress to the same next state). 
This ensures that requested structures are created by robot teams reliably. 

An example of a construction task performed by a team of 3-state FSR is shown in Figure 
\ref{FigRobotEx}. In this example, the initial environment consists of 
two parallel east-west-oriented walls of length $9$ and a structure seed square in the 
grid-column to the immediate east of two walls. The task is to construct an east-west 
oriented freespace-based linear structure (analogous to a painted lane-divider on a 
highway) extending westwards from the seed to the westmost edge of the environment.
The initial position of the $9$ robots on the team is immediately to the north of the 
southern wall. The subsequent operation of the team creates the requested structure by
``growing'' it westwards from the initial seed, with the robots progressing eastwards
along the seed-line into a holding area to the east (which for space reasons is not shown
in the diagram). The required sensing of squares immediately around a robot 
can be implemented with robot sensory radius $r = 1$ (under visual sensing) and
scalar fields with source-value 1 and decay-value 0.5 (under scalar-field sensing).
Note that this team operates correctly and deterministically as long as 
the seed square is not either immediately to the southeast of
the north wall or immediately to the northeast of the south wall. Otherwise, 
the eastmost robot on the team will have two transitions enabled on first
encountering the seed structure square to its immediate east (namely,
$\{(q0,(\#,N),*,stay,q1),(q0,(X,E),(X,U),goNorth,q2)\}$ in the first case and
$\{(q1,(\#,S),*,stay,q0),(q1,(X,E),(X,U),goSouth,q2)\}$
in the second case), which by the FSR operation
rules discussed earlier will cause the team's operation to terminate.

We can now formalize the computational problems that we will analyze in 
the remainder of this paper. Our problems relative to scalar-field
environments and sensing are as follows:

\vspace*{0.1in}

\noindent
{\sc Team / scalar-field environment verification} (TeamEnvVerSF) \\
{\em Input}: An environment $E$ based on grid $G$ and field-set $S$, an FSR 
              team $T$, a structure $X$, an initial positioning $p_I$ of $T$ in
              $E$, and a position $p_X$ of $X$ in $E$. \\
{\em Question}: Does $T$ started at $p_I$ in $E$ create $X$ at $p_X$?

\vspace*{0.1in}

\noindent
{\sc Controller design by library selection (Scalar-field)} (ContDesLSSF) \\
{\em Input}: An environment $E$ based on grid $G$ and field-set $S$, a requested team-size 
              $|T|$, an initial positioning $p_I$ of $T$ in $E$, a structure $X$, a 
              position $p_X$ of $X$ in $E$, a transition template library $L$, and
              positive integers $|Q|$ and $d$. \\
{\em Output}: An FSR controller $c$ with at most 
               $|Q|$ states and at most $d$ transitions chosen from $L$ out of any state 
               such that an FSR team with $|T|$ robots based on $c$ started at $p_I$ creates
               $X$ at $p_X$, if such a $c$ exists, and special symbol $\bot$ 
               otherwise.

\vspace*{0.1in}

\noindent
{\sc Team design by library selection (Scalar-field)} (TeamDesLSSF)  \\
{\em Input}: An environment $E$ based on grid $G$ and field-set $S$, a requested
              team-size $|T|$, an FSR library $L$, an initial region $E_I$
              of size $T$ in $E$, a structure $X$, and a position $p_X$ of $X$ in 
              $E$. \\
{\em Output}: An FSR team $T$ selected from $L$ such that $T$ started in $E_I$
               creates $X$ at $p_X$, if such a a $T$ exists,
               and special symbol $\bot$ otherwise.

\vspace*{0.15in}

\noindent
{\sc Scalar-field environment design} (EnvDesSF) \\
{\em Input}: An environment-grid $G$, a field-set $S$, an FSR team $T$ based on
              controller $c$, a structure $X$, an initial positioning $p_I$ of 
              $T$ in $G$, and a position $p_X$ of $X$ in $G$. \\
{\em Output}: An environment $E$ derived from $G$ and $S$ such that $T$ started
               at $p_I$ creates $X$ at $p_X$, if such an $E$ exists, and 
               special symbol $\bot$ otherwise.

\vspace*{0.1in}

\noindent
{\sc Team / scalar-field environment co-design by library selection} \\ 
     (TeamEnvDesLSSF) \\
{\em Input}: An environment-grid $G$, a field-set $S$, a team-size $|T|$, an FSR
             library $L$, a region $E_I$ of size $|T|$ in $G$, a 
             structure $X$, and a position $p_X$ of $X$ in $G$. \\
{\em Output}: A team $T$ of size $|T|$ selected from $L$ and an environment $E$
               derived from $G$ and $S$ such that $T$ started at $E_I$ in $E$ 
               creates $X$ at $p_X$, if such a $T$ exists, and special symbol 
               $\bot$ otherwise .\\
	
\vspace*{0.1in}

\noindent
The above correspond to the following previously-analyzed problems that
are relative to square-type environments and visual sensing:

\vspace*{0.1in}

\noindent
{\sc Team / square-type environment verification} (TeamEnvVerST) \\
{\em Input}: An environment $E$ based on square-type set $E_T$,
              a structure $X$, an initial positioning $p_I$ of $T$ in
              $E$, and a position $p_X$ of $X$ in $E$. \\
{\em Question}: Does $T$ started at $p_I$ in $E$ create $X$ at $p_X$?

\vspace*{0.1in}

\noindent
{\sc Controller design by library selection (Square-type)} (ContDesLSST) \\
{\em Input}: An environment $E$ based on square-type set $E_T$, a requested team-size 
              $|T|$, an initial positioning $p_I$ of $T$ in $E$, a structure $X$, a 
              position $p_X$ of $X$ in $E$, a transition template library $L$, and
              positive integers $r$, $|Q|$, and $d$. \\
{\em Output}: An FSR controller $c$ with sensory radius $r$, at most $|Q|$ 
               states, and at most $d$ transitions chosen from $L$ out of any state such 
               that an FSR team with $|T|$ robots based on $c$ started at $p_I$ creates
               $X$ at $p_X$, if such a $c$ exists, and special symbol $\bot$ 
               otherwise.

\vspace*{0.1in}

\noindent
{\sc Team design by library selection (Square-type)} (TeamDesLSST)  \\
{\em Input}: An environment $E$ based on square-type set $E_T$, a requested
              team-size $|T|$, an FSR library $L$, an initial region $E_I$
              of size $T$ in $E$, a structure $X$, and a position $p_X$ of $X$ in 
              $E$. \\
{\em Output}: An FSR team $T$ selected from $L$ such that $T$ started in  $E_I$
               creates $X$ at $p_X$, if such a a $T$ exists,
               and special symbol $\bot$ otherwise.

\vspace*{0.1in}

\noindent
{\sc Square-type environment design} (EnvDesST) \\
{\em Input}: An environment-grid $G$, a square-typed set $E_T$, an FSR team $T$ 
              based on
              controller $c$, a structure $X$, an initial positioning $p_I$ of 
              $T$ in $G$, and a position $p_X$ of $X$ in $G$. \\
{\em Output}: An environment $E$ derived from $G$ and $E_T$ such that $T$ started
               at $p_I$ creates $X$ at $p_X$, if such an $E$ exists, and 
               special symbol $\bot$ otherwise.

%\vspace*{0.1in}
\newpage

\noindent
{\sc Team / square-type environment co-design by library selection} \\ 
     (TeamEnvDesLSST) \\
{\em Input}: An environment-grid $G$, a square-type set $E_T$, a team-size $|T|$,
             an FSR library $L$, a region $E_I$ of size $|T|$ in 
             $G$, a structure $X$, and a position $p_X$ of $X$ in $G$. \\
{\em Output}: A team $T$ of size $|T|$ selected from $L$ and an environment $E$
               derived from $G$ and $E_T$ such that $T$ started at $E_I$ in $E$ 
               creates $X$ at $p_X$, if such a $T$ exists, and special symbol 
               $\bot$ otherwise .\\
	
\vspace*{0.1in}

\noindent
Problems TeamDesLSSF and EnvDesSF are from \cite{WV21},
problems TeamEnvVerST, TeamDesLSST, EnvDesST, and TeamEnvDesST are the
previously-analyzed problems ContEnvVer \cite{WV18_SI}, DesCon \cite{War19}, EnvDes
\cite{WV18_SI}, and CoDesignLS \cite{TW19} and problem ContDesLSST is 
essentially problem ContDesLS \cite{War22} which was in turn a variation
on previously-analyzed problem ContDes \cite{WV18_SI}.
Without loss of generality, we will assume that each member of $T$ starts
operating in the initial state $q_0$ of its associated controller.
Following \cite{War19}, we shall also assume for problems TeamDesLSST and TeamDesLSSF
that (1) all FSR in $L$ are behaviorally distinct and (2) when tasks complete
successfully, they do so relative to {\em any} positioning $p_I$ of the members of
$T$ in $E_I$.

All of our problems above are simplifications of their associated real-world
design problems. Many of these simplifications with respect to FSRs, FSR teams,
and square-type environments have already been noted and discussed in 
\cite{War19,WV18_SI}. Our conception of scalar-field 
environments introduces additional simplifications (e.g., quantity-spread from a field
source is instantaneous, decays as a linear function of Manhattan distance from the
source, and is not affected by obstacles).
The extent to which these simplifications affect the applicability of our results to 
real-world design problems in scalar-field environments can be addressed using
arguments analogous to those presented previously in \cite{War19,WV18_SI} relative to 
visual sensing (see also Section \ref{SectDiscImpRWR}).
That being said, recall from Section \ref{SectIntro} that our intent here is not so much 
to provide results of immediate use to real-world robots using visual and scalar-field 
sensors but rather to provide a simple setting in which to examine core computational 
characteristics of both types of sensing independent of conflating issues arising from 
errors in robot perception, control, and movement and more complex and realistic models of
visual and scalar-field environments. This will be done in the next section.

\section{Results}

\label{SectRes}

In this section, we first describe two types of efficient solvability, polynomial-time
exact solvability and fixed-parameter tractability, and sketch the techniques by
which unsolvability results are proven for these types (Section \ref{SectResDesSolv}).
Then, in Sections \ref{SectResTeamEnvVer}--\ref{SectResTeamEnvDesLS}, we consider the five
verification and design problems listed in Section \ref{SectIntro}. For each problem, we 
review previously obtained results for that problem relative to visual sensing in 
square-type environments as well as the techniques used to prove these results, and then
describe what results and techniques carry over for that problem relative to
scalar-field sensing and environments. To focus on the patterns in and implications 
of these results in the main text, all proofs of previously unpublished results are
given in an appendix. 

\subsection{Types of Efficient Solvability}

\label{SectResDesSolv}

There are two basic questions when analyzing a problem $\Pi$ computationally: (1) is $\Pi$
efficiently solvable in general (i.e., for all possible inputs)?; and, if not, (2) under 
which restrictions (if any) is $\Pi$ efficiently solvable? These questions are based on
the following two types of efficient solvability:

\begin{enumerate}
\item{
\textbf{Polynomial-time exact solvability}: An exact 
polynomial-time algorithm is a deterministic algorithm whose runtime is 
upper-bounded by $c_1|x|^{c_2}$, where $|x|$ is the size of the input $x$
and where $c_1$ and $c_2$ are constants, and
is always guaranteed to produce the correct output for all inputs. A problem 
that has a polynomial-time algorithm is said to be \textbf{polynomial-time 
tractable}; otherwise, a problem that does not have such an algorithm is
said to be \textbf{polynomial-time intractable}. Polynomial-time tractability is desirable 
because runtimes increase
slowly as input size increases, and hence allow the solution of larger inputs.
}
\item{
\textbf{Effectively polynomial-time exact restricted solvability}: Even if
a problem is not solvable in the sense above, a restricted version
of that problem may be exactly solvable in close-to-polynomial time.
Let us characterize restrictions on problem inputs in terms of a 
set $K = \{k_1, k_2, \ldots, k_{|K|} \}$ of aspects of the input. 
For example, possible restrictions on the inputs of ContDesLSSF could
be the number of possible states and outgoing transitions per state in
a controller and the number of transitions in the given transition
template library (see also Table \ref{TabPrm}). Let $\la K \ra$-$\Pi$
denote a problem $\i$ so restricted relative to an aspect-set $K$.

One of the most
popular ways in which an algorithm can operate in close-to-polynomial
time relative to restricted inputs is {\bf fixed-parameter (fp-)} \linebreak {\bf tractability} 
\cite{DF99}. Such an algorithm runs in time that is non-polynomial purely in 
terms of the aspects in $K$, {\em i.e.}, in time $f(K)|x|^c$ where $f()$ is 
some function, $|x|$ is the size of input $x$, and $c$ is a constant. A problem
with such an algorithm for aspect-set $K$ is said to be \textbf{fixed-parameter
(fp-)tractable relative to $K$}. Fixed-parameter tractability
generalize polynomial-time exact solvability by allowing the leading constant 
$c_1$ of the input size in the runtime upper-bound of an algorithm to be a 
function of $K$. Though such algorithms run in non-polynomial time in general, 
for inputs in which all the aspects in $K$ have very small constant values 
and $f(K)$ thus collapses to a possibly large but nonetheless constant 
value, such algorithms (particularly if $f()$ is suitably well-behaved,
({\em e.g}, $(1.2)^{k_1 + k_2}$) may be acceptable.
}
\end{enumerate}

\noindent
The second type is useful in isolating and investigating sources of computational
intractability in problems. Following \cite{vREM+08}, for a set of aspects $K$ of a problem 
$\Pi$, we say that {\bf $K$ is a source of intractability in $\Pi$} if
$\la K \ra$-$\Pi$ is fp-tractable; if $K$ is such that $\la K' \ra$-$\Pi$ is not
fp-tractable for any subset $K' \subset K$, then this source of computational
intractability is also {\bf minimal}. Such sources of 
intractability (particularly if they are minimal) are very useful in highlighting
those mechanisms in a problem that interact and can (if allowed to operate in an 
unrestricted manner) yield intractability (Section \ref{SectDiscImpSen}).

\begin{table}[t]
\caption{Parameters for Design Problems in Square-type (ST) and Scalar-field (SF) 
          Environments. 
These parameters are divided into four groups:
          (1) parameters characterizing robot teams ($|T|$, $h$);
          (2) parameters characterizing individual robots ($|Q|$, $d$, $|f|$, $r$,);
          (3) parameters characterizing the controller and team design process ($|L|$); and
          (4) parameters charactering environments and requested structures
               ($|E|$, $|E_T|$, $|S|$, $|S_E|$, $|q_E|$, $|X|$).
           The problems to which each parameter is applicable are indicated in the third
           column of the table.
}
\vspace*{0.1in}
\label{TabPrm}
\centering
\begin{tabular}{| c || l | c |}
\hline
Parameter   & Description & Applicability  \\
\hline\hline
$|T|$       & \# robots in team & All \\
\hline
$h$         & \# robot-types in team & All \\
\hline\hline
$|Q|$       & Max \# states per robot & All \\
\hline
$d$       & Max \# outgoing transitions per state & All \\
\hline
$|f|$     & Max length transition-trigger formula & All \\
\hline
$r$       & Sensory radius of robot               & ST only \\
\hline\hline
$|L|$       & \# entities in design library & ContDesLSS*, \\
            &                      & TeamDesLSS*, \\
            &                      & TeamEnvDesLSS* \\
\hline\hline
$|E|$       & \# squares in environment & All \\
\hline
$|E_T|$       & \# environment square-types& ST only \\
\hline
$|S|$       & \# scalar-field types & SF only \\
\hline
$|S_E|$       & Max \# scalar-fields in environment & SF only \\
\hline
$|q_E|$       & Max \# scalar-fields of type $q$ in environment & SF only \\
\hline
$|X|$       & \# squares in structure & All \\
\hline
\end{tabular} 
\end{table}

To show solvability of a problem $\Pi$ relative to one of the above types, one need only
give an algorithm of that type for $\Pi$. To show unsolvability, we use
reductions between pairs of problems, where a reduction from a problem $\Pi$ to a problem 
$\Pi'$ is essentially an efficient algorithm $A$ for solving $\Pi$ which uses a 
hypothetical algorithm for solving $\Pi'$. Reductions are useful in two ways:
 
\begin{enumerate}
\item If $\Pi$ reduces to $\Pi'$ and $\Pi'$ is efficiently solvable by
       algorithm $B$ then $\Pi$ is efficiently solvable (courtesy of
       the algorithm $A'$ that invokes $A$ relative to $B$).
\item If $\Pi$ reduces to $\Pi'$ and $\Pi$ is not efficiently solvable 
       then $\Pi'$ is not efficiently solvable (as otherwise, by the logic of (1)
       above, $\Pi$ would be efficiently solvable, which would be a
       contradiction).
\end{enumerate}

\noindent
Each of our types of efficient solvability has its own associated type of reducibility
designed to pass that type of solvability backwards along a reduction by the logic of (1) 
above. These are the standard reducibilities for polynomial-time exact and
fixed-parameter solvability.

\begin{definition}
\cite[Section 3.1.2]{Gol08}
Given decision problems $\Pi$ and $\Pi'$, i.e., problems whose answers
are either ``Yes'' or ``No'', $\Pi$ {\bf polynomial-time (Karp) reduces
to} $\Pi'$ if there is a polynomial-time computable function $f()$ such that for
any instance $x$ of $\Pi$, the answer to $\Pi$ for $x$ is ``Yes'' if and only 
if the answer to $\Pi'$ for $f(x)$ is ``Yes''.
\end{definition}

\begin{definition}
\cite{DF99}\footnote{
Note that this definition given here is actually Definition 6.1 in 
\cite{vRB+19}, which modifies that in \cite{DF99} to accommodate
parameterized problems with multi-parameter sets.
}
Given parameterized decision problems $\Pi$ and $\Pi'$, $\Pi$ {\bf parameterized
reduces to} $\Pi'$ if there is a function $f()$ which transforms instances
$\langle x, K \rangle$ of $\Pi$ into instances $\langle x', K' \rangle$ of
$\Pi'$ such that $f()$ runs in $f'(K)|x|^c$ time for some function $f'()$ and
constant $c$, $k' = g_{k'}(K)$ for each $k' \in K$ for some function $g_{k'}()$,
and for any instance $\langle x, K \rangle$ of $\Pi$, the answer to $\Pi$ for 
$\langle x, K\rangle$ is ``Yes'' if and only if the answer to $\Pi'$ for 
$f(\langle x, K \rangle)$ is ``Yes''.
\label{DefRedParam}
\end{definition}

\noindent
For technical reasons, our reducibilities operate on and establish unsolvability of
decision problems. Of the problems defined in Section \ref{SectForm}, only TeamEnvVerSF 
and TeamEnvVerST are currently decision problems.
However, this is not an issue if decision versions of the other non-decision problems are
formulated such that they can be solved efficiently using algorithms for those
non-decision problems, as such algorithms allow unsolvability results
for the decision problems to propagate to their associated non-decision problems
(see the appendix for details).

Our reductions will be from the following decision problems:

\vspace*{0.1in}

\noindent
{\sc Compact Deterministic Turing Machine computation} (CDTMC) \\ \cite[Problem AL3]{GJ79} \\
{\em Input}: A deterministic Turing Machine $M$ with state-set $Q$, tape alphabet
              $\Sigma$, and transition-function $\delta$, an input $x$, and a
              positive integer $k$. \\
{\em Question}: Does $M$ accept $x$ in a computation that uses at most $k$
              tape squares?

\vspace*{0.1in}

\noindent
{\sc Dominating set} \cite[Problem GT2]{GJ79} \\
{\em Input}: An undirected graph $G = (V, E)$ and a positive integer $k$. \\
{\em Question}: Does $G$ contain a dominating set of size $k$,
              i.e., is there a subset $V' \subseteq V$, $|V'| = k$, such
              that for all $v \in V$, either $v \in V'$ or there is at least one
              $v' \in V'$ such that $(v, v') \in E$?

\vspace*{0.1in}

\noindent
{\sc 3-Satisfiability} (3SAT) \cite[Problem LO2]{GJ79} \\
{\em Input}: A set $U$ of variables and a set $C$ of disjunctive clauses over $U$
              such that each clause $c \in C$ has $|c| = 3$. \\
{\em Question}: Is there a satisfying truth assignment for $C$?

\vspace*{0.1in}

\noindent
{\sc Clique} \cite[Problem GT19]{GJ79} \\
{\em Input}: An undirected graph $G = (V, E)$ and a positive integer $k$. \\
{\em Question}: Does $G$ contain a clique of size $k$,
              i.e., is there a subset $V' \subseteq V$, $|V'| = k$, such
              that for all $u, v \in V'$, $(u,v) \in E$?

\vspace*{0.1in}

\noindent
In addition to the four problems above, we shall also make use of 
{\sc Dominating set$^{PD3}$}, the version of {\sc Dominating set} in which the 
given graph $G$ is planar and each vertex in $G$ has degree at most 3. 
For each vertex $v \in V$ in a graph $G$, let the complete neighbourhood 
$N_C(v)$ of $v$ be the set composed of $v$ and the set of all vertices in $G$ 
that are adjacent to $v$ by a single edge, i.e., $v \cup \{ u ~ | ~ u ~ \in 
V ~ \rm{and} ~ (u,v) \in E\}$. 
We assume below for each instance of {\sc Clique}
and {\sc Dominating set} an arbitrary ordering on the vertices of $V$ such that
$V = \{v_1, v_2, \ldots, v_{|V|}\}$. Versions of these problems are only known to be
tractably unsolvable modulo the conjectures $P \neq NP$ and $FPT \neq W[1]$;
however, this is not a problem in practice as both of these conjectures are widely 
believed within computer science to be true \cite{DF13,For09}.

There are a variety of techniques for creating reductions from a problem $\Pi$ to a 
problem $\Pi'$ (Section 3.2 of \cite{GJ79}; see also Chapters 3 and 6 of \cite{vRB+19}).
One of these techniques, component design (in which a constructed instance of $\Pi'$ is 
structured as components that generate candidate solutions for the given instance
of $\Pi$ and check these candidates to see if any are actual solutions), will be used
extensively in the following subsections. Creating such reductions can be difficult. 
However, with a bit of forethought, a single reduction can often be constructed such that
it yields multiple results. For example, as algorithms that run in polynomial time
also run in fixed-parameter tractable time, provided the appropriate 
relationships specified by the functions $g_{k'}()$ in Definition \ref{DefRedParam} hold 
between parameters of interest, a polynomial-time reduction can imply both polynomial-time 
exact and fixed-parameter intractability results. Moreover, given a fp-tractability or 
intractability result, additional fp-tractability and intractability results can
often be derived using the following easily-proven lemmas.

\begin{lemma}
\cite[Lemma 2.1.30]{War99}
If problem $\Pi$ is fp-tractable relative to parameter-set $K$ then $\Pi$ is
       fp-tractable for any parameter-set $K'$ such that $K \subset K'$.
\label{LemPrmProp1}
\end{lemma}

\begin{lemma}
\cite[Lemma 2.1.31]{War99}
If problem $\Pi$ is fp-intractable relative to parameter-set $K$ then $\Pi$ is
       fp-intractable for any parameter-set $K'$ such that $K' \subset K$.
\label{LemPrmProp2}
\end{lemma}

\noindent
As we shall see later in this paper, these lemmas are of use use both in establishing the 
minimality of sources of intractability and in more easily establishing the parameterized 
complexity of a problem relative to all combinations of parameters in a given parameter-set,
i.e., performing a systematic parameterized complexity analysis \cite{War99} .

In the following subsections, we shall, in addition to our tractability and
intractability results, discuss the algorithms and reductions by which we 
derive these results. To highlight such details is admittedly unusual 
outside of the theoretical computer science literature. However, we believe
that even a basic appreciation of the ways in which a problem's mechanisms
interact in algorithms to yield tractability and in reductions to yield
intractability can give useful insights into
that problem, with the latter also providing ready targets for 
subsequent restrictions that may yield additional efficient algorithms. 

\subsection{Team / Environment Verification}

\label{SectResTeamEnvVer}

In a team / environment verification problem, both the robot team and the environment
are given as part of the problem input, so there is no opportunity to use an unspecified
part of the constructed instance of that problem in a candidate-solution generation 
component for the given instance of $\Pi$ in a reduction from $\Pi$ to that problem. 
Rather, the operation of 
the given robot team in its given environment must be structured to effectively simulate
the mechanisms in $\Pi$ itself. Such is the case in the following result for TeamEnvVerST.

\begin{description}
\item[{\bf Result A.ST.1}] \cite[Result A]{WV18_SI}: TeamEnvVerST is not polynomial-time exact \linebreak solvable.
\end{description}

\noindent
This result was proved by a polynomial-time reduction from CDTMC to \linebreak TeamEnvVerST,
in which an instance of TeamEnvVerST is constructed such that the environment $E$
corresponds to the DTM tape, the square-types in $E_T$ correspond to the tape alphabet 
$\Sigma$, and the single robot in $T$ corresponds to the DTM $M$'s deterministic controller
and tape read/write head. The operation of this robot in $E$ thus effectively simulates
the computation of $M$ on $x$.

A most useful aspect of this reduction is that as the TM read/write head only views a single
tape-square at a time, the robot in $T$ has sensory radius $r = 0$. Given this, we can
easily modify the reduction sketched above to show the polynomial-time intractability of 
TeamEnvVerSF --- as $r = 0$, we can simulate the tape-squares in $E_T$ using point-fields 
with source-value and decay 1 (i.e., the field-quantity in each such point-field is only 
detectable in the square at which the point-field is positioned) and thus trivially
modify the visual sensing robot to create an equivalent scalar-field sensing robot. This
yields the following result.

\begin{description}
\item[{\bf Result A.SF.1}:] TeamEnvVerSF is not polynomial-time exact solvable.
\end{description}

\noindent
Given this, we will restrict our four scalar-field
design problems to operate in polynomial time; this is denoted adding the superscript
$fast$ to the problem name, e.g., ContDesLSSF$^{fast}$. This notion of time-bounded robot 
team operation, introduced in \cite{WV18_SI} as {\bf $(c_1,c_2)$-completability}. 
requires that each robot team complete its task within $c_1|E|^{c_2}$ timesteps for 
constants $c_1$ and $c_2$. We here make two modifications to this notion. First, as was
done in \cite{TW19}, we broaden the timestep bound to $c_1(|E| + |Q|)^{c_2}$ to 
accommodate FSR that make a number of internal state-changes without moving. Second, 
as was done in \cite{WdH+Sub}, $c_1$ and $c_2$ are no longer part of problem inputs but 
are rather fixed beforehand, which allows us to avoid certain technical issues.
To ensure generous but nonetheless low-order polynomial runtime bounds,
we will assume that $c_1 = 10$ and $c_2 = 3$.

The above demonstrates that even though scalar-field sensing seems simpler than 
visual sensing, it still has the same degree of computational intractability when it
comes to controller / environment verification. Moreover, we have also seen that the
two types of sensing are equivalent (in the sense that robots with one type of
sensing can simulate robots with the other type) when $|T| = 1$ and $r = 0$ given 
appropriately restricted point fields. That being said, as we shall see in the
following subsections, this equivalence most decidedly does not hold in general.

\subsection{Controller Design by Library Selection} 

\label{SectResContDesLS}

Unlike the team / environment verification problems examined in Section 
\ref{SectResTeamEnvVer}, a controller design by library selection problem
gives the general form but not the exact structure of the robot controller as part of 
the problem input. Hence, the transition-template library $L$ and the values of $|Q|$ and 
$d$ can be specified to create a candidate-solution generation component and the
environment can be specified to serve as a candidate-solution checking component
for the given instance of $\Pi$ in a reduction from $\Pi$ to our controller design by
library selection problem.

\begin{figure}[p]
\begin{center}
\includegraphics[width=5.0in]{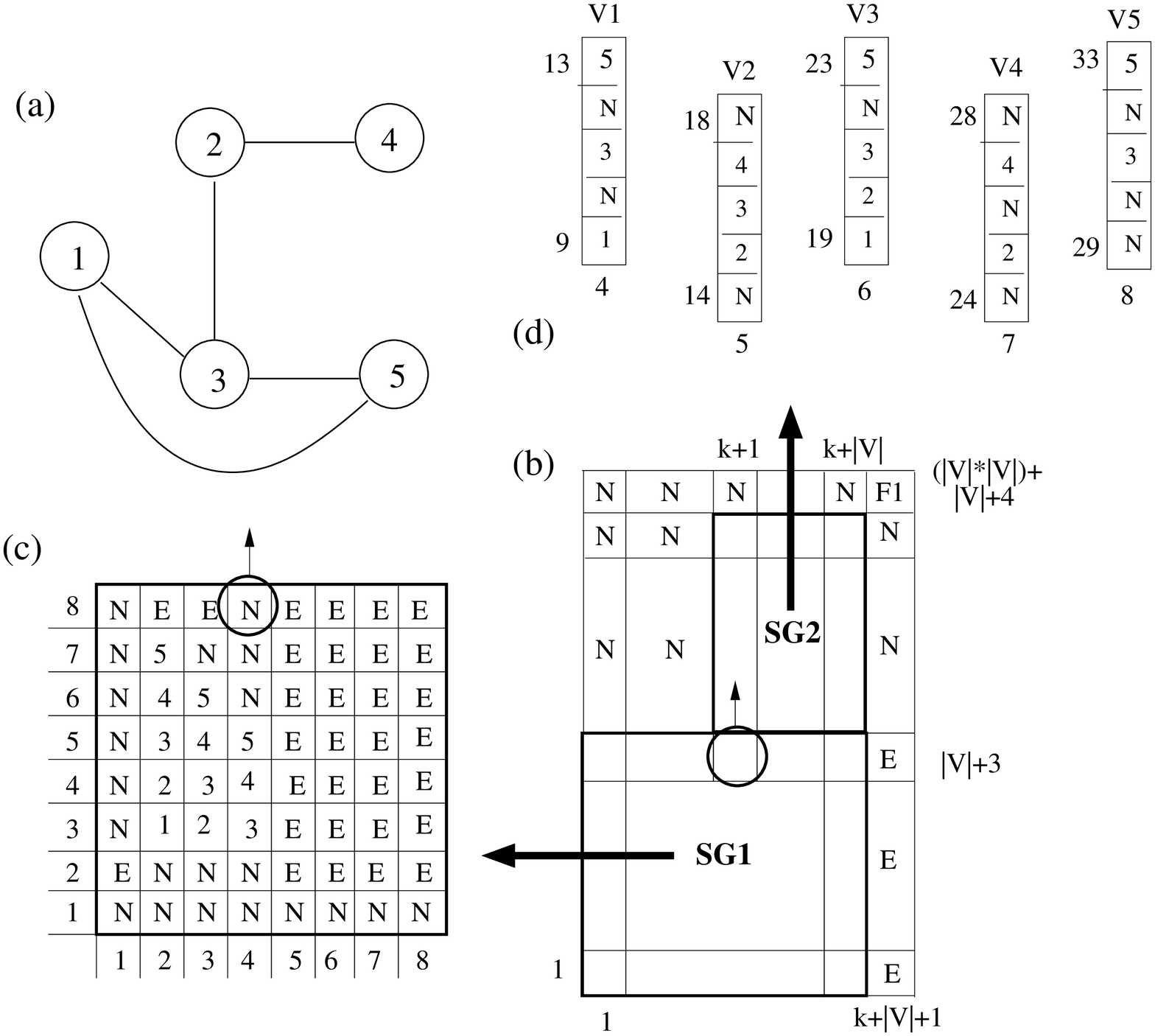}
\end{center}
\vspace*{0.1in}
\caption{A sample graph and its associated team-environment (Modified from Figure 2 in
          \cite{WV18_AT}). a) An undirected graph on five vertices. b) The 
          general structure of the team-environment $E$ constructed by the reduction in 
          the proof of Result B in \cite{WV18_SI} 
          from a given instance $(G = (V,E), k)$ of {\sc Dominating set}. Note that this
          environment is not drawn to scale but rather to illustrate the large-scale
          features of the environment. c) The subgrid SG1 constructed from the graph in
          part (a) when $k = 3$. d) The column-fragments of the subgrid SG2 constructed
          from the graph in part (a) that correspond to the vertex-neighborhoods of the
          vertices in that graph. }
\label{FigRed}
\end{figure}

Such is the case in previous results for controller design that were proved relative to a 
problem ContDes$^{fast}$ \cite{WV18_SI}, in which controller transitions were designed
from scratch by including $|f|$ as part of the problem input. The polynomial-time
exact intractability result for this problem \cite[Result B]{WV18_SI} used a
reduction from {\sc Dominating set} to create a somewhat complex environment
for a team composed of a single-state robot (see Figure \ref{FigRed}(b)). In order to 
force the transitions in such a robot to encode a candidate dominating set of size
$k$ in the given graph $G$, the robot had to navigate from the southwestmost
corner of the environment to the top of the $(k+1)$st column in subgrid SG1 (see Figure 
\ref{FigRed}(c)). From there, the robot navigated the $|V|$ columns of subgrid SG2 (see 
Figure \ref{FigRed}(d)), where each column represented the vertex neighbourhood of a 
particular vertex in $G$ and the robot could progress eastward from one column to the next
if and only if that robot had a transition corresponding to a vertex in the neighbourhood 
encoded in the first column. Subgrid SG2 thus checked if the robot encoded an actual
dominating set of size $k$ in $G$, such that the robot could enter the northeastmost square 
of the environment and build the requested structure there if and only if the $k$ 
east-moving transitions in the robot encoded a dominating set of size $k$ in $G$.

By using an appropriate transition template library $L$, we can simplify
the reduction sketched above as in \cite{War22} such that we no longer need the northwest $e_N$-based
or SG1 subgrids in the environment or the restriction on $|f|$ in the problem input. This 
yields the following result.

\begin{description}
\item[{\bf Result B.ST.1} (Modified from \protect{\cite[Result B]{War22}}):] ContDesLSST$^{fast}$ is not \linebreak polynomial-time exact solvable.
\end{description}

\noindent
As $|T| = 1$ and $r = 0$ in this reduction, we can easily employ the techniques
sketched in the previous subsection to derive our next result.

\begin{description}
\item[{\bf Result B.SF.1}:] ContDesLSSF$^{fast}$ is not polynomial-time exact solvable.
\end{description}

Given the above, it is natural to wonder under which restrictions efficient solvability
is possible. Let us first consider fixed-parameter intractability results, starting with
problem ContDesLSST$^{fast}$. Our first result follows directly from the reduction by
which we proved polynomial-time intractability.

\begin{description}
\item[{\bf Result B.ST.2}:] 
$\la |T|, h, |Q|, d, |f|, r, |X|\ra$-ContDesLSST$^{fast}$ is fp-intractable.
\end{description}

\begin{figure}[p]
\begin{center}
\includegraphics[width=2.8in]{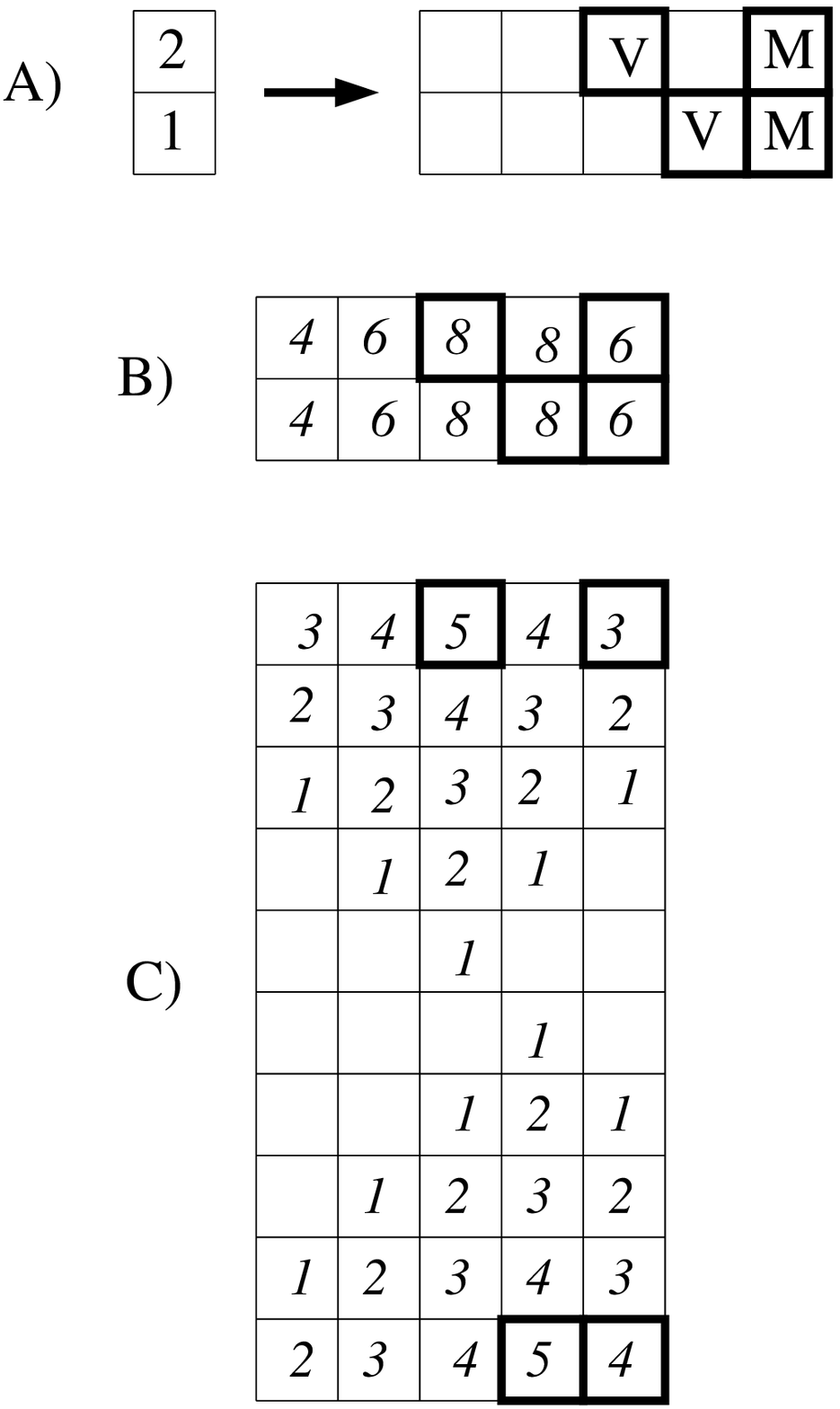}
\end{center}
\vspace*{0.1in}
\caption{Reducing the number of square-types and scalar-field types in complex environments. 
          a) Replacing multiple vertex-symbols with pairs of vertex- and marker-symbols 
          (see main text and proof of Result B.ST.3 for details). This is done relative to 
          a vertex-neighbourhood containing $v_1$ and $v_2$ when $|V| = 4 $. b) Interference
          problems with respect to field-quantity $fq_V$ created by straightforward 
          replacement of vertex- and marker-symbols
          with vertex and marker point-fields (vertex-neighborhood and vertices as in (a)).
          Note that courtesy of interference, both of these vertices in these rows would be misperceived as
          non-existent vertex $v_6$ at the marker-field positions. c) Proper replacement of 
          vertex- and marker-symbols with vertex and marker point-fields under appropriate 
          row-separation (i.e., $2|V|$ separating rows) to avoid interference 
          (vertex-neighbourhood and vertices as in (a)) (see main text
          and proof of Result B.SF.4 for details). 
          Vertex and marker positions in all subfigures are indicated by boldfaced boxes.
}
\label{FigRedEnvFix}
\end{figure}

\noindent
We can in turn restrict $E_T$ at the cost of unrestricted $r$ by replacing each
occurrence of a vertex-symbol in a column of the environment with a $(|V| +1)$-length
row, in which the eastmost marker-symbol in the row indicates that a vertex is present
in that column's neighbourhood and the position of a vertex-symbol in one of the
first $|V|$ squares indicates which vertex it is (see Figure \ref{FigRedEnvFix}(a)).

\begin{description}
\item[{\bf Result B.ST.3}:] 
$\la |T|, h, |Q|, d, |f|, |E_T|, |X|\ra$-ContDesLSST$^{fast}$ is fp-intractable.
\end{description}

\noindent
Result B.ST.2 also holds for problem ContDesLSSF$^{fast}$ courtesy of the reduction
in the proof of Result B.SF.1. Moreover, if we in turn modify that reduction to reduce 
from {\sc Dominating set$^{PD3}$} (in which each vertex appears in at most four vertex 
neighborhoods, those of its three neighbors and its own) instead of {\sc Dominating set}, 
we get a second result for free.

\begin{description}
\item[{\bf Result B.SF.2}:] 
$\la |T|, h, |Q|, d, |f|, |X|\ra$-ContDesLSSF$^{fast}$ is fp-intractable.
\item[{\bf Result B.SF.3}:] 
$\la |T|, h, |Q|, |f|, |q_E|, |X|\ra$-ContDesLSSF$^{fast}$ is fp-intractable.
\end{description}

\noindent
The second of our fp-intractability results for ContDesLSST$^{fast}$, however,
does not translate over so directly. This is because, for the first time, we
must use point-fields whose influence propagates beyond the square in which
they are placed --- indeed, to encode the vertex-symbols, we need a corresponding
vertex point-field that is detectable up to $|V|$ squares from its source (Figure 
\ref{FigSFSChar}(c)). Straightforward implementations of such point-fields in adjacent rows 
of the environment can interfere to destroy the encodings of vertices associated with these 
point-fields (Figure \ref{FigRedEnvFix}(b)). This interference can be mitigated by increasing
the separations between the rows in which vertex point-fields occur so that these 
point-fields do not interfere at all (see Figure \ref{FigRedEnvFix}(c)), at the cost of
increasing the size of the resulting environment.

\begin{description}
\item[{\bf Result B.SF.4}:] 
$\la |T|, h, |Q|, d, |f|, |S|, |X|\ra$-ContDesLSSF$^{fast}$ is fp-intractable.
\end{description}

Let us now consider fixed-parameter tractability. In this case, we once again have 
equivalence of results because the proofs of these results rely only on the combinatorics of
combining states and transitions from transition template library $L$ to create 
controllers, and these combinatorics are identical for problems ContDesLSST$^{fast}$ and
ContDesLSSF$^{fast}$.

\begin{description}
\item[{\bf Result B.ST.4}:] 
$\la |Q|, |L|\ra$-ContDesLSST$^{fast}$ is fp-tractable.
\item[{\bf Result B.SF.5}:] 
$\la |Q|, |L|\ra$-ContDesLSSF$^{fast}$ is fp-tractable.
\end{description}

\noindent
Note that both of these results collapse to fp-tractability relative to $|L|$ alone
when $|Q| = 1$. 

A summary of all of our fixed-parameter results derived in this subsection is given in Table
\ref{TabPrmResContDesLS}. Given our fp-intractability results at this time, neither of
our fp-tractability results are minimal in  the sense described in Section
\ref{SectResDesSolv}.

\begin{table}[t]
\caption{A Detailed Summary of Our Fixed-parameter Results for Controller Design by Library
          Selection.  Each column in this table is a result which holds relative to the 
          parameter-set consisting of all parameters with a @-symbol in that column. If the
          result holds when a particular parameter has a constant value $c$, that is 
	  indicated by $c$ replacing @ for that parameter in that result's column.
          Note that within each result-group, intractability results are first and 
          tractability results (shown in bold) are last. 
}
\label{TabPrmResContDesLS}
\vspace*{0.1in}
\centering
\begin{tabular}{ | p{0.6cm}  || p{0.7cm}  p{0.7cm}  p{0.7cm} || p{0.7cm}  p{0.7cm} p{0.7cm} p{0.8cm} | }
\hline
& \multicolumn{3}{c ||}{Square-type (ST)}
 & \multicolumn{4}{c |}{Scalar-field (SF)} \\
& ST.2 & ST.3 & {\bf ST.4} 
& SF.2 & SF.3 & SF.4 & {\bf SF.5} \\
\hline\hline
$|T|$       & 1  & 1  & {\bf --}
            & 1  & 1  & 1  & {\bf --} \\
\hline
$h$         & 1  & 1  & {\bf --}
            & 1  & 1  & 1  & {\bf --} \\
\hline
$|Q|$       & 1  & 1  & {\bf @ }
            & 1  & 1  & 1  & {\bf @ } \\
\hline
$d$         & @  & @  & {\bf --}
            & @  & -- & @  & {\bf --} \\
\hline
$|f|$       & 1  & 3  & {\bf --}
            & 1  & 1  & 3  & {\bf --} \\
\hline
$r$         & 0  & -- & {\bf --}
            & N/A & N/A & N/A & {\bf N/A} \\
\hline\hline
$|L|$       & -- & -- & {\bf @ }
            & -- & -- & -- & {\bf @ } \\
\hline\hline
$|E|$       & -- & -- & {\bf --}
            & -- & -- & -- & {\bf --} \\
\hline
$|E_T|$     & -- & 5  & {\bf --}
            & N/A & N/A & N/A & {\bf N/A} \\
\hline
$|S|$       & N/A & N/A & {\bf --}
            & -- & -- & 5  & {\bf --} \\
\hline
$|S_E|$     & N/A & N/A & {\bf --}
            & -- & -- & -- & {\bf --} \\
\hline
$|q_E|$     & N/A & N/A & {\bf --}
            & -- & 4  & -- & {\bf --} \\
\hline
$|X|$       & 1  & 1  & {\bf --}
            &  1 &  1 &  1 & {\bf --} \\
\hline
\end{tabular}
\end{table}

The above demonstrates that results derived by utilizing perception at a distance 
relative to visual sensing (i.e., $r > 0$) can still hold relative scalar-field sensing; 
however, this comes at the cost of larger environments to avoid interference between 
separate fields based on the same field-quantity. This problem with interference will
re-occur in many of the results described in the following subsections.
 
\subsection{Team Design by Library Selection} 

\label{SectResTeamDesLS}

Team design by library selection is unique among the design problems considered in this 
paper, in that it is the only one that is currently known to have a restricted case that is
solvable in polynomial time --- namely, for homogeneous robot teams whose members
all have the same controller, i.e., $h = 1$.

\begin{description}
\item[{\bf Result C.ST.1}] \cite[Result A]{War19}: TeamDesLSST$^{fast}$ is
polynomial-time exact solvable when $h = 1$.
\end{description}

\noindent
This does not, however, hold when teams are heterogeneous, i.e., $h > 1$.
 
\begin{description}
\item[{\bf Result C.ST.2}] \cite[Result B]{War19}: TeamDesLSST$^{fast}$ is not exact
polynomial-time \linebreak solvable when $h > 1$.
\end{description}

\noindent
The reduction underlying this result exploits the fact that the general form but
not the exact structure (in terms of individual member FSRs) of the team is given as
part of the problem input. Hence, analogous to the case with controller design by
library selection in Section \ref{SectResContDesLS}, the FSR library $L$ and 
the value of $|T|$ can be specified to create a candidate-solution generation component and
the environment can be specified to serve as a candidate-solution checking component
for the given instance of a suitable problem, which is once again {\sc Dominating set}.

The reduction works as follows \cite{War19}: for a given instance of {\sc Dominating set}, 
we construct an environment 
like that in part (a) of Figure \ref{FigEnvType} such that a team of $k + 1$ robots will be 
able to co-operatively construct a single-square structure at the square with type $e_T$
if and only if there is a dominating
set of size at most $k$ in the graph in the given instance of {\sc Dominating set}.
The robots are initially positioned in the top-left corner of the movement-track and move 
in a counter-clockwise fashion around this track. A functional robot team of $|T| = k + 1$
robots, each with sensory radius $r = 1$, selected from $L$ consists of at most $k$ 
``vertex neighbourhood'' robots which 
attempt to fill in a scaffolding of squares in the center of the central line of squares and
at least one ``checker'' robot which determines if all squares in this scaffolding are 
eventually filled in (which only occurs if the vertex-neighborhood robots encode a 
dominating set of size at most $k$ in $G$) and, if so, places the single square 
corresponding to the requested structure at the square with type $e_T$.
Note that as the two types of robots in this reduction only communicate
indirectly via additions to and the sensing of the central scaffolding, this
is a good example of communication by stigmergy \cite{BDT99}.

The reduction above in turn yields three fixed-parameter intractability results.
The first of these follows directly from this reduction, and the second 
by modifying this reduction to collapse a set of short-triggering-formula 
transitions between a pair of states into a single long-triggering-formula transition 
between those states.

\begin{description}
\item[{\bf Result C.ST.3}] \cite[Result E]{War19}: 
$\la |T|, h, |Q|, |f|, r, |X|\ra$-TeamDesLSST$^{fast}$ is \linebreak fp-intractable.
\item[{\bf Result C.ST.4}:] 
$\la |T|, h, |Q|, d, r, |X|\ra$-TeamDesLSST$^{fast}$ is fp-intractable.
\end{description}

\vspace*{0.1in}

\noindent
The third of our fp-intractability results for TeamDesLSST$^{fast}$ makes two modifications
to the reduction in the proof of Result C.ST.2. First, each of the vertex-symbols $e_{Vi}$ 
in the southmost row of $E$ is replaced with a column of height $\lceil \log_2 |V|\rceil$ 
giving a binary encoding of $i$ using symbols $e_0$ and $e_1$; as such, this is an 
alternate scheme to that based on vertex- and marker-symbols used in the proof of 
Result B.ST.3 for restricting $|E_T|$. Second, the tradeoff between transition
trigger-formula length and number of outgoing transitions per state described above is
applied.

\begin{description}
\item[{\bf Result C.ST.5}:] 
$\la |T|, h, |Q|, d, |E_T|, |X|\ra$-TeamDesLSST$^{fast}$ is fp-intractable.
\end{description}

\noindent
We also have the following fp-tractability results.

\begin{description}
\item[{\bf Result C.ST.6}] \cite[Result G]{War19}: 
$\la |T|, |L|\ra$-TeamDesLSST$^{fast}$ is fp-tractable.
\item[{\bf Result C.ST.7}] \cite[Result I]{War19}: 
$\la |T|, |Q|, r, |E_T|\ra$-TeamDesLSST$^{fast}$ is fp-tractable.
\end{description}

\noindent
The first result is analogous to that in Result B.ST.4, in that it relies only on the
combinatorics of making $|T|$ choices from $L$ to make teams. The second result
in turns modifies the first by using combinatorics based on $|Q|$, $r$, and $|E_T|$ 
to bound the number of possible behaviorally-distinct FSRs in $L$.

Let us now consider to what extent the results above for TeamDesLSST$^{fast}$
transfer over to TeamDesLSSF$^{fast}$. As the algorithm underlying Result C.ST.1 relies 
only on the timestep-bound on successful team task completion imposed by $(c_1,c_2)$ 
completability, the same result holds for TeamDesLSSF$^{fast}$. 

\begin{description}
\item[{\bf Result C.SF.1}] \cite[Result A.1]{WV21}: TeamDesLSSF$^{fast}$ is exact polynomial-time solvable when $h = 1$.
\end{description}

\noindent
An analogue of the polynomial-time exact intractability in Result C.ST.2 when teams are 
heterogeneous also holds. 

\begin{description}
\item[{\bf Result C.SF.2}] \cite[Result A.2]{WV21} TeamDesLSSF$^{fast}$ is not exact polynomial-time \linebreak solvable 
when $h > 1$.
\end{description}

\noindent
There are, however, complications due to point-field interference. Oddly
enough, this is not due to environmental point-fields. As $r = 1$ in the reduction
above, given the structure of the environment shown in Figure \ref{FigEnvType}(a), all
squares in both the outer ring and the central scaffolding can readily have their
symbols replaced with corresponding point-fields with source-value 1 and decay 0.5. 
Provided appropriate care is taken (see Figure \ref{FigSFSChar}(d)), no robot traveling 
around the movement-track will ever misperceive any field-quantity related to those squares. 

\begin{figure}[p]
\begin{center}
\includegraphics[width=3.5in]{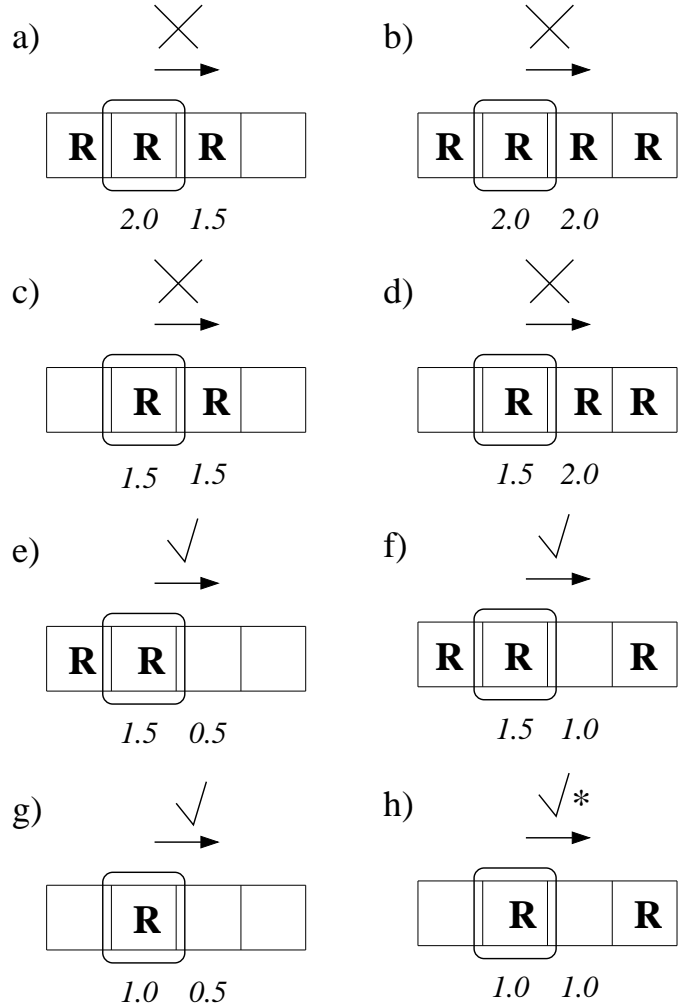}
\end{center}
\vspace*{0.1in}
\caption{Scalar-field sensing of adjacent robots on a linear movement-track
          (Figure 2 in \cite{WV21}). This figure 
          shows the eight possible situations when a scalar-field sensing robot on a linear
          movement-track (indicated by the bold-faced rounded square) is trying to determine
          if there is already another robot in a square which it is considering moving into.
          Subfigures (a)-(d) and (e)-(h) depict situations in which movement is illegal 
          and legal, respectively.
          For simplicity, we only show the case of eastward movement, though movement
          in all other directions are analogous. We also collapse movement in the middle
          of tracks and into corners as these too are analogous; the rightmost square 
          in each situation is thus either on the same line as the others squares or one
          above the third 
          square. In each situation, we show the values of field-quantity $fq_{robot}$
          in both the square currently occupied by the robot and the square which that robot
          is considering moving into.
}
\label{FigAdjRobSF}
\end{figure}

The problem this time (summarized succinctly in Figure \ref{FigAdjRobSF}) is interference 
generated by the point-fields associated with the robots themselves. The visual sensing 
robots in the reduction in the proof of Result C.ST.2 need only ensure that there is no robot
in a square they want to move into to safely move into that square. Scalar-field
sensing robots cannot do this --- indeed, as we cannot be sure what other robots
are nearby and in what positions, Figure \ref{FigAdjRobSF} shows that it is difficult even
on a linear movement-track for
a scalar-field sensing robot to determine whether or not there is a robot already in a 
square to which it wishes to move. Figures \ref{FigAdjRobSF}(e)-(h) suggest that when
$fval(fq_{robot}, \leq, 1.5)$ is $True$, legal movement to be possible. However, this alone
is not sufficient as it would allow illegal movement in the situations shown in Figures 
\ref{FigAdjRobSF}(c) and (d). We can add a gradient-sensing predicate
in the direction $dir$ in which movement is wanted, but there are again problems. If we
use $fgrd(fq_{robot}, \geq, dir)$, we allow not only the legal movement in Figures 
\ref{FigAdjRobSF}(e)-(h) but also the illegal movement in Figures \ref{FigAdjRobSF}(b) and 
(c); on the other hand, if we use use $fgrd(fq_{robot}, >, dir)$, we disallow the illegal 
movement in Figures \ref{FigAdjRobSF}(b) and (c) at the cost of also disallowing the
legal movement in Figure \ref{FigAdjRobSF}(h). There does not seem to be a perfect solution.
Hence, we here adopt the second alternative, consoling ourselves with the thought
that, as the environments constructed in our reduction are big enough that there will 
always be at least one pair of robots separated by two or more spaces on the movement 
track, any situation like that shown in Figure \ref{FigAdjRobSF}(h) eventually becomes one
of the situations in Figures \ref{FigAdjRobSF}(e) or (g) that are covered by our adopted 
solution.
 
The reduction above in turn yields several fixed-parameter intractability results. The
first pair of fp-intractability results follows directly from this reduction, with the 
second result in the pair using the tradeoff between transition trigger formula length and 
number of transitions per state sketched for Result C.ST.4.

\begin{description}
\item[{\bf Result C.SF.3}] \cite[Result A.3]{WV21}: $\la |T|, h, |Q|, |f|, |X|\ra$-TeamDesLSST$^{fast}$ is \linebreak fp-intractable.
\item[{\bf Result C.SF.4}] \cite[Result A.4]{WV21}: $\la |T|, h, |Q|, d, |X|\ra$-TeamDesLSST$^{fast}$ is \linebreak fp-intractable.
\end{description}

\noindent
The second pair of fp-intractability results modifies the reduction in the proof of
Result C.SF.2  along the lines in the proof of Result B.SF.4 to both (1) replace each of 
the vertex point-fields $s_{Vi}$ in the lower row with a column of height $|V| + 1$ and 
appropriately-positioned vertex and marker point-fields  and (2) place $2|V|$ blank columns
between each pair of vertex-columns to prevent vertex point-field interference. Once again,
the second result in the pair uses the tradeoff between transition trigger formula length 
and number of transitions per state sketched above.

\begin{description}
\item[{\bf Result C.SF.5}] \cite[Result A.5]{WV21}: $\la |T|, h, |Q|, |f|, |S|, |X|\ra$-TeamDesLSST$^{fast}$ is \linebreak fp-intractable.
\item[{\bf Result C.SF.6}] \cite[Result A.6]{WV21}: $\la |T|, h, |Q|, d, |S|, |X|\ra$-TeamDesLSST$^{fast}$ is \linebreak fp-intractable.
\end{description}

\noindent
We also have the following fixed-parameter tractability result.

\begin{description}
\item[{\bf Result C.SF.7}] \cite[Result A.7]{WV21}: $\la |T|, |L|\ra$-TeamDesLSST$^{fast}$ is fp-tractable.
\end{description}

\noindent
This result follows directly from the algorithm in the proof of Result C.ST.6.
Unfortunately, we do not have an analogue for TeamDesLSSF$^{fast}$ of Result C.ST.7
as there does not appear to be a way to combinatorially bound the number of
possible behaviorally-distinct scalar-field sensing FSR using the problem-aspects
in Table \ref{TabPrm}.

A summary of all of our fixed-parameter results derived in this subsection is given in Table
\ref{TabPrmResTeamDesLS}. Given our fp-intractability results at this time, none of
our fp-tractability results are minimal in  the sense described in Section
\ref{SectResDesSolv}. 

The above demonstrates that problems with point-field interference can occur with even the 
simplest non-trivial point-fields that are perceptible out to only Manhattan distance one 
from the point-field source if point-fields are dynamic (i.e.,
associated with robots) rather than static (i.e., associated with the environment).
Nonetheless, these problems are resolvable. As we shall see in the next section,
they are also resolvable (at the cost of more complex robots and environments, including
our first use of edge-fields) in cases involving perception at Manhattan distances 
arbitrarily larger than one.

\begin{table}[t]
\caption{A Detailed Summary of Our Fixed-parameter Results for Team Design by Library
          Selection. This table is interpreted as described in the caption of
          Table \ref{TabPrmResContDesLS}.
}
\label{TabPrmResTeamDesLS}
\vspace*{0.1in}
\centering
\begin{tabular}{ | p{0.6cm}  || p{0.7cm} p{0.7cm}  p{0.7cm}  p{0.8cm} p{0.8cm} ||  p{0.7cm} p{0.7cm} p{0.7cm} p{0.7cm} p{0.8cm} | }
\hline
& \multicolumn{5}{c ||}{Square-type (ST)}
 & \multicolumn{5}{c |}{Scalar-field (SF)} \\
& ST.3 & ST.4 & ST.5 & {\bf ST.6} & {\bf ST.7} 
& SF.3 & SF.4 & ST.5 & ST.6 & {\bf SF.7} \\
\hline\hline
$|T|$       & @  & @  & @  & {\bf @ } & {\bf @ }
            & @  & @  & @ & @ & {\bf @ } \\
\hline
$h$         & @  & @  & @  & {\bf --} & {\bf --}
            & @  & @  & @ & @ & {\bf --} \\
\hline
$|Q|$       & 3  & 3  & 3  & {\bf --} & {\bf @ }
            & 3  & 3  & 3 & 3 & {\bf --} \\
\hline
$d$         & -- & 7  & 7  & {\bf --} & {\bf --}
            & -- & 7  & -- & 7  & {\bf --} \\
\hline
$|f|$       & 16 & -- & -- & {\bf --} & {\bf --}
            & 27 & -- & 31 & -- & {\bf --} \\
\hline
$r$         & 1  & 1  & -- & {\bf --} & {\bf @ }
            & N/A & N/A & N/A & N/A & {\bf N/A} \\
\hline\hline
$|L|$       & -- & -- & -- & {\bf @ } & {\bf --}
            & -- & -- & -- & -- & {\bf @ } \\
\hline\hline
$|E|$       & -- & -- & -- & {\bf --} & {\bf --}
            & -- & -- & -- & -- & {\bf --} \\
\hline
$|E_T|$     & -- & -- & 13 & {\bf --} & {\bf @ }
            & N/A & N/A & N/A & N/A & {\bf N/A} \\
\hline
$|S|$       & N/A & N/A & N/A & {\bf N/A} & {\bf N/A}
            & -- & -- & 13 & 13 & {\bf --} \\
\hline
$|S_E|$     & N/A & N/A & N/A & {\bf N/A} & {\bf N/A}
            & -- & -- & -- & -- & {\bf --} \\
\hline
$|q_E|$     & N/A & N/A & N/A & {\bf N/A} & {\bf N/A}
            & -- & -- & -- & -- & {\bf --} \\
\hline
$|X|$       & 1  & 1  & 1  & {\bf --} & {\bf --}
            & 1  & 1  & 1  & 1  & {\bf --} \\
\hline
\end{tabular}
\end{table}

\subsection{Environment Design}

\label{SectResEnvDes}

Unlike all problems examined so far in this paper, an environment design problem
gives the general form but not the exact structure of the environment as part of 
the problem input. Hence, the environment size $|E|$ and the environment-square contents
(either square-type set $E_T$ or field-set $S$ in the case of EnvDesST$^{fast}$ or
EnvDesSF$^{fast}$, respectively) can be specified to create a candidate-solution generation
component and the robot team $T$ can be specified to serve as a candidate-solution checking
component for the given instance of $\Pi$ in a reduction from $\Pi$ to our environment
design problem.

\begin{figure}[p]
\begin{center}
\includegraphics[width=4.7in]{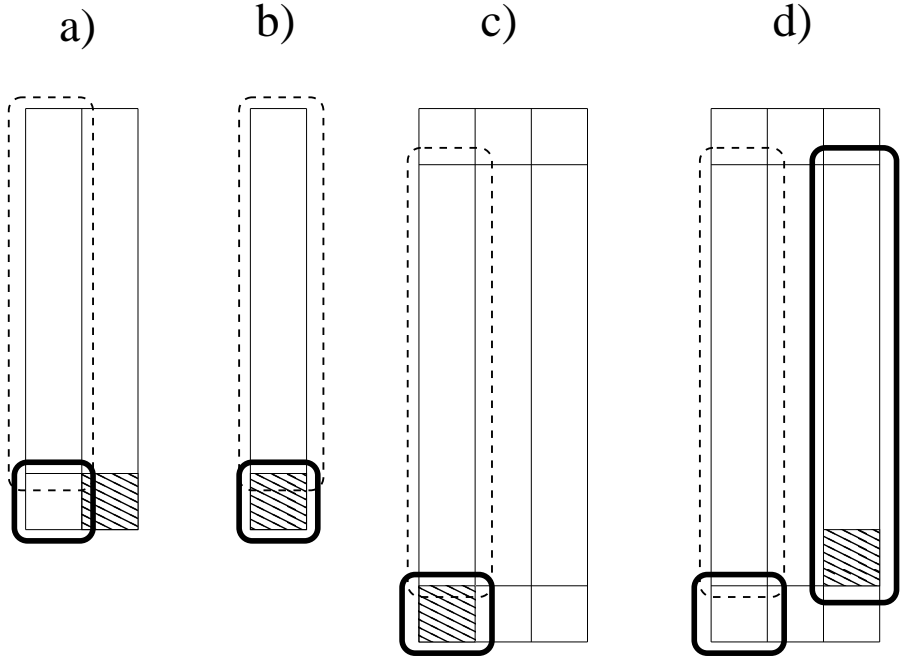}
\end{center}
\vspace*{0.1in}
\caption{General structures of environments used to encode candidate solutions in
          environment design problems (Modified from Figure 3 in \cite{WV21}). a) 
          Environment structure for reductions to
          EnvDesST$^{fast}$. b) Environment structure for reductions to EnvDesSF$^{fast}$
          involving single-robot teams (Results D.SF.1--D.SF.3). c) Environment structure
          for reductions to EnvDesSF$^{fast}$ involving single-robot teams (Results D.SF.4 
          and D.SF.5). d) Environment structure for reduction to EnvDesSF$^{fast}$ involving
          multi-robot teams (Results D.SF.6). In each environment,
          the region encoding the candidate solution is indicated by a dashed box,
          the initial positioning of the robots is indicated by boldface boxes, and the
          position of the requested structure is indicated by a cross-hatched box. 
}
\label{FigEnvDesEnv}
\end{figure}

This was done in \cite{WV18_SI} for EnvDesST$^{fast}$ using reductions from 3SAT and 
{\sc Clique} in which the environment had the two-column structure shown in Figure 
\ref{FigEnvDesEnv}(a). In the reductions from 3SAT, the candidate solution to the given 
instance of 3SAT was encoded in the northmost $|V|$ squares of the first column and the sole
robot in $T$ was in the southmost square of that column. If this robot determined that
the candidate solution so encoded was an actual solution to the given instance of 
3SAT, the robot moved one square to the east and created the requested structure $X$ in
the southmost square of the second column. This yielded the following result.

\begin{description}
\item[{\bf Result D.ST.1}] \cite[Result C]{WV18_SI}: EnvDesST$^{fast}$ is not exact
polynomial-time solvable.
\end{description}

\noindent
This reduction in turn gave the first two parameterized intractability results, the second 
of these by modifying the reduction to apply the tradeoff between transition trigger-formula
length and the number of outgoing transitions per state sketched in Section
\ref{SectResTeamDesLS}.

\begin{description}
\item[{\bf Result D.ST.2}] \cite[Result H]{WV18_SI}: 
$\la |T|, |f|, d, |E_T|, |X|\ra$-EnvDesST$^{fast}$ is fp-intractable.
\item[{\bf Result D.ST.3}] \cite[Result I]{WV18_SI}: 
$\la |T|, |Q|, d, |E_T|, |X|\ra$-EnvDesST$^{fast}$ is fp-intractable.
\end{description}

\noindent
The remaining two previous parameterized intractability results instead used a
reduction from {\sc Clique} in which the northmost $k$ squares in the first
column encoded the vertices in a candidate clique of size $k$. The number of
states in the single checking robot was also bounded by a function of $k$ as the number of 
states required to implement the required solution-checks ($k$ states to verify that the 
vertices in the encoded candidate solution are all different and $k(k - 1)/2$ 
states to verify that there is an edge between each pair of vertices in the candidate 
solution in the graph $G$ in the given instance of {\sc Clique}) was also a function 
of $k$.

\begin{description}
\item[{\bf Result D.ST.4}] \cite[Result J]{WV18_SI}: 
$\la |T|, |Q|, |f|, r, |E|, |X|\ra$-EnvDesST$^{fast}$ is fp-intractable.
\item[{\bf Result D.ST.5}] \cite[Result K]{WV18_SI}: 
$\la |T|, |Q|, d, r, |E|, |X|\ra$-EnvDesST$^{fast}$ is fp-intractable.
\end{description}

\noindent
The second of these results used the tradeoff between transition trigger-formula
length and number of outgoing transitions per state sketched above. There was a single
previous parameterized tractability result,

\begin{description}
\item[{\bf Result D.ST.6}] \cite[Result N]{WV18_SI}: 
$\la |E|, |E_T|\ra$-EnvDesST$^{fast}$ is fp-tractable.
\end{description}

\noindent
This result was based on the brute-force enumeration of all possible environments and 
checking of robot team operation in each of those environments.

The intractability results above do not translate over to problem EnvDesSF$^{fast}$.
This is because the complex sensing at a distance between pairs of elements in
the candidate solution encoded in $E$ required by a reduction from {\sc Clique}
is impossible to implement with scalar-field sensing. That being said, the
scalar-field sensing at a distance illustrated in Figure \ref{FigSFSChar}(c) can
work for a reduction from {\sc Dominating set} using an environment such as that
in Figure \ref{FigEnvDesEnv}(b). By using different point-fields for each vertex
in given graph $G$, a suitably configured scalar-field sensing robot positioned at the 
southmost end of the single column in the environment can step through a set of
single-element checks to see if the encoded candidate solution contains at least
one vertex in each vertex neighborhood in given graph $G$. The size of the environment
ensures that there are at most $k$ vertices in the candidate solution, and we don't
care if the same element occurs multiple times as (1) this will not interfere with
the detection of that element by the robot (i.e., we only care about the presence
and not the position of an element) and (2) we are interested in the presence in $G$ of
dominating sets with at most and not necessarily exactly $k$ vertices. This reduction
yields the following initial result.

\begin{description}
\item[{\bf Result D.SF.1}] \cite[Result B.1]{WV21}: EnvDesSF$^{fast}$ is not polynomial-time exact \linebreak solvable.
\end{description}

\noindent
It also yields our first pair of fixed-parameter intractability results, with the
second in the pair using the tradeoff between transition trigger-formula
length and number of outgoing transitions per state sketched above. 

\begin{description}
\item[{\bf Result D.SF.2}] \cite[Result B.2]{WV21}: 
$\la |T|, h, |f|, |E|, |S_E|, |q_E|, |X|\ra$-EnvDesSF$^{fast}$ is \linebreak fp-intractable.
\item[{\bf Result D.SF.3}]  \cite[Result B.3]{WV21}:
$\la |T|, h, |Q|, d, |E|, |S_E|, |q_E|, |X|\ra$-EnvDesSF$^{fast}$ is \linebreak fp-intractable.
\end{description}

\noindent
The sensing at a distance approach sketched above will not work if we want to restrict the 
number of types of point-fields in $S$, as a static checker-robot cannot verify the
positions of sensed point-fields encoding only the presence of vertices in a column of 
$|V|$ squares. However, taking our cue from the dynamic robots in Sections 
\ref{SectResContDesLS} and \ref{SectResTeamDesLS}, if the encoded candidate solution
will not come to the checking robot, the checking robot can go to the encoded
solution. An environment suited to this strategy is shown in Figure \ref{FigEnvDesEnv}(c),
in which the movement-track for the checker robot is the outermost ring of squares.
Given a suitable edge-field positioned at the northmost edge of $E$, a checker robot
can now assess the positions as well as the presence of vertex point-fields in the
candidate solution encoded in the middle $|V|$ squares of the westmost column;
moreover, given appropriately placed direction-indicating point-fields on the remainder
of the movement-track, the checker robot can move around the movement-track in
clockwise fashion as many times as necessary to check that the encoded candidate solution 
contains at least one vertex in each of the vertex neighborhoods in given graph $G$.
This reduction yields our first pair of fixed-parameter intractability results, with 
the second in the pair once again using the tradeoff between transition trigger-formula
length and number of outgoing transitions per state sketched above. 

\begin{description}
\item[{\bf Result D.SF.4}]  \cite[Result B.4]{WV21}:
$\la |T|, h, |f|, |S|, |S_E|, |q_E|, |X|\ra$-EnvDesSF$^{fast}$ is \linebreak fp-intractable.
\item[{\bf Result D.SF.5}]  \cite[Result B.5]{WV21}:
$\la |T|, h, d, |S|, |S_E|, |q_E|, |X|\ra$-EnvDesSF$^{fast}$ is \linebreak fp-intractable.
\end{description}

At this point, it may appear that $|Q|$, $d$, $|f|$, and $|S|$ cannot be restricted
simultaneously to get fp-intractability. This is, however, possible if we split
the functions in the single robot above over multiple robots --- namely, a set of
$|V|$ vertex-neighborhood robots (each of which, for a specific vertex $v_i$, checks if 
there is a vertex in
the encoded candidate solution that is in neighbourhood of vertex $v_i$ in $G$, and
if so, places a point-field of type $s_X$ at square $E_{2,i+1}$ in the central
column of $E$) and a checker robot (which first verifies that $E$ has the proper
initial format and then checks if the vertex neighborhood robots have fully
filled in middle $|V|$ squares of the central column with point-fields of type $s_X$;
if so, the encoded candidate solution is in fact a dominating set of size $k$ in
$G$ and the requested structure $X$ is created at $p_X = E_{3,2}$). Collision avoidance
between the various robots in $T$ as they move clockwise around the movement-track
in the environment shown in Figure \ref{FigEnvDesEnv}(d) is guaranteed by using
the formula derived in Section \ref{SectResTeamDesLS}. This reduction yields the
following result.

\begin{description}
\item[{\bf Result D.SF.6}]  \cite[Result B.6]{WV21}:
$\la |Q|, |f|, d, |S|, |X|\ra$-EnvDesSF$^{fast}$ is fp-intractable.
\end{description}

We also have one fp-tractability result.

\begin{description}
\item[{\bf Result D.SF.7}]  \cite[Result B.7]{WV21}:
$\la |E|, |S|\ra$-EnvDesSF$^{fast}$ is fp-tractable.
\end{description}

\noindent
This result is based on the brute-force algorithm underlying Result D.ST.6.

A summary of all of our fixed-parameter results derived in this subsection 
is given in Table \ref{TabPrmResEnvDes}. Unlike our previously-examined problems,
our fp-intractability results for EnvDesST$^{fast}$ and EnvDesSF$^{fast}$ in conjunction
with Lemma \ref{LemPrmProp2} establish that both of our fp-tractability results above are 
minimal in  the sense described in Section \ref{SectResDesSolv}. 

The above demonstrates that problems with point-field interference and sensing at a 
distance are resolvable (albeit at the cost of more complicated robots and environments, 
including the first use of edge-fields) if, as in Section \ref{SectResTeamDesLS}, robots 
are allowed to move around (in this case, over the encoded candidate solution itself). 
Moreover, the state-complexity of the robots can be dramatically reduced if we also allow
robots to modify the environment to communicate with each other indirectly via stigmergy 
(by the filling in of the central scaffolding). Both of these strategies will 
underlie our final set of results derived in the next subsection.

\begin{table}[t]
\caption{A Detailed Summary of Our Fixed-parameter Results for Environment Design.
          Selection. This table is interpreted as described in the caption of
          Table \ref{TabPrmResContDesLS}.
}
\label{TabPrmResEnvDes}
\vspace*{0.1in}
\centering
\begin{tabular}{ | p{0.6cm}  || p{0.7cm}  p{0.7cm}  p{0.7cm} p{0.7cm} p{0.8cm} || p{0.7cm}  p{0.7cm} p{0.7cm} p{0.7cm} p{0.7cm} p{0.8cm} | }
\hline
& \multicolumn{5}{c ||}{Square-type (ST)}
 & \multicolumn{6}{c |}{Scalar-field (SF)} \\
& ST.2 & ST.3 & ST.4 & ST.5 & {\bf ST.6} 
& SF.2 & SF.3 & SF.4 & SF.5 & SF.6 & {\bf SF.7} \\
\hline\hline
$|T|$       & 1  & 1  & 1  & 1  & {\bf --}
            & 1  & 1  & 1  & 1  & -- & {\bf --} \\
\hline
$h$         & 1  & 1  & 1  & 1  & {\bf --}
            & 1  & 1  & 1  & 1  & -- & {\bf --} \\
\hline
$|Q|$       & -- & 1  & @  & 1  & {\bf --}
            & -- & 1  & -- & -- & @  & {\bf --} \\
\hline
$d$         & 2  & 2  & -- & 2  & {\bf --}
            & -- & 1  & -- & 3  & 4  & {\bf --} \\
\hline
$|f|$       & 5  & -- & 3  & -- & {\bf --}
            & 1  & -- & 3  & -- & 12 & {\bf --} \\
\hline
$r$         & -- & -- & @  & @  & {\bf --}
            & N/A & N/A & N/A & N/A & N/A & {\bf N/A} \\
\hline\hline
$|L|$       & N/A & N/A & N/A & N/A & {\bf N/A}
            & N/A & N/A & N/A & N/A & N/A & {\bf N/A} \\
\hline\hline
$|E|$       & -- & -- & @  & @  & {\bf @ }
            & @  & @  & -- & -- & -- & {\bf @ } \\
\hline
$|E_T|$     & 5  & 5  & -- & -- & {\bf @ }
            & N/A & N/A & N/A & N/A & N/A & {\bf N/A} \\
\hline
$|S|$       & N/A & N/A & N/A & N/A & {\bf N/A}
            & -- & -- & 8  & 8  & 8  & {\bf @ } \\
\hline
$|S_E|$     & N/A & N/A & N/A & N/A & {\bf N/A}
            & @  & @  & @  & @  & -- & {\bf --} \\
\hline
$|q_E|$     & N/A & N/A & N/A & N/A & {\bf N/A}
            & @  & @  & @  & @  & -- & {\bf --} \\
\hline
$|X|$       & 1  & 1  & 1  & 1  & {\bf --}
            & 1  & 1  & 1  & 1  & 1  & {\bf --} \\
\hline
\end{tabular}
\end{table}

\subsection{Team / Environment Co-design by Library Selection}

\label{SectResTeamEnvDesLS}

In team / environment co-design by library selection, the final type of problem examined in 
this paper, the general form but not the exact structure of {\em both} the environment 
{\em and} the robot team is part of the problem input. This may initially seem too
unconstrained to encode anything, as we always had at least one of these entities
specified in full in each of our previously examined problems. However, with
a bit of forethought and care, the environment size $|E|$ and the environment-square contents
(either square-type set $E_T$ or field-set $S$ in the case of EnvDesST$^{fast}$ or
EnvDesSF$^{fast}$, respectively) can be specified to create a candidate-solution generation
component and the robot team size $|T|$  and the robot controller library $L$ can be 
specified to create a candidate-solution checking component for the given instance of $\Pi$ 
in a reduction from $\Pi$ to our team / environment co-design problem.

This was done in \cite{WV18_SI} for TeamEnvDesLSST$^{fast}$ using reductions from 3SAT and 
{\sc Clique} in \cite{TW19}. The basic reduction used was actually that from {\sc Clique}
seen above in the proof of Result D.ST.4, modified such that the sole robot in $T$ was
the only robot in $L$ --- given this, the only choice for $E_I$ was that robot, and
the reduction functioned and was proved correct as was done for Result D.ST.4. This yielded
the following result.

\begin{description}
\item[{\bf Result E.ST.1}] \cite[Theorem 1]{TW19}: TeamEnvDesLSST$^{fast}$ is not polynomial-time \linebreak exact solvable.
\end{description}

\noindent
This reduction was directly implied the first of the following three fixed-parameter
intractability results. The other two results follow by more complex versions
of the tradeoff between transition trigger-formula length and the number of outgoing
transitions per state described in previous subsections.

\begin{description}
\item[{\bf Result E.ST.2}] \cite[Theorem 3]{TW19}: 
$\la |T|, |Q|, d, r, |L|, |E|, |X|\ra$-TeamEnvDesLSST$^{fast}$ is fp-intractable.
\item[{\bf Result E.ST.3}] \cite[Theorem 4]{TW19}: 
$\la |T|, |Q|, |f|, r, |L|, |E|, |X|\ra$-TeamEnvDesLSST$^{fast}$ is fp-intractable.
\item[{\bf Result E.ST.4}] \cite[Theorem 5]{TW19}: 
$\la |T|, d, |f|, r, |L|, |E|, |X|\ra$-TeamEnvDesLSST$^{fast}$ is fp-intractable.
\end{description}

\noindent
Fixed-parameter intractability when $|Q|$, $d$, and $|f|$ were simultaneously
restricted was done via a reduction from 3SAT in which a candidate variable assignment
is encoded in the environment and this assignment is checked by $|C|$ robots,
one for each clause in $C$. To ensure that the $|C|$ robots corresponding to all
of the given clauses were chosen from $L$, the robot corresponding to clause $i$,
$2 \leq i \leq |C| - 1$, was specified to require that it was located in a square of
type $e_{ci}$ and that the squares to its immediate west and east were of type
$e_{c(i-1)}$ and $e_{c(i+1)}$ (the robots corresponding to the first and last
clauses in $C$ were encoded analogously). This not only forced the choice 
from $L$ of the robots corresponding to all $|C|$ given clauses in the given
instanced of 3SAT but also placed each-clause robot in a unique position from which
the location of the three variable-assignments in the candidate variable assignment
that it needed to consult to check clause satisfaction were readily known (something
that would not have been possible if the clause robots were at arbitrary positions
within $E_I$).  This yielded the following result.

\begin{description}
\item[{\bf Result E.ST.5}] \cite[Theorem 6]{TW19}: 
$\la |Q|, d, |f|, |X|\ra$-TeamEnvDesLSST$^{fast}$ is \linebreak fp-intractable.
\end{description}

\noindent
There was also a single fp-tractability result.

\begin{description}
\item[{\bf Result E.ST.6}] \cite[Theorem 7]{TW19}: 
$\la |L|, |E|, |E_T|\ra$-TeamEnvDesLSST$^{fast}$ is fp-tractable.
\end{description}

\noindent
This result was based on the brute-force enumeration and checking of both all possible 
environments and all possible selections of $|T|$ robots from $L$ and all possible
orderings of these robots in $E_I$.

\vspace*{0.1in}

Given the observation above that environment design and environment / team
co-design by library selection are effectively equivalent when $|T| = 1$,
the first five results derived in Section \ref{SectResEnvDes} for EnvDesSF$^{fast}$
directly give us our first five results for TeamEnvDesLSSF$^{fast}$.

\begin{description}
\item[{\bf Result E.SF.1}:] TeamEnvDesSF$^{fast}$ is not polynomial-time exact solvable.
\item[{\bf Result E.SF.2}:] 
$\la |T|, h, |f|, |L|, |E|, |S_E|, |q_E|, |X|\ra$-TeamEnvDesSF$^{fast}$ is \linebreak fp-intractable.
\item[{\bf Result E.SF.3}:] 
$\la |T|, h, |Q|, d, |L|, |E|, |S_E|, |q_E|, |X|\ra$-TeamEnvDesSF$^{fast}$ is \linebreak fp-intractable.
\item[{\bf Result E.SF.4}:] 
$\la |T|, h, |f|, |L|, |S|, |S_E|, |q_E|, |X|\ra$-TeamEnvDesSF$^{fast}$ is \linebreak fp-intractable.
\item[{\bf Result E.SF.5}:] 
$\la |T|, h, d, |L|, |S|, |S_E|, |q_E|, |X|\ra$-TeamEnvDesSF$^{fast}$ is \linebreak fp-intractable.
\end{description}

\noindent
The sixth result derived in Section \ref{SectResEnvDes} for EnvDesSF$^{fast}$ 
also holds and by the same reduction, modulo the placing of all robots in $T$
in $L$. We do not need to invoke the interlocked-symbol strategy used in the proof of 
Result E.ST.5 to choose and position the robots in $L$ to create $T$ because, in the 
reduction in the proof of Result D.SF.6, (1) courtesy of its transition structure,
the checker robot must be selected from $L$ and initially positioned in $E'_{1,1}$ in 
order to both progress around the movement-track and (if the central scaffolding is filled 
in with point-fields of $s_X$) create $X$ at $p_X$, and (2) in order for all squares in the 
central scaffolding to be filled in, all $|V|$ vertex neighbourhood robots from $L$ must be 
selected from $L$ and can be initially positioned in an arbitrary order in the middle $|V|$ 
squares of the eastmost column in $E'$. 
 
\begin{description}
\item[{\bf Result E.SF.6}:] 
$\la |Q|, |f|, d, |S|, |X|\ra$-TeamEnvDesSF$^{fast}$ is fp-intractable.
\end{description}

\noindent
We also have an fp-tractability result.

\begin{description}
\item[{\bf Result E.SF.7}:] 
$\la |L|, |E|, |S|\ra$-TeamEnvDesSF$^{fast}$ is fp-tractable.
\end{description}

\noindent
This result is based on the brute-force enumeration and checking algorithm in the 
proof of Result E.ST.6.

A summary of all of our fixed-parameter results derived in this subsection is given in Table
\ref{TabPrmResTeamEnvDesLS}. Given the fp-intractability results we have at this time,
only our fp-tractability result for TeamEnvDesLSSF$^{fast}$ is 
minimal in  the sense described in Section \ref{SectResDesSolv}. 

The above demonstrates that even apparently totally unconstrained problems may, under
certain circumstances, be constrained enough to encode candidate solution generating
and checking components in a reduction to show intractability.

\begin{table}[t]
\caption{A Detailed Summary of Our Fixed-parameter Results for Environment / Team Co-design
          by Library Selection
          Selection. This table is interpreted as described in the caption of
          Table \ref{TabPrmResContDesLS}.
}
\label{TabPrmResTeamEnvDesLS}
\vspace*{0.1in}
\centering
\begin{tabular}{ | p{0.6cm}  || p{0.7cm}  p{0.7cm}  p{0.7cm} p{0.7cm} p{0.8cm} || p{0.7cm}  p{0.7cm} p{0.7cm} p{0.7cm} p{0.7cm} p{0.8cm} | }
\hline
& \multicolumn{5}{c ||}{Square-type (ST)}
 & \multicolumn{6}{c |}{Scalar-field (SF)} \\
& ST.2 & ST.3 & ST.4 & ST.5 & {\bf ST.6} 
& SF.2 & SF.3 & SF.4 & SF.5 & SF.6 & {\bf SF.7} \\
\hline\hline
$|T|$       & 1  & 1  & 1  & -- & {\bf --}
            & 1  & 1  & 1  & 1  & -- & {\bf --} \\
\hline
$h$         & 1  & 1  & 1  & -- & {\bf --}
            & 1  & 1  & 1  & 1  & -- & {\bf --} \\
\hline
$|Q|$       & 1  & 5  & @  & 4  & {\bf --}
            & -- & 1  & -- & -- & @  & {\bf --} \\
\hline
$d$         & 3  & -- & 3  & 3  & {\bf --}
            & -- & 1  & -- & 3  & 4  & {\bf --} \\
\hline
$|f|$       & -- & 3  & 3  & 7  & {\bf --}
            & 1  & -- & 3  & -- & 12 & {\bf --} \\
\hline
$r$         & @  & @  & @  & -- & {\bf --}
            & N/A & N/A & N/A & N/A & N/A & {\bf N/A} \\
\hline\hline
$|L|$       & 1  & 1  & 1  & -- & {\bf @ }
            & 1  & 1  & 1  & 1  & -- & {\bf @ } \\
\hline\hline
$|E|$       & @  & @  & @  & -- & {\bf @ }
            & @  & @  & -- & -- & -- & {\bf @ } \\
\hline
$|E_T|$     & -- & -- & -- & -- & {\bf @ }
            & N/A & N/A & N/A & N/A & N/A & {\bf N/A} \\
\hline
$|S|$       & N/A & N/A & N/A & N/A & {\bf N/A}
            & -- & -- & 8  & 8  & 8  & {\bf @ } \\
\hline
$|S_E|$     & N/A & N/A & N/A & N/A & {\bf N/A}
            & @  & @  & @  & @  & -- & {\bf --} \\
\hline
$|q_E|$     & N/A & N/A & N/A & N/A & {\bf N/A}
            & @  & @  & @  & @  & -- & {\bf --} \\
\hline
$|X|$       & 1  & 1  & 1  & 1  & {\bf --}
            & 1  & 1  & 1  & 1  & 1  & {\bf --} \\
\hline
\end{tabular}
\end{table}

\section{Discussion}

\label{SectDisc}
 
The time has now come for a broader look at what the results derived in Section 
\ref{SectRes} mean. We will first consider what our results say about the algorithmic 
options for the verification and design problems defined in Section \ref{SectForm} and 
what this implies about mechanism interactions in these problems and the computational 
characteristics of visual and scalar-field sensing (Section \ref{SectDiscImpSen}). We 
then look at what our results say about real-world verification
and design problems (Section \ref{SectDiscImpRWR}). Finally, given the insights
we have gained doing a comparative computational complexity analysis 
and the need to make such analyses more accessible, we give some 
initial thoughts on a metaphor for organizing and reasoning about results derived in such 
analyses (Section \ref{SectDiscThink}).

\subsection{Implications for Our Defined Problems}

\label{SectDiscImpSen}

Let us first consider what the results given in Sections 
\ref{SectResTeamEnvVer}--\ref{SectResTeamEnvDesLS} have to say in general about 
verification and design problems for teams of visual and scalar-field sensing robots.

\begin{itemize}
\item{
{\bf What types of efficient algorithms are (not) possible?} As was the case
with visual sensing robots in \cite{TW19,War19,WV18_SI}, all of our design problems
relative to scalar-field sensing robots are polynomial-time intractable in general,
even if the robot teams are restricted to finishing their construction tasks in
low-order polynomial time (Results B.SF.1, C.SF.2, D.SF.1, and E.SF.1; see also 
Result A.SF.1). Moreover, as was also the case in \cite{TW19,War19,WV18_SI},
this intractability continues to hold when almost all parameters in Table \ref{TabPrm}
are restricted individually and simultaneously in many combinations, and often when
these restrictions are to small constant values (see Tables 
\ref{TabPrmResContDesLS}--\ref{TabPrmResTeamEnvDesLS}). We also have few fixed-parameter
tractability results, though there is at least one for each problem examined. All known 
fixed-parameter algorithms rely on brute-force enumeration and checking of alternatives. It
is possible that better types of fixed-parameter algorithms exist relative to known and
other combinations of parameters, derivable by techniques described in 
\cite{CF+15,FL+19,Nie06}. It is also possible that they do not. 

Such questions may be
resolved in time by establishing the parameterized status of all of our problems
relative to all possible combinations of parameters in Table \ref{TabPrm}, i.e.,
perform systematic parameterized complexity analyses for our problems relative to
the parameters in Table \ref{TabPrm} \cite{War99}. Such analyses are typically made much
easier by exploiting Lemmas \ref{LemPrmProp1} and \ref{LemPrmProp2} relative to
fp-tractability results based on the smallest parameter-sets possible and fp-intractability 
results based on the largest parameter-sets possible, For now, parameters of immediate
interest are:

\begin{itemize}
\item
$|L|$ for problems ContDesLSST$^{fast}$, ContDesLSSF$^{fast}$, 
TeamDesLSST$^{fast}$, and TeamDesLSSF$^{fast}$ (to determine whether or not the observed
\linebreak fp-tractability of these problems (Results B.ST.5, B.ST.5, C.ST.6,
and C.SF.7) is minimal in the sense described
in Section \ref{SectResDesSolv});
\item 
$|E|$ for the same problems (to determine whether
or not environment size can be restricted in some fashion to give fp-tractability, as
it can be for our environment design (Results D.ST.6 and D.ST.7)  and team / environment 
co-design by library selection (Results E.ST.6 and E.SF.7) problems); and 
\item
$|E_T|$ for
TeamEnvDesST$^{fast}$ (to determine whether or not the observed fp-tractability
of this problem (Result E.ST.6) is minimal). 
\end{itemize}

\noindent
The fp-status of our controller and team
design problems under scalar-field sensing results relative to $|E|$, $|S|$, $|S_E|$, and 
$|q_E|$ is also pressing, as our intractability proofs to date for these problems 
require large environments with a large number and/or variety of scalar fields. It would 
be most interesting to see if fp-tractability holds for smaller environments with a small 
number and/or variety of scalar fields, as these situations may be more typical in 
real-world instances of scalar-field sensing by robot teams performing certain types
of distributed construction \cite{Var18,Var20}.
}
\item{
{\bf How do mechanisms interact, and what are the tradeoffs between mechanisms that 
preserve (in)tractability?} The best evidence for mechanism interactions that we currently
have are 
our three minimal fp-tractability results --- $\{|E|, |E_T|\}$ for EnvDesST$^{fast}$
(Result D.ST.6), $\{|E|, |S|\}$ for \linebreak EnvDesSF$^{fast}$ (Result D.SF.7), and $\{|L|, |E|, 
|S|\}$ for TeamEnvDesLSSF$^{fast}$ (Result E.SF.7).
In each of these results, it seems that restrictions in any one (or in the case of the
third result, two) of the parameters involved leaves the problem fp-intractable but
restricting all parameters involved causes a collapse to fp-intractability. These
sources of intractability should be investigated in more detail by future research. 
There are hints of other interactions in our other intractability results.
Foremost among these is the interaction between $|f|$ and $d$ which occurs in 
our team (result-pairs C.ST.3 and C.ST.4, C.SF.3 and C.SF.4, and C.SF.5 and C.SF.6) and
environment (results pairs D.ST.3 and D.ST.4, D.SF.2 and D.SF.3, and D.SF.4 and D.SF.5)
design problems, which (as was first pointed out in Section
\ref{SectResTeamDesLS}) allows us to preserve fp-intractability
by trading short transition-trigger formulas over multiple transitions between a pair of 
states for a single transition with a long transition-trigger formula between that 
state-pair. There are also others, notably a possible interaction between $|T|$ and
$\{|Q|, d, |f|\}$ seen in result-pairs D.SF.5 and D.SF.6, E.ST.4 and E.ST.5,
and E.SF.5 and E.SF.6. Whether these are artifacts of our incomplete knowledge or
indicative of actual minimal sources of intractability remains to be seen, and is yet
another reason for doing systematic parameterized analyses of our problems
relative to the parameters in Table \ref{TabPrm}.
}
\end{itemize}

\noindent
Given the above, we can now do an initial comparison of the computational characteristics
of visual and scalar-field sensing. With respect to the verification and design problems 
considered here, our proofs suggest that intractability results for scalar-field sensing 
robots require larger number of states (to accommodate the need to travel to rather than 
scrutinize from afar aspects of their environments) and larger environments when $|S|$ is
reduced (to avoid interference between point-fields) then analogous results for visual 
sensing robots.  That being said, known 
fp-(in)intractability results under both types of sensing given in Tables 
\ref{TabPrmResContDesLS}--\ref{TabPrmResTeamEnvDesLS} are 
nonetheless very similar in terms of the parameters involved and the values of these 
parameters under which the results hold. It will be interesting to see if 
this apparent similarity in the computational power of visual and scalar-field sensing 
increases or decreases as more results are derived in future.

Quite aside from the insights gained above, a comparison of groups of results for related
problems like that given in this paper
is invaluable methodologically. This is so because research on problems and derivation of
results is often done over a period of time, for different venues with different
concerns. The gathering together and comparison of results for related problems will
thus not only highlight gaps in these sets of results but also suggest techniques
for filling these gaps with respect to one problem that were developed with respect
to another. For example, two 3SAT-based results for EnvDesST$^{fast}$ in \cite{WV18_SI}
(Results D.ST.2 and D.ST.3), while not referenced in \cite{TW19}, are actually
also applicable to TeamEnvDesLSST$^{fast}$ --- indeed, they supply fp-intractability
relative to $|E_T|$ which shows that the fp-tractability result for this problem 
(Result E.ST.6) is in fact minimal. It would also be interesting to see if the
techniques used to show fp-intractability of EnvDesSF$^{fast}$, TeamEnvDesLSST$^{fast}$,
and TeamEnvDesLSSF$^{fast}$ relative to $\{|Q|, d, |f|\}$ here and in \cite{TW19}
(Results D.SF.6, E.ST.5, and E.SF.6, respectively) could be applied with equal effect to
EnvDesST$^{fast}$ (whose results were derived two years earlier in \cite{WV18_SI}).
These and other such examples will greatly aid in the systematic parameterized complexity
analyses that need to be done for our problems.

\subsection{Implications of Results for Real-World Robotics}

\label{SectDiscImpRWR}

The results in this paper have been derived under basic models of square-type and scalar 
field environments and a basic model of robot team operation where knowledge of the 
environment is complete and sensing, movement, and action execution are error-free. As
shown in Section \ref{SectDiscImpSen}, this has enabled insights into core computational
aspects of robot team verification and design under visual and scalar-field sensing. 
However, these insights have been gained at the cost of ignoring real-world applications 
in which verification and design occur in the face of partial or
even untrustworthy knowledge about complex environments when sensing, motion, and action 
execution are error-ridden. Though accommodating real-world robotics was not a
goal of our work reported here, it is nonetheless natural to ask the
following:

\begin{enumerate}
\item Are our results directly applicable to real-world
       robotics?
\item If not, can our results be extended to
       fully accommodate real-world robotics?
\end{enumerate}

\noindent
In this subsection, we will address each of these questions in turn,
as well as the extent to which conclusions drawn from our results
in Section \ref{SectDiscImpSen} are applicable to real-world robotics.

It turns out that our intractability results, while not directly applicable
to real-world robotics, have a surprisingly broad 
applicability. This is because (as is discussed in more detail in
Section 5 of \cite{War19})  our models are special cases of 
more realistic models, e.g., 

\begin{itemize}
\item{
Our 2D grid-based environments without obstacles are special cases of 2D
grid-based environments with obstacles (as the class of all such environments
includes the class of environments with no obstacles as a special case).
}
\item{
Our 2D grid-based environments are special cases of 3D grid-based environments
(as 2D grid-based environments can be recoded as 
special cases of 3D grid-based environments \cite[Section 2.1]{War19}).
}
\item{
Our fully-known 2D grid-based environments are special cases of partially known 
and totally unknown 2D and 3D grid-based environments (as the class of all such 
environments include fully known environments as a special case).
}
\item{
Our deterministic FSR model is a special case of probabilistic FSR models
\cite[Footnote 1]{War19}.
}
\item{
Our exact model of FSR motion and sensing is a special case of
probabilistic models allowing imprecise FSR motion and sensing (as the class
of all such models includes the model with exact motion and sensing as a
special case).
}
\item{
Our team operation model which does not allow direct communication between
robots is a special case of team operation models that do allow some
form or degree of direct communication between robots (as the class
of all such models includes the model with no direct communication as a special case).
}
\end{itemize}

\noindent
More realistic versions of our problems can be created by replacing any
combination of simple special-case models with the appropriate combination of more 
general models, e.g., 
deterministic robot team operation in a 2D grid-based environment without obstacles
$\Rightarrow$ 
probabilistic robot team operation in a 3D grid-based environment with obstacles.
Courtesy of the special-case relationship, any automated system that solves 
such a more realistic problem $\Pi'$ can also solve the original problem $\Pi$ 
defined relative to the simple special-case models. Intractability
results for $\Pi$ then also apply to $\Pi'$ as well as the operation of any 
automated system solving $\Pi'$. Tractability results typically do not propagate
from special cases to more general problems, as algorithms often exploit 
particular details of the inputs and outputs in their associated problems to 
attain efficiency or even work at all. That being said, all of our tractability
results also have a broad applicability because the algorithms described in 
their proofs depend only on the combinatorics of the number of possible 
controller, team, or environment choices. 

The special-case relationship exploited above has other limitations as well, in that it 
cannot accommodate aspects of real-world robotics that are not generalizations
of the simple models considered here, e.g., continuous environments, scalar fields with 
continuous Gaussian decay over space and/or source-decay over time, continuous 
asynchronous robot team operation. However, there is no reason to expect that 
computational complexity proofs cannot be done for fully realistic verification and design
problems that incorporate these aspects.  A useful aid in this would be a research 
framework that builds from results for simpler models to results for more complex ones. 
Such a framework (sketched previously in \cite{WV21}) could be based on a fusion of the 
natural language complexity game \cite{Ris93} and the tractable design cycle
\cite{vRB+19}. In such a framework, more complex aspects 
would be incrementally built on top of initially simple models in repeated rounds, where 
each round adds an aspect to the current model, uses computational complexity analysis to
find those restrictions under which the augmented model is tractable, and then uses the 
restricted augmented model as input to the next round. In this way, rather than 
addressing the computational intractability in real-world applications in an
ad hoc top-down manner, this intractability would be dealt with systematically in a
reasoned bottom-up manner. Similar frameworks have already been applied to good
effect in the development of tractable versions of various natural language processing
\cite{Ris93} and cognitive \cite{vRB+19} tasks. It seems reasonable to conjecture that 
the framework above could be successfully applied to tasks in robotics such as those 
investigated in this paper. If so, the models and results given here would be a good 
starting point.

What does all this imply for the applicability of conclusions we have 
drawn from our results in Section \ref{SectDiscImpSen} to real-world 
robotics? The hard truth is that we don't know yet. The results on which 
these conclusions
are based are not only incomplete with respect to the simplified problems 
examined in this paper but could also potentially change when we
examine more realistic versions of these problems. For example, as noted 
previously in Section 6.2 of \cite{War19}, there are examples of problems 
where allowing entities to be continuous rather than discrete can cause 
computational complexity to decrease (e.g., 0/1 integer programming 
\cite[Problem MP1]{GJ79} vs.\ linear programming \cite{Kar84}) 
or increase (e.g., finding Steiner trees in graphs 
\cite[Problem ND12]{GJ79} vs.\ finding Steiner trees in the 2D Euclidean 
plane \cite[Problem ND13]{GJ79}). In the absence of proofs (which would
be produced by future research), we cannot
be certain which (if any) of our results will survive in more realistic
settings. We are not alone in this, however, as such uncertainty is 
characteristic of the initial stages of many scientific investigations.
Hence, with respect to the question beginning this paragraph, we will
for now advise cautious optimism and conjecture that (in the absence of
proof to the contrary) all conclusions drawn in Section
\ref{SectDiscImpSen} are also applicable to real-world robotics. 

\subsection{Metaphors to Aid Thinking About Comparative \\ Computational Complexity Analyses}

\label{SectDiscThink}

Quite aside from the technical difficulties associated with deriving individual
computational complexity results for problems, there are more general conceptual
difficulties associated with organizing and reasoning about the sets of results for
related problems underlying comparative analyses such as that presented in this paper. 
Computing is replete with frameworks that abstract away from low-level 
details to make hardware and software more accessible to non-specialists,
e.g., high-level programming languages like FORTRAN, C, and Python, the relational
database model, the desktop GUI metaphor. Given our goals of making the results of our 
analyses more accessible to and hence encouraging closer collaboration of CS theorists 
with robotics researchers, it would be worth considering similar frameworks for
thinking about comparative computational complexity analyses. To this end, we propose in 
this section a metaphor based on natural history studies in biology \cite{Bat50}.

What are the characteristics of natural history studies that would make them a suitable
metaphor for our purposes? In biology, natural histories have in common a focus on
 individual organisms in their natural settings and a reliance on observation of 
organisms and their behaviour as the most trustworthy tool for learning about them 
\cite[page 3]{Fle05}. A natural history of a biological species seeks to establish basic 
empirical facts by answering the following questions \cite[page 326]{Bar86}:

\begin{itemize}
\item What type of organism is it?
\item Where does it live?
\item How many are there?
\item How does it survive and reproduce?
\item How does it come to be like it is and live where it does?
\end{itemize}

\noindent
Natural histories for related species can be combined to give a natural history for that
group of species, e.g., a natural history of South American monkeys. Natural histories 
have many uses in human health, food security, and conservation and management 
\cite{TA+14} and are invaluable in comparative studies of species. Given appropriately 
detailed natural histories, such comparative studies can focus not only on characterizing
groups of related organisms, e.g., the finches of the Galapagos islands, but can also be 
used to investigate the behaviour of the same mechanism or mechanisms in related 
organisms, e.g., beak structures in the finches of the Galapagos islands. As such,
observation-based biological natural histories are a useful precursor to the development
of subsequent higher-level biological theories like ecology, 
taxonomy, and evolutionary theory.

Given the above, what would a computational analogue of biological natural history
studies look like? A computational natural history (for want of a better term)
would focus on individual computational problems and their basic properties, which would
be established by answering the following questions:

\begin{itemize}
\item What type of problem is it?
\item Where does the problem occur, and in what situations?
\item What types of efficient algorithms are (not) possible for the problem?
\item How do mechanisms in the problem interact, and what are the 
       tradeoffs between mechanisms that preserve (in)tractability?
\end{itemize}

\noindent
In such a framework, algorithm and reduction details for a problem of interest can be
seen as analogous to the particulars of species behavior in biological natural histories.
Computational natural histories for related problems can be combined to give a 
computational natural history for that group of problems, e.g., verification and design 
problems for teams of scalar-field sensing robots. Computational natural histories have 
many uses in selecting the best possible algorithms and developing new algorithms for 
real-world applications and are invaluable in comparative studies of problems. Given 
detailed computational natural histories structured along the lines described above, 
comparative studies such as those presented in this paper focus not only on characterizing
the algorithmic options for groups of related problems, e.g., the verification and design
of of teams of scalar-field sensing robots, but can also be used to investigate the 
behaviour of the same mechanism or mechanisms in related problems, e.g., visual 
vs.\ scalar-field sensing in robot team verification and design problems. As such,
result-based computational natural histories could be a useful precursor to the 
development of subsequent higher-level theories addressing issues like
the reasons behind tractability and intractability.

The correspondence underlying this metaphor is by no means ideal --- biological
organisms have a shared evolutionary history and exist in the real world while
computational problems (though inspired by real-world activities) are mathematical 
abstractions, and this must be remembered at all times.\footnote{
Oddly enough, this might be ensured by the awkwardness of the
term ``computational natural history''.
} 
Such cautions aside, much
as parameterized complexity analysis highlights the interactions of mechanisms
within an individual problem to produce tractability and intractability, the metaphor 
proposed above highlights the comparing of problems and their associated
computational characteristics in groups as a first step to gaining deeper understanding
of problem mechanisms in particular and the roots of tractability and intractability 
in general. It is possible that another metaphor may in the end be more appropriate
for this purpose --- we look forward to the thoughts of other researchers on this.

\section{Conclusions and Future Work}

\label{SectConc}

In this paper, we have given the first comparison of the computational characteristics of
visual and scalar-field sensing relative to five verification and design problems 
associated with distributed construction tasks performed by robot teams.
This was done relative to basic models of
square-type and scalar-field environments and a basic model of robot team operation in
which teams of robots with deterministic finite-state controllers perform
construction tasks in a non-continuous, synchronous, and error-free manner in 2D
grid-based environments. Our results show that for both types of sensing, all of our
problems are polynomial-time intractable in general and remain intractable under
a variety of restrictions on parameters characterizing robot controllers, teams, and
environments, both individually and in many combination and often when parameters
are restricted to small constant values. That being said, our results also include
restricted situations for each of our problems in which those problems are effectively 
polynomial-time tractable. Though there are differences, these results show that
both types of sensing have (at least in this stage of our investigation) roughly the same
patterns and types of intractability and tractability results.

There are several promising directions for future research. First among these
is to complete the analyses given here along the lines described in Section
\ref{SectDiscImpSen} by establishing the parameterized status of all of our problems
relative to all possible combinations of the parameters listed in Table \ref{TabPrm}.
The second is to extend our basic model of scalar-field sensing to allow the 
investigation of computational complexity issues associated with more general
types of scalar fields, including both fields not generated by discrete
environmental sources (see Footnote \ref{FnCSF}) and transient fields associated
with digital pheromones, as well 
as quantitative stigmergy \cite{TB99}. The third is to build on
the results derived in this paper to explore algorithmic options for more
complex models approximating real-world distributed construction and other tasks by 
teams of visual and scalar-field sensing robots. This is a longer-term project that
may best be accomplished using the research framework described at the end of
Section \ref{SectDiscImpRWR}.

\section*{Acknowledgments}
TW was supported by National Science and Engineering
Council (NSERC) Discovery Grant 228104-2015.

% Generated by IEEEtran.bst, version: 1.13 (2008/09/30)

\appendix

\section{Proofs of Results}

As mentioned in Section \ref{SectResDesSolv}, for technical reasons, all of our 
intractability results are proved relative to decision versions of problems, i.e., problems
whose solutions are either ``yes'' or ``no''. For example, problems TeamEnvVerSF and
TeamEnvVerST defined in Section \ref{SectForm} are decision problems. Though none of the
other problems defined in Section \ref{SectForm} are decision problems, each such
problem can be made into a decision problem by asking if that
problem's requested output exists; let the decision version for such a problem {\bf X}
be denoted by {\bf X}${}_D$. The following three easily-proven lemmas will be useful
below in transferring results from decision problems to their associated
non-decision and parameterized problems; these lemmas follow from the observation that
any algorithm for non-decision problem {\bf X} can be used to
solve {\bf X}${}_D$ and the definition of fp-tractability.

\begin{lemma}
If {\bf X}$_D$ is not solvable in polynomial time relative to conjecture \textbf{C} then
{\bf X} is not solvable in polynomial time relative to conjecture \textbf{C}.
\label{LemAppProp1}
\end{lemma}

\begin{lemma}
Given a parameter-set $K$ for problem {\bf X}, 
if $\la K \ra$-{\bf X}$_D$ is not fixed-parameter tractable relative to conjecture \textbf{C} then
$\la K \ra$-{\bf X} is not fixed-parameter tractable relative to conjecture \textbf{C}.
\label{LemAppProp2}
\end{lemma}

\begin{lemma}
Given a parameter-set $K$ for problem {\bf X}, if {\bf X}$_D$
is $NP$-hard when the value of every parameter $k \in K$ is fixed to a
constant value, then $\la K \ra$-{\bf X} $\not\in FPT$ unless $P = NP$.
\label{LemAppProp3}
\end{lemma}

\subsection{Proofs for Controller / Environment Verification}

\vspace*{0.1in}

\begin{description}
\item[{\bf Result A.SF.1}:] If TeamEnvVerSF is polynomial-time exact solvable then 
                      $P = NP$.
\end{description}
\begin{proof}
Consider the following reduction from CDTMC to TeamEnvVerSF, based on the reduction
from CDTMC to TeamEnvVerST given in Lemma 3 in the supplementary materials of
\cite{WV18_SI}: Given an instance $\la M = (Q, \Sigma, \delta), x, k\rangle$
of CDTMC, construct an instance $\langle E, S, T, X, p_I, p_X\rangle$ of TeamEnvVerSF
as follows: Let $S$ be a set of point-fields based on field-quantities in the set
$FQ = \{ fq_y ~ | ~ y \in \Sigma\} ~ \cup ~ \{fq_B, fq_{F1}, fq_{X}\}$ where each
point-field has source-value and decay 1, $E$ be a $\max(k, |x|) + 2
\times 1$ grid in which the first $|x|$ squares encode $x$ using the first
$|\Sigma|$
field-quantities in $FQ$, the next $\max(k - |x|,
1)$ squares host point-fields based on field-quantity $fq_B$, and the final square hosts a
point-field based on field-quantity $fq_{F1}$, $p_I = E_{1,1}$, and $X$ is a single square 
at $p_X = E_{\max(k,|x|)+2,1}$. Let $T$ consist of a single FSR such that $Q' =
Q \cup \{q_{F1}\}$ and $\delta'$ consists of the FSR analogues of all transitions in
$\delta$ (phrased now in terms of eastward and westward movements in $E$)
plus the transitions $\langle q_A, *, *, goEast, q_{F1}\rangle$, 
$\langle q_{F1}, fval(fq_{F1}, =, 0.0), *, goEast, q_{F1}\rangle$, $\langle q_{F1},
fval(fq_{F1},\geq,1), fmod(s_X,(0,0)), stay,$ \linebreak $q_{F1}\rangle$, and $\langle q_{F1},
fval(fq_X,\geq,1), *, stay, q_{F1}\rangle$ where $q_A$ is the accepting state of $M$.
The TM $M$ underlying the single FSR in $T$ is deterministic in the classical
automata-theoretic sense \cite{HMU01} and thus has at most one
transition enabled at any time; hence, the operation of that FSR in $E$ is 
deterministic in the sense defined in this paper.
Note that this instance of TeamEnvVerSF can be constructed in time polynomial
in the size of the given instance of CDTMC.

Let us now prove that this reduction is correct, in that the answer for the given
instance of CDTMC is ``Yes'' if and only if the answer for the constructed instance
of TeamEnvVerSF is ``Yes''.
If $M$ accepts $x$ using at most $k$ tape squares, the single FSR in $T$
starting from $p_I$ (as it is based on the same transitions as $M$) will
eventually enter state $q_A$ sitting on one of the first $k$ squares of $E$. The
extra transitions described above will then ensure that the FSR proceeds
to the eastmost square of $E$, i.e., $p_X$, replaces $s_{F1}$ with $s_{X}$,
i.e., $X$, and then stays there. Conversely, if the single FSR in $T$ can
proceed from $p_I$ to create $X$ at $p_X$ then at some point it must have entered
$q_A$ (as only the extra transitions added to $M$ originating from $q_A$ could have
moved eastwards over the second-last square in $E$ to reach $p_X$), which
would imply that $M$ accepts $x$.

Given that CDTMC is $PSPACE$-complete \cite[Problem Al3]{GJ79}, the reduction above
establishes that TeamEnvVerSF is $PSPACE$-hard when $|T| = 1$. The result then
follows from the fact that 
$NP \subseteq PSPACE$,
\end{proof}

\subsection{Proofs for Controller Design by Library Selection}

\vspace*{0.1in}

\begin{description}
\item[{\bf Result B.ST.1} (Modified from \protect{\cite[Result B]{War22}}):] If 
ContDesLSST$^{fast}$ is \linebreak polynomial-time exact solvable then $P = NP$.
\end{description}
\begin{proof}
Consider the following reduction from {\sc Dominating set} to ContDesLSST$^{fast}_D$, based 
on the reduction from {\sc Dominating set} to ContDes$^{fast}_D$ given in Lemma 5 in the 
supplementary materials of \cite{WV18_SI} as modified in the proof of 
Result B in the supplementary materials of \cite{War22}. Given an instance $\la G = (V,E), k\rangle$
of {\sc Dominating Set}, construct an instance $\langle E', E'_T, |T|, p_I, X, p_X, |L|,
r, |Q|, d\rangle$ of ContDesLSST$^{fast}_D$ as follows: Let $E'$ be the environment
constructed in Lemma 5 in the supplementary materials of \cite{WV18_SI} with the
northwest $e_N$-based and SG1 subgrids removed, $E'_T$ be the version of $E_T$ in that
same lemma with $e_N$ and $e_E$ replaced by $e_B$ and $e_{F2}$ replaced by $e_X$, $p'_I = E'_{1,1}$,
$X$ be a single square of type $e_X$ at $p_X = E'_{|V| + 1,|V|^2+1}$, 
$L = 
\{\la q, enval(y, (0,0)), *, goEast, q'\ra ~ | ~ y \in \{e_1, \ldots, e_{|V|}\}\} \cup
\{\la q, enval(e_{F1}, (0,0)),$ \linebreak $enmod(e_X, (0,0)),$ $stay, q'\ra,
  \la q, *, *, goNorth, q'\ra\}$, 
$|T| = |Q| = 1$, $r = 0$, and $d = k + 2$. This instance 
of ContDesLSST$^{fast}_D$ can be constructed in time polynomial in the size of the given 
instance of {\sc Dominating Set}. 

Observe that the use of $L$ means that we no longer need 
subgrid SG1 and the restrictions on $|f|$ posited in Lemma 5 above to force the created
FSR to have $k$ east-moving transitions corresponding to a candidate dominating set of
$k$ distinct vertices in $G$. Hence, by slight simplifications and modifications of the 
proof of correctness of the reduction in Lemma 5 above, it can be shown that there is a 
dominating set of size $k$ in graph $G$ in the given instance of {\sc Dominating set} if and
only if there is an FSR with the structure specified in the constructed instance of 
ContDesLSST$^{fast}_D$ such that (1) $X$ can be constructed at $p_X$ if this FSR starts at 
$p_I$ and (2) the $k + 2$ transitions in this FSR are $k$ east-moving transitions from $L$
whose activation-formula predicates correspond to the vertices in a dominating set of 
size $k$ in $G$, and the final two transitions in $L$. As each transition in this 
FSR has an activation-formula consisting of either $*$ or a single predicate evaluating
if that square has a particular square-type, there can be at most one transition 
enabled at a time and the operation of this FSR in $E'$ is deterministic.
As the single robot in $T$ can only move north or east and
does one of either in each move, the number of transitions executed in this construction
task is the Manhattan distance from $p_I$ to $p_X$ in $E$. This distance is $(|V| + 1) +
(|V|^2 + 1) < |E| = c_1|E|^{c_2} < c_1(|E| + |Q|)^{c_2}$ when $c_1 = c+2 = 1$ , which
means that the construction task is $(c_1,c_2)$-completable when $c_1 = c_2 = 1$. 

As {\sc Dominating set} is $NP$-complete \cite[Problem GT2]{GJ79}, the reduction above
establishes that ContDesLSST$^{fast}_D$ is $NP$-hard; our result then follows from Lemma 
\ref{LemAppProp1}. To complete the proof, note that in the constructed instance of 
ContDesLSST$^{fast}_D$, $|T| = h = |Q| = |f| = |X| = 1$, $r = 0$, and $d = k + 2$.
\end{proof}

\begin{description}
\item[{\bf Result B.SF.1}:] If ContDesLSSF$^{fast}$ is polynomial-time exact solvable then 
                      $P = NP$.
\end{description}
\begin{proof}
As $|T| = 1$ and $r = 0$ in the instance of ContDesLSST$^{fast}_D$ constructed by the
reduction in the proof of Result B.ST.1, we can use the techniques illustrated in the
proof of Result A.SF.1 to simulate all square-types in $E'_T$ with point-fields that
have source-value and delay 1 and replace the visual sensing robot in $T$ with an
equivalent scalar-field sensing robot. The result then follows by appropriate modifications
to the proof of Result B.ST.1. To complete the proof, note that in the constructed instance
of ContDesLSSF$^{fast}_D$, $|T| = h = |Q| = |f| = |X| = 1$ and $d = k + 2$.
\end{proof}

\begin{description}
\item[{\bf Result B.ST.2}:] If $\la |T|, h, |Q|, d, |f|, r, |X|\ra$-ContDesLSST$^{fast}$ is 
                      fp-tractable then \linebreak $FPT = W[1]$.
\end{description}
\begin{proof}
Follows from the $W[2]$-hardness of $\la k \ra$-{\sc Dominating set} \cite{DF99},
the reduction from {\sc Dominating set} to ContDesLSST$^{fast}_D$ in the proof
of Result B.ST.1, the fact that $W[1] \subseteq W[2]$, and Lemma \ref{LemAppProp2}.
\end{proof}

\begin{description}
\item[{\bf Result B.ST.3}:] If $\la |T|, h, |Q|, d, |f|, |E_T|, |X|\ra$-ContDesLSST$^{fast}$ is 
                      fp-tractable then $FPT = W[1]$.
\end{description}
\begin{proof}
Consider the following reduction from {\sc Dominating set} to 
ContDesLSST$^{fast}_D$, based on the reduction from {\sc Dominating set} to 
ContDesLSST$^{fast}_D$ given in the proof of Result B.ST.1. Given an instance 
$\la G = (V,E), k\rangle$ of {\sc Dominating Set}, construct an instance 
$\langle E', E'_T, |T|, p_I, X, p_X, |L|, r,$ $|Q|, d\rangle$ of ContDesLSST$^{fast}_D$ as 
follows: Let $E'$ be the environment constructed in the proof of Result B.ST.1 with
each symbol-entry $x$ in each vertex-neighbourhood column replaced by (1) a row of 
$|V| + 1$ squares in which the $((|V| - i) + 1)$st square has type $e_V$, the
$(|V| + 1)$st square has type $e_M$, and all other squares have type $e_B$ if $x =  
e_{Vi}$, and (2) a row of $|V| + 1$ squares of type $e_B$ if $x = e_B $ (see Figure
\ref{FigRedEnvFix}(a) in the main text), $E'_T = \{e_B, e_M, e_{F1}, e_{robot}, e_X\}$,
$p'_I = E'_{1,1}$, $X$ be a single square of type $e_X$ at $p_X = 
E'_{2|V| + 1,|V|^2+1}$, $L = \{\la q, enval(e_M, (0,0)) ~\rm{and}~ enval(e_V, (0,-i)), *,$ \linebreak
$goEast, q'\ra ~ | ~ 1 \leq i \leq | V|\} ~ \cup ~
\{\la q, enval(e_{F1}, (0,0)), enmod(e_X, (0,0)), stay, q'\ra,$ \linebreak
  $\la q, *, *, goNorth, q'\ra\}$, 
$|T| = |Q| = 1$, $r = |V|$, and $d = k + 2$. This 
instance of ContDesLSST$^{fast}_D$ can be constructed in time polynomial in the size of 
the given instance of {\sc Dominating Set}. By slight modifications to the proof of 
correctness of the reduction in the proof of Result B.ST.1, it can be shown that there is a 
dominating set of size $k$ in graph $G$ in the given instance of {\sc Dominating set} if and
only if there is an FSR with the structure specified in the constructed instance of 
ContDesLSST$^{fast}_D$ such that (1) $X$ can be constructed at $p_X$ if this FSR starts at 
$p_I$ and (2) the $k + 2$ transitions in this FSR are $k$ east-moving transitions from $L$
whose activation-formula predicates correspond to the vertices in a dominating set of 
size $k$ in $G$, and the final two transitions in $L$. By the same arguments as those
given in the proof of Result B.ST.1, the operation of this FSR in $E'$ is deterministic
and the construction task is $(c_1,c_2)$-completable when $c_1 = c_2 = 1$. 

The result follows from the $W[2]$-hardness of $\la k \ra$-{\sc Dominating set} \cite{DF99},
the reduction above, the fact that $W[1] \subseteq W[2]$, and Lemma \ref{LemAppProp2}.
To complete the proof, note that in the constructed instance of 
ContDesLSST$^{fast}_D$, $|T| = h = |Q| = |X| = 1$, $|f| = 3$, $|E_T| = 5$, and $d = k + 2$.
\end{proof}

\begin{description}
\item[{\bf Result B.SF.2}:] If $\la |T|, h, |Q|, d, |f|, |X|\ra$-ContDesLSSF$^{fast}$ is 
                      fp-tractable then \linebreak $FPT = W[1]$.
\end{description}
\begin{proof}
Follows from the $W[2]$-hardness of $\la k \ra$-{\sc Dominating set} \cite{DF99},
the reduction from {\sc Dominating set} to ContDesLSST$^{fast}_D$ in the proof
of Result B.SF.1, the fact that $W[1] \subseteq W[2]$, and Lemma \ref{LemAppProp2}.
\end{proof}

\begin{description}
\item[{\bf Result B.SF.3}:] If $\la |T|, h, |Q|, |f|, |q_E|, |X|\ra$-ContDesLSSF$^{fast}$ is 
                      fp-tractable then \linebreak $P = NP$.
\end{description}
\begin{proof}
If we replace {\sc Dominating set} with {\sc Dominating set$^{PD3}$} in the reduction in 
the proof of Result B.SF.1, as each vertex in $G$ has degree 3, each vertex
point-field can occur at most four times (in the vertex-neighbourhood columns of that
vertex and its three adjacent vertices) and $|q_E| = 4$. The result then follows from the
$NP$-hardness of {\sc Dominating set$^{PD3}$} \cite[Problem GT2]{GJ79} and Lemma
\ref{LemAppProp3}.
\end{proof}

\begin{description}
\item[{\bf Result B.SF.4}:] If $\la |T|, h, |Q|, d, |f|, |S|, |X|\ra$-ContDesLSSF$^{fast}$ is 
                      fp-tractable then $FPT = W[1]$.
\end{description}
\begin{proof}
Consider the following reduction from {\sc Dominating set} to 
ContDesLSSF$^{fast}_D$, based on the reduction from {\sc Dominating set} to 
ContDesLSST$^{fast}_D$ given in the proof of Result B.ST.3. Given an instance 
$\la G = (V,E), k\rangle$ of {\sc Dominating Set}, construct an instance 
$\langle E', S, |T|, p_I, X, p_X, |L|, r, |Q|,$ $d\rangle$ of ContDesLSSF$^{fast}_D$ as 
follows: Let $E'$ be the environment constructed in the proof of Result B.ST.3 with
the following modifications: (1) each row is now separated by $2|V|$ blank
rows; (2) each marker-symbol $e_M$ is replaced by a point-field $s_M$ with
source-value 1 and decay 1; (3) each vertex-symbol is replaced by a point-field
$s_V$ with source-value $|V| + 1$ and decay 1; and (4) symbol $e_{F1}$ is replaced 
by a point-field $s_{F1}$ with source-value and decay 1 (see Figure
\ref{FigRedEnvFix}(c) in the main text). Let $S = \{s_M, s_V, s_{F1}, s_{robot}, s_X\}$,
$p'_I = E'_{1,1}$, $X$ be a single point-field of type $s_X$ at $p_X = 
E'_{2|V| + 1,2|V|^3+1}$, 
$L = \{\la q, fval(fq_M, \geq, 1.0) ~\rm{and}~ 
fval(fq_V, =, (|V| + 1) - i), *, goEast, q'\ra ~ | ~ 1 \leq i \leq | V|\} ~\cup~
\{\la q, fval(e_{F1}, \geq, 1.0), fmod(s_X, (0,0)), stay, q'\ra,
\la q, *, *, goNorth, q'\ra\}$, 
$|T| = |Q| = 1$, and $d = k + 2$. This instance of 
ContDesLSSF$^{fast}_D$ can be constructed in time polynomial in the size of 
the given instance of {\sc Dominating Set}. By slight modifications to the proof of 
correctness of the reduction in the proof of Result B.ST.3, it can be shown that there is a 
dominating set of size $k$ in graph $G$ in the given instance of {\sc Dominating set} if and
only if there is an FSR with the structure specified in the constructed instance of 
ContDesLSSF$^{fast}_D$ such that (1) $X$ can be constructed at $p_X$ if this FSR starts at 
$p_I$ and (2) the $k$ east-moving transitions in this FSR correspond to the vertices in a 
dominating set of 
size $k$ in $G$, and the final two transitions in $L$. By the same arguments as those
given in the proof of Result B.ST.1, the operation of this FSR in $E'$ is deterministic
and the construction task is $(c_1,c_2)$-completable when $c_1 = c_2 = 1$. 

The result follows from the $W[2]$-hardness of $\la k \ra$-{\sc Dominating set} \cite{DF99},
the reduction above, the fact that $W[1] \subseteq W[2]$, and Lemma \ref{LemAppProp2}.
To complete the proof, note that in the constructed instance of 
ContDesLSSF$^{fast}_D$, $|T| = h = |Q| = |X| = 1$, $|f| = 3$, $|S| = 5$, and $d = k + 2$.
\end{proof}

\begin{description}
\item[{\bf Result B.ST.4}:] $\la |Q|, |L|\ra$-ContDesLSST$^{fast}$ is fp-tractable.
\end{description}
\begin{proof}
Consider the algorithm that generates all possible controllers and then for
each controller $c$ checks in polynomial time if a team $T$ composed of
$c$ started at $p_I$ will create $X$ at $p_X$. The number of controllers
is upper-bounded by $|L|^{d|Q|}|Q|^{d|Q|}$, which, as $c$ is deterministic and
hence $d \leq |L|$, is in turn upper-bounded by a function of $|Q|$ and $|L|$.
As the construction task is assumed to be $(c_1,c_2)$-completable, the checking 
of the team with any such controller can be done in $c_1(|E| + |Q|)^{c_2}$
multiplied by some polynomial of the input size (i.e., the time required to
simulate $T$ in $E$ for one timestep), completing the proof.
\end{proof}

\begin{description}
\item[{\bf Result B.SF.5}:] $\la |Q|, |L|\ra$-ContDesLSSF$^{fast}$ is fp-tractable.
\end{description}
\begin{proof}
Follows from the algorithm in the proof of Result B.ST.4.
\end{proof}

\subsection{Proofs for Team Design by Library Selection}

\vspace*{0.1in}

\begin{description}
\item[{\bf Result C.ST.4}:] 
If $\la |T|, h, |Q|, d, r, |X|\ra$-TeamDesLSST$^{fast}$ is fp-tractable then \linebreak $FPT = W[1]$.
\end{description}
\begin{proof}
Modify the reduction in the proof of Lemma A.2 in the appendix of \cite{War19}
such that in transition-groups
(5), (6), and (7) for vertex neighborhood robots, instead of having potentially
$|V|$ transitions in each group, have a single transition whose transition
formula ORs together all of the predicates of the form $enval(e_{vj}, (0,-1))$ to create
a parenthesis-enclosed subformula that is then ANDed with the other predicates in 
each existing transition trigger formula. Note that this modified reduction still runs
in time polynomial in the size of the given instance of {\sc Dominating set};
moreover, this reduction is correct by slight modifications of the arguments given
for the proof of correctness of the reduction in the proof of Lemma A.2.
The result then  follows from the $W[2]$-hardness of $\la k \ra$-{\sc Dominating set} 
\cite{DF99}, the modified reduction above, the fact that $W[1] \subseteq W[2]$, and
Lemma \ref{LemAppProp2}.  To complete the proof, note that in the constructed instance of 
TeamDesLSST$^{fast}_D$, $r = |X| = 1$, $|Q| = 3$, $d = 7$, and $|T| = h = k + 1$.
\end{proof}

\begin{description}
\item[{\bf Result C.ST.5}:] 
If $\la |T|, h, |Q|, d, |E_T|, |X|\ra$-TeamDesLSST$^{fast}$ is fp-tractable
then \linebreak $FPT = W[1]$.
\end{description}
\begin{proof}
Modify the reduction in the proof of Lemma A.3 in the appendix of \cite{War19} such that in
transition-groups (5), (6), and (7) for vertex neighborhood robots, instead of having 
potentially $|V|$ transitions in each group, have a single transition whose transition
formula ORs together all of the subformulas used to recognize $v_i$, $1 \leq i \leq |V|$,
to create a parenthesis-enclosed subformula that is then ANDed with the other predicates in 
each existing transition trigger formula. Note that this modified reduction still runs
in time polynomial in the size of the given instance of {\sc Dominating set};
moreover, this reduction is correct by slight modifications of the arguments given
for the proof of correctness of the reduction in the proof of Lemma A.3. The result then 
follows from the $W[2]$-hardness of $\la k \ra$-{\sc Dominating set} \cite{DF99}, the 
modified reduction above, the fact that $W[1] \subseteq W[2]$, and Lemma \ref{LemAppProp2}.
To complete the proof, note that in the constructed instance of TeamDesLSST$^{fast}_D$, 
$|X| = 1$, $|Q| = 3$, $d = 7$, $|E_T| = 13$, and $|T| = h = k + 1$.
\end{proof}

\begin{description}
\item[{\bf Result C.SF.1}] \cite[Result A.1]{WV21}: TeamDesLSSF$^{fast}$ is exact polynomial-time \linebreak solvable when $h = 1$.
\end{description}
\begin{proof}
Follows from the algorithm in the proof of Result A in \cite{War19}.
\end{proof}

\begin{description}
\item[{\bf Result C.SF.2}] \cite[Result A.2]{WV21}: If TeamDesLSSF$^{fast}$ is exact polynomial-time \linebreak solvable when $h > 1$
                      then $P = NP$.
\end{description}
\begin{proof}
Consider the following reduction from {\sc Dominating set} to \linebreak 
TeamDesLSSF$^{fast}_D$, based on the reduction from {\sc Dominating set} to 
DesCon$^D_{syn}$ given in the proof of Lemma A.2 in \cite{War19}.
Given an instance $\langle G = (V, E), k\rangle$ of {\sc Dominating set}, construct an 
instance $\langle E', S, |T|, L, E_I, X, p_X\rangle$ of TeamDesLSSF$^{fast}_D$ 
as follows: Let $E'$ be a $(|V| + 8) \times 5$ grid containing point-fields from the set
$S = \{s_N, s_S, s_E, s_W, s_{v1}, s_{v2}, \ldots s_{v|V|}, s_T, 
s_{M1}, s_{M2}, s_{M3}, s_{robot}, s_X\}$, each with source-value 1 and decay 0.5, as follows:

\begin{itemize}
\item $E'_{1,2} = s_E$.
\item $E'_{|V|+8,2} = E'_{|V|+8,3} = s_N$.
\item For $3 \leq i \leq (|V| + 7)$, $E'_{i,5} = s_W$.
\item $E'_{1,3} = E'_{1,4} = s_S$.
\item $E'_{3,3} = s_{M1}$.
\item $E'_{|V|+4,3} = s_{M2}$.
\item $E'_{|V|+5,3} = s_T$.
\item $E'_{|V|+6,3} = s_{M3}$.
\item For $1 \leq i \leq |V|$, $E'_{i+3,1} = s_{vi}$.
\end{itemize}

\noindent
An example environment $E'$ when $|V| = 5$ and $k = 2$ is shown in Figure 
\ref{FigEnvType}(a) in the main text. Let $|T| = k + 1$, 
$E_I = \{ E'_{j,4} ~ | ~ 3 \leq j \leq k + 3\}$, 
and FSR library $L = \{r_{v1}, r_{v2}, \ldots, r_{v|V|}, 
r_{chk}\}$ consist of $|V|$ vertex neighbourhood robots $r_{vi}$, $1 \leq i \leq |V|$, and 
a checker robot $r_{chk}$ that are specified as follows:

\begin{itemize}
\item{
{\bf Vertex neighbourhood robot}: Each vertex neighbourhood robot $r_{vi}$ is based on a single 
state $q_0$ and has the following transitions:

\begin{enumerate}
\item $\langle q_0, (fval(fq_E, =, 0.5)$ or $fval(fq_{M1}, =, 0.5)$ or \linebreak
                     $fval(fq_{M2}, =, 0.5)$ or $fval(fq_T, =, 0.5)$ or 
                     $fval(fq_{M3}, =, 0.5)$ or $fval(fq_X, =, 0.5))$ and 
	       $(fval(fq_N, =, 0.0)$ and $fval(fq_S, =, 0.0)$ and $fval(fq_W, =, 0.0))$ and
               $(fval(fq_{robot}, \leq, 1.5)$ and $fgrd(fq_{robot}, >, East)),
	       *, goEast, q_0\rangle$.
\item $\langle q_0, fval(fq_N, =, 0.5)$ and 
               $(fval(fq_{robot}, \leq, 1.5)$ and $fgrd(fq_{robot}, >, North)),$ \linebreak
               $*, goNorth, q_0\rangle$.
\item $\langle q_0, fval(fq_W, =, 0.5)$ and 
               $(fval(fq_{robot}, \leq, 1.5)$ and $fgrd(fq_{robot}, >, West)),$
               \linebreak $*, goWest, q_0\rangle$.
\item $\langle q_0, fval(fq_S, =, 0.5)$ and 
               $(fval(fq_{robot}, \leq, 1.5)$ and $fgrd(fq_{robot}, >, South)),$
               \linebreak $*, goSouth, q_0\rangle$.
\item For $v_j \in N_C(v_i)$, $\langle q_0, fval(fq_{vj}, =, 0.5)$ and 
                                             $fval(s_X, =, 0.0)$ and \linebreak
               $(fval(fq_{robot}, \leq, 1.5)$ and $fgrd(fq_{robot}, >, East)),
               fmod(s_X,(0,1)), goEast, q_0\rangle$.
\item For $v_j \in N_C(v_i)$, $\langle q_0, fval(fq_{vj}, =, 0.5)$ and 
                                             $fval(s_X, =, 0.5)$ and \linebreak
               $(fval(fq_{robot}, \leq, 1.5)$ and $fgrd(fq_{robot}, >, East)),
               *, goEast, q_0\rangle$.
\item For $v_j \not\in N_C(v_i)$, $\langle q_0, (fval(fq_{vj}, =, 0.5)$ and  \linebreak
               $(fval(fq_{robot}, \leq, 1.5)$ and $fgrd(fq_{robot}, >, East)),
               *, goEast, q_0\rangle$.
\end{enumerate}

\noindent
The transitions in (2), (3), and (4) allow $r_{vi}$ to progress northwards along the
west, westwards along the north, and southwards along the east sides of the movement-track, 
respectively. All remaining transitions cover movement on the south side of the 
movement-track. The transitions in (5) allow $r_{vi}$ to put a point-field of type $s_X$ in
the scaffolding to its immediate north and progress eastwards if there is a point-field to 
its immediate south whose type corresponds to a vertex in the complete neighbourhood of 
$v_i$ in $G$, while the transitions in (6) and (7) allow eastward progression if either such
an $s_X$ is already in place or the type of the point-field to the immediate south 
corresponds to a vertex that is not in the complete neighbourhood of $v_i$ in $G$. Finally,
transition (1) allows $r_{vi}$ to progress eastwards over all remaining positions on 
the south side of the movement-track. 
}
\item{
{\bf Checker robot}:
The checker robot $r_{chk}$ is based on three states $q_0$, $q_1$, and $q_2$ and has 
the following transitions:

\begin{enumerate}
\item $\langle q_0, fval(fq_E, =, 0.5)$ and 
                    $(fval(fq_{robot}, \leq, 1.5)$ and $fgrd(fq_{robot}, >, East)),$ \linebreak
                    $*, goEast, q_0\rangle$.
\item $\langle q_0, fval(fq_N, =, 0.5)$ and 
                    $(fval(fq_{robot}, \leq, 1.5)$ and $fgrd(fq_{robot}, >, North)),$ \linebreak
                    $*,$ $ goNorth, q_0\rangle$.
\item $\langle q_0, fval(fq_W, =, 0.5)$ and 
                    $(fval(fq_{robot}, \leq, 1.5)$ and $fgrd(fq_{robot}, >, West)),$ \linebreak
                    $*,$ $ goWest, q_0\rangle$.
\item $\langle q_0, fval(fq_S, =, 0.5)$ and 
                    $(fval(fq_{robot}, \leq, 1.5)$ and $fgrd(fq_{robot}, >, South)),$ \linebreak
                    $*,$ $ goSouth, q_0\rangle$.
\item $\langle q_0, fval(fq_{M1}, =, 0.5)$ and 
                    $(fval(fq_S, =, 0.0)$ and $fval(fq_W, =, 0.0))$ and 
                    $(fval(fq_{robot}, \leq, 1.5)$ and $fgrd(fq_{robot}, >, East)),
                    *, goEast, q_1\rangle$.
\item $\langle q_1, fval(fq_X, =, 0.5)$ and 
                    $(fval(fq_{robot}, \leq, 1.5)$ and $fgrd(fq_{robot}, >, East)),$ \linebreak
                    $*, goEast, q_1\rangle$.
\item $\langle q_1, fval(fq_{M2}, =, 0.5)$ and 
                    $(fval(fq_{robot}, \leq, 1.5)$ and $fgrd(fq_{robot}, >, East)),$ \linebreak
                    $*, goEast, q_1\rangle$.
\item $\langle q_1, fval(fq_T, =, 0.5)$ and 
                    $(fval(fq_{robot}, \leq, 1.5)$ and $fgrd(fq_{robot}, >, East)),$ \linebreak
                    $fmod(s_X, (0,1)), goEast, q_1\rangle$.
\item $\langle q_1, fval(fq_{M3}, =, 0.5)$ and 
                    $(fval(fq_{robot}, \leq, 1.5)$ and $fgrd(fq_{robot}, >, East)),$ \linebreak
                    $*, goEast, q_0\rangle$.
\item $\langle q_1, fval(fq_X, =, 0.0)$ and 
                    $(fval(fq_{robot}, \leq, 1.5)$ and $fgrd(fq_{robot}, >, East)),$ \linebreak
                    $*, goEast, q_2\rangle$.
\item $\langle q_2, (fval(fq_X, =, 0.5)$ or $fval(fq_X, =, 0.0)$ or
                     $fval(fq_{M2}, =, 0.5)$ or \linebreak $fval(fq_T, =, 0.5))$ and
                    $(fval(fq_{robot}, \leq, 1.5)$ and $fgrd(fq_{robot}, >, East)),$ \linebreak
                    $*, goEast, q_2\rangle$.
\item $\langle q_2, fval(fq_{M3}, =, 0.5)$ and 
                    $(fval(fq_{robot}, \leq, 1.5)$ and $fgrd(fq_{robot}, >, East)),$ \linebreak
                    $*, goEast, q_0\rangle$.
\end{enumerate}

\noindent
The transitions in (2), (3), and (4) allow $r_{chk}$ to progress northwards along 
the west, westwards along the north, and southwards along the east sides of the 
movement-track, respectively. All of this movement is done when $r_{chk}$ is
in state $q_0$. All remaining transitions cover movement on the south side of the 
movement-track. On entering the south side, $r_{chk}$ enters state $q_1$ (transition 
(5)) and remains there if the scaffolding has been completely filled in by the vertex 
neighbourhood robots (transitions in (6--7)). This then allows $r_{chk}$ to create $X$ 
at $p_X$ (transition (8)) and then return to state $q_0$ (transition (9)). If at
any point $r_{chk}$ finds that part of the scaffolding has not been filled in 
(transition (10)), it enters state $q_2$, skips over the remainder of the scaffolding
(transition (11)) and then returns to state $q_0$ (transition (12)). Finally,
transition (1) allows $r_{chk}$ to progress eastwards from the westmost square on
the south side of the movement track.
}
\end{itemize}

\noindent
Note the use in all of the transitions described above of the robot collision-avoidance 
formula $(fval(fq_{robot}, \leq, 1,5)$ and $fgrd(fq_{robot}, >, dir))$ derived in Section
\ref{SectResTeamDesLS} that allows moves in direction $dir$ along the movement-track.
Finally, let $X$ be a single point-field of type $s_X$ at position $p_X = E'_{|V|+5,3}$.
Note that this instance of TeamDesLSSF$^{fast}_D$ can be constructed in 
time polynomial in the size of the given instance of {\sc Dominating set}. 

By slight modifications to the proof of correctness of the reduction in the proof of
Lemma A.2 in \cite{War19}, it can be shown that there is a dominating set of size at most 
$k$ in $G$ in the given instance of {\sc Dominating set} if and only if there is robot team
$T$ consisting of $k + 1$ robots from $L$ in the constructed instance of 
TeamDesLSSF$^{fast}_D$ such that, when started at any positioning of the team members in 
$E_I$, $T$ constructs $X$ at $p_X$. By the same arguments as those given in the proof of 
Lemma A.2 in \cite{War19}, the operation of $T$ in $E'$ is deterministic.  With
respect to $(c_1, c_2)$-completability, observe that in the worst case, all 
vertex neighbourhood robots have to pass below $s_{M1}$ and $s_{M2}$ on the south side 
of the movement-track (in order to fill in all point-fields of type $s_X$ in the
scaffolding) and then the checker robot must similarly pass below $s_{M1}$ and
$s_{M2}$ on the south side of the movement-track (to ensure that all positions in
the scaffolding have been filled in). As the positions of the members of $T$ in $E_I$ are
not guaranteed, this means that all robots need to progress around the movement-track at
most twice relative to their starting positions. As the movement-track is of length
$2|V| + 14$ and at least one robot moves forward in each timestep (as there will always
be at least one robot with a free square in front of it on the movement track that is not 
situation (h) in Figure \ref{FigAdjRobSF} in the main text), this means that in the worst 
case at most $2(2|V| + 14)(k + 1) \leq (4|V| + 28)(|V| + 1)$ timesteps
are required for $T$ to construct $X$ at $p_X$. However, as $|E| = 5|V| + 40$,
%$(4|V| + 28)(|V| + 1) < (5|V| + 40)(|5|V| + 40) = c_1|E|^{c_2} < c_1(|E| + |Q|)^{c_2}$
when $c_1 = 1$ and $c_2 = 2$, which means that this construction task is 
$(c_1,c_2)$-completable wrt $T$ when $c_1 = 1$ and $c_2 = 2$.

The result follows from the $NP$-hardness of {\sc Dominating set} \cite[Problem GT2]{GJ79},
the reduction above, and Lemma \ref{LemAppProp1}. To complete the proof, note that in the 
constructed instance of TeamDesLSSF$^{fast}_D$, $|X| = 1$, $|Q| = 3$, $|f| = 27$, and $|T| = h = k + 1$.
\end{proof}

\begin{description}
\item[{\bf Result C.SF.3}] \cite[Result A.3]{WV21}: If $\la |T|, h, |Q|, |f|, |X|\ra$-TeamDesLSST$^{fast}$ is \linebreak fp-tractable
                      then $FPT = W[1]$.
\end{description}
\begin{proof}
Follows from the $W[2]$-hardness of $\la k \ra$-{\sc Dominating set} \cite{DF99},
the reduction from {\sc Dominating set} to TeamDesLSSF$^{fast}_D$ in the proof
of Result C.SF.2, the fact that $W[1] \subseteq W[2]$, and Lemma \ref{LemAppProp2}.
\end{proof}

\begin{description}
\item[{\bf Result C.SF.4}] \cite[Result A.4]{WV21}: If $\la |T|, h, |Q|, d, |X|\ra$-TeamDesLSST$^{fast}$ is \linebreak fp-tractable
                      then $FPT = W[1]$.
\end{description}
\begin{proof}
Modify the reduction in the proof of Result C.SF.2 such that in transition-groups
(5), (6), and (7) for vertex neighborhood robots, instead of having potentially
$|V|$ transitions in each group, have a single transition whose transition
formula ORs together all of the predicates of the form $fval(fq_{vj}, =, 0.5)$ to create
a parenthesis-enclosed subformula that is then ANDed with the other three predicates in 
each existing transition trigger formula. Note that this modified reduction still runs
in time polynomial in the size of the given instance of {\sc Dominating set};
moreover, this reduction is correct by slight modifications of the arguments given
for the proof of correctness of the reduction in the proof of Result C.SF.2.
The result then  follows from the $W[2]$-hardness of $\la k \ra$-{\sc Dominating set} 
\cite{DF99}, the modified reduction above, the fact that $W[1] \subseteq W[2]$, and
Lemma \ref{LemAppProp2}.  To complete the proof, note that in the constructed instance of 
TeamDesLSSF$^{fast}_D$, $|X| = 1$, $|Q| = 3$, $d = 7$, and $|T| = h = k + 1$.
\end{proof}

\begin{description}
\item[{\bf Result C.SF.5}] \cite[Result A.5]{WV21}: If $\la |T|, h, |Q|, |f|, |S|, |X|\ra$-TeamDesLSST$^{fast}$ is \linebreak fp-tractable
                      then $FPT = W[1]$.
\end{description}
\begin{proof}
Consider the reduction in the proof of Result C.SF.2 relative to the following
modifications:

\begin{enumerate}
\item{
Delete all point-fields in the set 
$\{ s_{vi} ~ | ~ 1 \leq i \leq |V|\}$ from $S$
and add to $S$ point-fields $s_M$ and $s_B$ with source-value 1 and decay 0.5 and a 
point-field $s_V$ with source-value $|V| + 1$ and decay 1.
}
\item{
Expand the environment in the north-south direction by $|V|$ squares to replace each 
square with point-field $s_{vi}$, $1 \leq i \leq |V|$ with a vertex column consisting
of $|V| + 1$ squares with a point-field of type $s_M$ in the northmost square of that column
and a point-field of type $s_V$ $i$ squares below $s_M$ to the south.
}
\item{
Expand the environment created in (2) above in the east-west direction by $2|V|$ \linebreak $(|V| - 1)$ 
squares to place between each pair of adjacent vertex-columns an additional $2|V|$
blank columns, each consisting of $|V| + 1$ squares, in which the northmost square in the
column has a point-field of type $s_B$. The northmost square in the environment above
each such column will have a point-field $s_W$. 
}
\item{
Alter the positions of all other point-fields in $E'$ and the positions in
$E_I$, and $p_X$ accordingly given the environment-expansions in (2) and (3) above.
}
\item{
In the transition-set of each vertex neighborhood robot, replace the term \linebreak
$fval(fq_{vi}, = 0.5)$, $1 \leq i \leq |V|$, with the formula 
$(fval(fq_M, =, 0.5)$ and $fval(fq_V,$ $=, (|V| + 1) - i))$.
}
\item{
Add the transition $\la q_0, fval(fq_B, =, 0.5), fmod(s_X, (0,1)), goEast\ra$
to the \linebreak transition-set of each neighbourhood robot.
}
\end{enumerate}

\noindent
Modifications (1), (2), and (3) above are analogous to those developed in the proof
of Result B.SF.4 to reduce the size of $S$. Modifications (5) and (6) are required
to make the vertex neighborhood robots fill in the portions of the central scaffolding 
above the new blank columns and thus allow the checker robot to function as in the proof of
Result C.SF.2.  Note that this instance of TeamDesLSSF$^{fast}_D$ can be constructed in 
time polynomial in the size of the given instance of {\sc Dominating set}. 

By slight modifications to the proof of correctness of the reduction in the proof of
Lemma A.2 in \cite{War19}, it can be shown that there is a dominating set of size at most 
$k$ in $G$ in the given instance of {\sc Dominating set} if and only if there is robot team
$T$ consisting of $k + 1$ robots from $L$ in the constructed instance of 
TeamDesLSSF$^{fast}_D$ such that, when started at any positioning of the team members in 
$E_I$, $T$ constructs $X$ at $p_X$. By the same arguments as those given in the proof of 
Lemma A.2 in \cite{War19}, the operation of $T$ in $E'$ is deterministic.  With
respect to $(c_1, c_2)$-completability, as all robots need to progress around the 
movement-track twice relative to their starting positions, the movement-track is now of 
length
$2(|V| + 2|V|(|V| - 1)) + 14 =
 2(|V| + 2|V|^2 - 2|V|) + 14 =
 2|V| + 4|V|^2 - 2|V| + 14 =
 4|V|^2 + 14$, 
and at least one robot moves forward in each timestep,
this means that in the worst 
case at most $2(4|V|^2 + 14)(k + 1) 
              \leq (8|V|^2 + 28)(|V| + 1)
              =    8|V|^3 + 8|V|^2 + 28|V| + 28$ timesteps
are required for $T$ to construct $X$ at $p_X$. However, as 

\begin{eqnarray*}
|E| & = & (|V| + 5)(|V| + 2|V|(|V| - 1) + 8) \\
    & = & (|V| + 5)(2|V|^2 - 2|V| + 8) \\
    & = & (2|V|^3 - 2|V|^2 + 8|V| + 10|V|^2 -10|V| + 40) \\
    & = & 2|V|^3 + 8|V|^2 - 2|V| + 40 
\end{eqnarray*}

\noindent
then

\begin{eqnarray*}
8|V|^3 + 8|V|^2 + 28|V| + 28 & < & 8|V|^3 + 36|V|^2 + 28  \\
                             & < & 6(2|V|^3 + 8|V|^2 - 2|V|^2 + 40)  \\
                             & = & 6(2|V|^3 + 6|V|^2  + 40)  \\
                             & = & 12|V|^3 + 36|V|^2 + 240 \\ 
                             & < & c_1|E|^{c_2} \\
                             & < & c_1(|E| + |Q|)^{c_2} 
\end{eqnarray*}

\noindent
when $c_1 = 6$ and $c_2 = 1$, which means that this construction task is 
$(c_1,c_2)$-completable wrt $T$ when $c_1 = 6$ and $c_2 = 1$.

The result follows from the $W[2]$-hardness of $\la k \ra$-{\sc Dominating set} 
\cite{DF99}, the reduction above, the fact that $W[1] \subseteq W[2]$, and Lemma 
\ref{LemAppProp2}. To complete the proof, note that in the constructed instance of 
TeamDesLSSF$^{fast}_D$, $|X| = 1$, $|Q| = 3$, $|S| = 13$, $|f| = 31$, and $|T| = h = k + 1$.
\end{proof}

\begin{description}
\item[{\bf Result C.SF.6}] \cite[Result A.6]{WV21}: If $\la |T|, h, |Q|, d, |S|, |X|\ra$-TeamDesLSST$^{fast}$ is \linebreak fp-tractable 
                      then $FPT = W[1]$.
\end{description}
\begin{proof}
Modify the reduction in the proof of Result C.SF.5 such that in the modified 
transition-groups (5), (6), and (7) for vertex neighborhood robots, instead of having 
potentially $|V|$ transitions in each group, have a single transition whose transition
formula ORs together all of the subformulas of the form $(fval(fq_M, =, 0.5) ~\rm{and}~
fval(fq_V, = (|V| + 1)  - i))$ to create a parenthesis-enclosed subformula that is then 
ANDed with the other three predicates in each existing transition trigger formula. Note that
this modified reduction still runs in time polynomial in the size of the given instance of 
{\sc Dominating set}; moreover, this reduction is correct by slight modifications of the 
arguments given for the proof of correctness of the reduction in the proof of Result C.SF.5.
The result then  follows from the $W[2]$-hardness of $\la k \ra$-{\sc Dominating set} 
\cite{DF99}, the modified reduction above, the fact that $W[1] \subseteq W[2]$, and
Lemma \ref{LemAppProp2}.  To complete the proof, note that in the constructed instance of 
TeamDesLSSF$^{fast}_D$, $|X| = 1$, $|Q| = 3$, $d = 7$, $|S| = 13$, and $|T| = h = k + 1$.
\end{proof}

\begin{description}
\item[{\bf Result C.SF.7}] \cite[Result A.7]{WV21}: $\la |T|, |L|\ra$-TeamDesLSST$^{fast}$ is fp-tractable.
\end{description}
\begin{proof}
Follows from the algorithm in the proof of Result G in \cite{War19}.
\end{proof}

\subsection{Proofs for Environment Design}

\vspace*{0.1in}

\begin{description}
\item[{\bf Result D.SF.1}] \cite[Result B.1]{WV21} If EnvDesSF$^{fast}$ is polynomial-time exact solvable then $P = NP$.
\end{description}
\begin{proof}
Consider the following reduction from {\sc Dominating set} to EnvDesSF$^{fast}_D$.
Given an instance $\la G = (V,E), k\ra$ of {\sc Dominating set}, construct
instance $\la G', S, T,$ \linebreak $p_I, X, p_X\ra$ of EnvDesSF$^{fast}_D$ as follows: Let
$G'$ be a $1 \times (|V| + 1)$ grid, $S = \{s_S, s_{v1}, s_{v2},$ \linebreak $\ldots,
s_{v|V|}, s_{robot}, s_X\}$ be
a set of point-fields such that $s_S$ has source-value and decay 1 and $s_{vi}$,
$1 \leq i \leq |V|$, has source-value $|V| + 1$ and decay 1, and $T$ consist of
a single robot $r$ based on set $Q = \{q_0, q_{v0}, q_{v1}, q_{v2}, \ldots, q_{v|V|}\}$
with the following transitions:

\begin{enumerate}
\item $\la q_0, fval(fq_S, =, 1.0)), *, stay, q_{v0}\ra$
\item $\{ \la q_{v(i-1)}, fval(fq_{vj}, \geq, 1.0), *, stay,  q_{vi}\ra ~ | ~ 
             1 \leq i \leq |V| ~\rm{and}~ v_j \in N_C(v_i) \}$
\item $\la q_{v|V|}, fval(fq_S, =, 1.0), fmod(s_X, (0,0)), q_{v|V|}\ra$
\end{enumerate}

\noindent
Finally, let $p_I = E'_{1,1}$ and $X$ be a single point-field of type $s_X$ at
$p_X = E'_{1,1}$. Note that this instance of EnvDesSF$^{fast}_D$ can be constructed
in time polynomial in the size of the given instance of {\sc Dominating set}.

Let us now prove that this reduction is correct, in that the answer to the given
instance of {\sc Dominating set} is ``Yes'' if and only if the answer for the
constructed instance of EnvDesSF$^{fast}_D$ is ``Yes''. 

\begin{itemize}
\item{Suppose there is 
dominating set of size at most $k$ in the given graph $G$. Let the set $V' \subseteq V$
be the vertices in this dominating set; if $|V'| < k$, add an arbitrary set of
$k - |V'|$ vertices in $G$ to $V'$. Construct an environment $E'$ such that
square $E'_{1,1}$ contains point-field $s_S$ and square $E'_{1,i+1}$, $1 \leq i \leq k$,
contains the point-field $s_j$, where $v_j$ is the $i$th vertex in $V'$. In such
an environment, observe that $r$ started at $p_I$ will progress from $q_0$ to $q_{v|V|}$ and
construct $X$ at $p_X$. At any point in the operation of $r$, all enabled transitions
relative to current state $q$ do the same thing, which means that the operation of $r$
in $E'$ is deterministic in the sense described in Section \ref{SectForm}. With
respect to $(c_1,c_2)$-completability, observe that $r$ needs $|Q| + 1$ timesteps to
progress from $q_0$ to $q_{v|V|}$ and then construct $X$ at $p_X$. As $|Q| + 1 <
c_1(|E| + |Q|)^{c_2}$ when $c_1 = c_2 = 1$, this means that the task of constructing
$X$ at $p_X$ is $(c_1,c_2)$-completable when $c_1 = c_2 = 1$.
}
\item{
Conversely, suppose there is an environment $E'$ based on
$G'$ and $S$ such that robot $r$ started at $p_I$ constructs $X$ at $p_X$. In order
to progress from $q_0$ to $q_{v0}$, $E'_{1,1}$ must contain the point-field $s_S$; 
moreover, for $1 \leq i \leq |V|$, by the structure of the transition-set of $r$,
$r$ can only progress from $q_{v(i-1)}$ to $q_i$ (and hence eventually trigger the
transition from $q_{v|V|}$ that constructs $X$ at $p_X$) if there is a vertex $v$ in the
dominating set encoded in $E'$ such that $v \in N_C(v_i)$ in $G$.\footnote{
Note that in this case, a vertex $v$ may appear multiple times in the dominating set
encoded in $E'$. This is not a problem here, as (1) the $fval()$ predicates in the
transition trigger formulas in $r$ will still register the presence of $v$ in the encoded 
candidate solution even if the associated point-fields interfere to boost the value of 
$fq_v$ at $E'_{1,1}$ and (2) the vertices in the encoded dominating set need not be distinct
as we are are only interested in the presence of a dominating set of size {\em at most}
$k$ in $G$.
}
Thus, $r$ started at $p_I$ can construct $X$ at $p_X$ in $E'$ only if a dominating set of 
size at most $k$ in $G$ is encoded in the northmost $k$ squares of the first
column in $E'$.
}
\end{itemize}

\noindent
This proves the correctness of the reduction. The result then
follows from the $NP$-hardness of {\sc Dominating set} \cite[Problem GT2]{GJ79},
the reduction above, and Lemma \ref{LemAppProp1}. To complete the proof, note that in
the constructed instance of EnvDesSF$^{fast}_D$, $|T| = h = |f| = |X| = 1$ and
$|E| = |S_E| = |q_E| = 2(k + 1)$.
\end{proof}

\begin{description}
\item[{\bf Result D.SF.2}] \cite[Result B.2]{WV21}
If $\la |T|, h, |f|, |E|, |S_E|, |q_E|, |X|\ra$-EnvDesSF$^{fast}$ is \linebreak fp-tractable
                      then $FPT = W[1]$.
\end{description}
\begin{proof}
Follows from the $W[2]$-hardness of $\la k \ra$-{\sc Dominating set} \cite{DF99}, the
reduction in the proof of Result D.SF.1, the fact that $W[1] \subseteq W[2]$, and
Lemma \ref{LemAppProp2}.
\end{proof}

\begin{description}
\item[{\bf Result D.SF.3}] \cite[Result B.3]{WV21}: 
If $\la |T|, h, |Q|, d, |E|, |S_E|, |q_E|, |X|\ra$-EnvDesSF$^{fast}$ is fp-tractable
                      then $FPT = W[1]$.
\end{description}
\begin{proof}
Modify the reduction in the proof of Result D.SF.1 such that for $1 \leq i \leq |V|$,
the set of transitions 
$\{ \la q_{v(i-1)}, fval(fq_{vj}, \geq, 1.0), *, stay,  q_{vi}\ra 
~ | ~ v_j \in N_C(v_i) \}$ 
in transition-set (2) and the single transitions in sets
(1) and (3) is replaced by a single transition $\la q_0, f, fmod(s_X, (0,0)), stay, 
q_0\ra$, where $f$ is the formula AND-ing the subformulas OR-ing all $fval()$ predicates 
in the trigger formulas of the transitions in the sets for the various $v_i$ in 
transition-set (2) and then AND-ing this
with the predicate $fval(fq_S, =, 1.0)$. Note that this modified reduction still runs
in time polynomial in the size of the given instance of {\sc Dominating set};
moreover, this reduction is correct by slight modifications of the arguments given
for the proof of correctness of the reduction in the proof of Result D.SF.1.
The result then  follows from the $W[2]$-hardness of $\la k \ra$-{\sc Dominating set} 
\cite{DF99}, the modified reduction above, the fact that $W[1] \subseteq W[2]$, and
Lemma \ref{LemAppProp2}.  To complete the proof, note that in
the constructed instance of EnvDesSF$^{fast}_D$, $|T| = h = |Q| = d = |X| = 1$ and
$|E| = |S_E| = |q_E| = 2(k + 1)$.
\end{proof}

\begin{description}
\item[{\bf Result D.SF.4}] \cite[Result B.4]{WV21}: If $\la |T|, h, |f|, |S|, |S_E|, |q_E|, |X|\ra$-EnvDesSF$^{fast}$ is 
                      fp-tractable then $FPT = W[1]$.
\end{description}
\begin{proof}
Consider the following reduction from {\sc Dominating set} to EnvDesSF$^{fast}_D$.
Given an instance $\la G = (V,E), k\ra$ of {\sc Dominating set}, construct
instance $\la G', S, T,$ \linebreak $p_I, X, p_X\ra$ of EnvDesSF$^{fast}_D$ as follows: Let
$G'$ be a $3 \times (|V| + 2)$ grid, $S = \{s_P, s_S, s_N,$ \linebreak $s_E, s_W, 
s_L, s_{robot}, s_X\}$ be a set of fields such that $s_P$, $s_S$, $s_N$, $s_E$,
and $s_W$ are all point-fields with source-value and 
decay 1 and $s_L$ is an edge-field with source-value $|V| + 2$ and decay 1,
and $T$ consist of a single robot $r$ based on set 
$Q = \{q_{c0}, q_{c10}, q_{c11} \ldots,
q_{c1k},$ \linebreak $q_{c2}, q_{c3}, q_{c4}\} ~ \cup ~ \{q_{vi0}, q_{vi1},  \ldots, q_{vi4} ~ | 
~ 0 \leq i \leq |V|\}$ with the following transitions:

\begin{enumerate}
\item $\la q_{c0}, fval(fq_N, =, 1.0) ~\rm{and}~ fval(fq_L, =, 1,0), 
           *, goNorth, q_{c10}\ra$
\item $\{ \la q_{c1(i-1)}, fval(fq_P, =, 1.0), *, goNorth, q_{c1i}\ra,
          \la q_{c1(i-1)}, fval(fq_P, =, 0.0)$ \newline \hspace*{0.1in} and $ fval(fq_E, =, 0.0), 
              *, goNorth, q_{c1(i-1)}\ra ~ | ~ 1 \leq i \leq k\}$
\item $\la q_{c1k}, fval(fq_E, =, 1.0)) ~\rm{and}~ fval(fq_L, =, |V| + 2,0), 
           *, goEast, q_{c2}\ra$
\item $\la q_{c2}, fval(fq_E, =, 1.0)) ~\rm{and}~ fval(fq_L, =, |V| + 2,0), 
           *, goEast, q_{c2}\ra$
\item $\la q_{c2}, fval(fq_S, =, 1.0)) ~\rm{and}~ fval(fq_L, =, |V| + 2,0), 
           *, goSouth, q_{c3}\ra$
\item $\la q_{c3}, fval(fq_L, <, |V| + 2,0) ~\rm{and}~ fval(fq_L, >, 1.0), 
           *, goSouth, q_{c3}\ra$
\item $\la q_{c3}, fval(fq_W, =, 1.0) ~\rm{and}~ fval(fq_L, =, 1,0), 
           *, goWest, q_{c4}\ra$
\item $\la q_{c4}, fval(fq_W, =, 1.0) ~\rm{and}~ fval(fq_L, =, 1,0), 
           *, goWest, q_{c4}\ra$
\item $\la q_{c4}, fval(fq_N, =, 1.0)) ~\rm{and}~ fval(fq_L, =, 1,0), 
           *, goNorth, q_{v10}\ra$

\item $\{ \la q_{v(i-1)0}, fval(fq_P, =, 1.0) ~\rm{and}~ fval(fq_L, =, j + 1.0), 
               *, goNorth, q_{v(i-1)1}\ra,$ \newline \hspace*{0.1in}
         $\la q_{v(i-1)0}, fval(fq_P, =, 1.0) ~\rm{and}~ fval(fq_L, =, k + 1.0), 
               *, goNorth, q_{v(i-1)0}\ra,$ \newline \hspace*{0.1in}
         $\la q_{v(i-1)1}, fval(fq_E, =, 0.0), 
               *, goNorth, q_{v(i-1)1}\ra,$ \newline \hspace*{0.1in}
         $\la q_{v(i-1)0}, fval(fq_P, =, 0.0) ~\rm{and}~ fval(fq_E, =, 0.0), 
               *, goNorth, q_{v(i-1)0}\ra$ \newline \hspace*{0.1in}
           $~ | ~ 1 \leq i \leq |V|, v_j \in N_C(v_1), ~\rm{and}~ v_k \not\in N_C(v_i)\}$
\item $\{ \la q_{v(i-1)1}, fval(fq_E, =, 1.0) ~\rm{and}~ fval(fq_L, =, |V| + 2,0), 
              *, goEast,$ \linebreak $q_{v(i-1)2}\ra,$
         $\la q_{v(i-1)2},$ $fval(fq_E, =, 1.0) ~\rm{and}~ fval(fq_L, =, |V| + 2,0), 
              *, goEast, q_{v(i-1)2}\ra$ \linebreak $ ~ | ~ 1 \leq i \leq |V|\}$
\item $\{ \la q_{v(i-1)2}, fval(fq_S, =, 1.0) ~\rm{and}~ fval(fq_L, =, |V| + 2,0), 
           *, goSouth, q_{v(i-1)3}\ra$ \newline \hspace*{0.1in}
          $~ | ~ 1 \leq i \leq |V|\}$
\item $\{ \la q_{v(i-1)3}, fval(fq_L, <, |V| + 2,0) ~\rm{and}~
                                                    fval(fq_L, >, 1.0),
            *, goSouth, q_{v(i-1)3}\ra$ \newline \hspace*{0.1in}
           $~ | ~ 1 \leq i \leq |V|\}$
\item $\{ \la q_{v(i-1)3}, fval(fq_W, =, 1.0) ~\rm{and}~ fval(fq_L, =, 1,0), 
           *, goWest, q_{v(i-1)4}\ra,$ \newline \hspace*{0.1in}
         $\la q_{v(i-1)4}, fval(fq_W, =, 1.0) ~\rm{and}~ fval(fq_L, =, 1,0), 
           *, goWest, q_{v(i-1)4}\ra$ \newline \hspace*{0.2in}
         $ ~ | ~ 1 \leq i \leq |V|\}$
\item $\{ \la q_{v(i-1)4}, fval(fq_N, =, 1.0) ~\rm{and}~ fval(fq_L, =, 1,0), 
           *, goNorth, q_{vi0}\ra$ \newline \hspace*{0.1in} $ ~ | ~ 1 \leq i < |V|\}$

\item $\la q_{v|V|4}, fval(fq_N, =, 1.0) ~\rm{and}~ fval(fq_L, =, 1,0), 
           fmod(s_X, (0,0)), stay, q_{v|V|}\ra$
\end{enumerate}

\noindent
The transitions in (1)--(9) ensure that the environment $E'$ initially has the proper 
form, i.e.,

\begin{itemize}
\item Square $E'_{1,1}$ contains a point-field of type $s_N$;
\item Squares $E'_{1,i}$, $2 \leq i \leq |V| + 1$, encode a candidate dominating set
       consisting of $k$ distinct vertices in $G$ (indicated by point-fields
       of type $s_P$ at the appropriate positions);
\item Squares $E'_{1,|V|+2}$ and $E'_{2,|V|+2}$ contain point-fields of type $s_W$;
\item Square $E'_{3,|V|+2}$ contains a point-fields of type $s_S$; and
\item Squares $E'_{2,1}$ and $E'_{3,1}$ contain point-fields of type $s_W$.
\item The north edge of $E'$ has an edge-field $s_L$.
\end{itemize}

\noindent
Point-fields in other squares can be eliminated by setting $|S_E| = k + 7$ and $|q_E| = 
\max(2,k)$. The transitions in (10)--(15) check if the encoded dominating set contains at 
least one vertex in each complete vertex-neighbourhood in $G$, i.e., the encoded
dominating set is an actual dominating set of size $k$ in $G$; if this is so,
transition (16) constructs $X$ at $p_X$.
Finally, let $p_I = E'_{1,1}$ and $X$ be a single point-field of type $s_X$ at
$p_X = E'_{1,1}$. Note that this instance of EnvDesSF$^{fast}_D$ can be constructed
in time polynomial in the size of the given instance of {\sc Dominating set}.

Let us now prove that this reduction is correct, in that the answer to the given
instance of {\sc Dominating set} is ``Yes'' if and only if the answer for the
constructed instance of EnvDesSF$^{fast}_D$ is ``Yes''. 

\begin{itemize}
\item{
Suppose there is 
dominating set of size at most $k$ in the given graph $G$. Let the set $V' \subseteq V$
be the vertices in this dominating set; if $|V'| < k$, add an arbitrary set of
$k - |V'|$ vertices in $G$ to $V'$. Construct an environment $E'$ as described above.
In such an environment, observe that $r$ started at $p_I$ will progress from $q_0$ to 
$q_{v|V|4}$ and construct $X$ at $p_X$; at any point in the operation of $r$, the
current state of $r$ indicates how many vertex-neighborhoods in $G$ have
been evaluated against the dominating set encoded in $E'$. Also at any point in the 
operation of $r$, at most one transition is enabled, which means that the operation of $r$
in $E'$ is deterministic in the sense described in Section \ref{SectForm}. With respect to 
$(c_1,c_2)$-completability, observe that in the worst case, $r$ needs to do at most
$|V| + 1$ complete passes around the movement-track (the first to ensure that $E'$
is in the proper initial form and then $|V|$ passes to ensure that each of the 
complete vertex-neighborhoods in $G$ has at least one vertex in the dominating set
encoded in $E'$). As there are $2|V| + 6$ squares in the movement track, this means that
$r$ requires at most $(|V| + 1)(2|V| + 6)$ timesteps to construct $X$ at $p_X$.
As in turn $(|V| + 1)(2|V| + 6) < (3|V| + 9)^2 = |E|^2 < c_1(|E| + |Q|)^{c_2}$ when
$c_1 = 1$ and $c_2 = 2$, this means that the task of constructing
$X$ at $p_X$ is $(c_1,c_2)$-completable when $c_1 = 1$ and $c_2 = 2$.
}
\item{
Conversely, suppose there is an environment $E'$ based on
$G'$ and $S$ such that robot $r$ started at $p_I$ constructs $X$ at $p_X$. In order
to progress from $q_0$ to $q_{v00}$, $E'$ must be in the initial form described above;
moreover, for $1 \leq i \leq |V|$, by the structure of the transition-set of $r$,
$r$ can only progress from $q_{v(i-1)0}$ to $q_{v(i-1)4}$ (and hence eventually trigger the
transition from $q_{v|V|}$ that constructs $X$ at $p_X$) if there is a vertex $v$ in the
dominating set encoded in $E'$ such that $v \in N_C(v_i)$ in $G$.\footnote{
Note that here, as in the reduction in the proof of Result D.SF.1, a vertex $v$ may appear 
multiple times in the dominating set encoded in $E'$. Once again, this is not a problem
--- the vertices in the encoded dominating set need not be distinct as we are are only 
interested in the presence of a dominating set of size {\em at most} $k$ in $G$.
}
Thus, $r$ started at $p_I$ can construct $X$ at $p_X$ in $E'$ only if a dominating set of 
size at most $k$ in $G$ is encoded in the middle $|V|$ squares of the first
column in $E'$.
}
\end{itemize}

\noindent
This proves the correctness of the reduction. The result then follows from the 
$W[2]$-hardness of {\sc Dominating set} \cite{DF99}, the reduction above, the fact that
$W[1] \subseteq W[2]$, and Lemma \ref{LemAppProp2}. To complete the proof, note that in
the constructed instance of EnvDesSF$^{fast}_D$, $|T| = h = |X| = 1$, 
$|f| = 3$, $|S| = 8$, $|S_E| = k + 7$, and $|q_E| = \max(2,k)$.
\end{proof}

\begin{description}
\item[{\bf Result D.SF.5}] \cite[Result B.5]{WV21}: 
If $\la |T|, h, d, |S|, |S_E|, |q_E|, |X|\ra$-EnvDesSF$^{fast}$ is \linebreak fp-tractable then
$FPT = W[1]$.
\end{description}
\begin{proof}
Modify the reduction in the proof of Result D.SF.4 such that for $1 \leq i \leq |V|$,
the sets of transitions 
$\{ \la q_{v(i-1)0}, fval(fq_P, =, 1.0) ~\rm{and}~ 
fval(fq_L, =, j + 1.0), *, goNorth$ \linebreak $q_{v(i-1)1}\ra ~ | ~ v_j \in N_C(v_i)\}$ 
and
$\{ \la q_{v(i-1)0}, fval(fq_P, =, 1.0) ~\rm{and}~ 
fval(fq_L, =, k + 1.0), *,$ \linebreak $goNorth, q_{v(i-1)0}\ra ~ | ~ v_k \not\in N_C(v_i)\}$ 
in transition-set (10) are replaced by the pair of transitions 
$\la q_{v(i-1)0}, fval(fq_P, =, 1.0) ~\rm{and}~ f, *, goNorth,$ $q_{v(i-1)1}\ra$ and
$\la q_{v(i-1)0}, fval(fq_P, =, 1.0) ~\rm{and}~ f', *, goNorth, q_{v(i-1)0}\ra$, respectively,
where $f$ and $f'$ are the formulas OR-ing all formulas of the form 
$fval(fq_L, =, j + 1.0)$ and 
$fval(fq_L, =, k + 1.0)$ (each such Or-ed group now enclosed by parentheses)
in the trigger formulas of the transitions in their respective sets. Note that this modified 
reduction still runs in time polynomial in the size of the given instance of {\sc Dominating
set}; moreover, this reduction is correct by slight modifications of the arguments given
for the proof of correctness of the reduction in the proof of Result D.SF.1.
The result then  follows from the $W[2]$-hardness of $\la k \ra$-{\sc Dominating set} 
\cite{DF99}, the modified reduction above, the fact that $W[1] \subseteq W[2]$, and
Lemma \ref{LemAppProp2}.  To complete the proof, note that in
the constructed instance of EnvDesSF$^{fast}_D$, $|T| = h = |X| = 1$, $d = 3$,
$|S| = 8$, $|S_E| = k + 7$, and $|q_E| = \max(2,k)$.
\end{proof}

\begin{description}
\item[{\bf Result D.SF.6}]  \cite[Result B.6]{WV21}:
If $\la |Q|, |f|, d, |S|, |X|\ra$-EnvDesSF$^{fast}$ is fp-tractable then $FPT = W[1]$.
\end{description}
\begin{proof}
Consider the following reduction from {\sc Dominating set} to EnvDesSF$^{fast}_D$, based
on the reduction from {\sc Dominating set} to EnvDesSF$^{fast}_D$ given in the proof
of Result D.SF.4. Given an instance $\la G = (V,E), k\ra$ of {\sc Dominating set}, construct
instance $\la G', S, T, p_I, X, p_X\ra$ of EnvDesSF$^{fast}_D$ as follows: Let
$G'$ be a $3 \times (|V| + 2)$ grid, $S = \{s_P, s_S, s_N, s_E, s_W, 
s_L, s_{robot}, s_X\}$ be a set of fields such that $s_P$, $s_S$, $s_N$, $s_E$,
and $s_W$ are all point-fields with source-value and 
decay 1 and $s_L$ is an edge-field with source-value $|V| + 2$ and decay 1. The
environment $E'$ based on $G'$ and $S$ can be split into two regions: a movement track
consisting of the outer squares on all edges of $E'$ and a central scaffolding composed
of squares $E'_{2,2}, E'_{2,3}, \ldots, E'_{2,|V|+1}$ in the second column of $E'$. Let
$p_I = E'_{1,1} ~ \cup ~ \{ E'_{3,i} ~ | ~ 2 \leq i \leq |V| + 1\}$, $X$ 
be a single 
point-field of type $s_X$ at $p_X = E'_{3,2}$, and $T$ consist of $|V| + 1$ robots 
of two basic types:

\begin{description}
\item[1)]{
{\bf Vertex neighborhood robots}: Each vertex neighborhood robot $r_{vi}$, $1 \leq i \leq
|V|$, moves clockwise around the movement-track to check if the candidate dominating set 
encoded in $E'$ contains at least one vertex in the complete neighborhood in $G$ of the 
vertex associated with that robot; if so, $r_{vi}$ places a point-field of type $e_X$ at 
position $i$ in the central scaffolding. For $1 \leq i \leq |V|$, robot $r_{vi}$ is 
initially positioned at $E'_{3,i+1}$, is based on state-set $Q = \{q_0, q_{v1}, q_{v1f}, 
q_{v2}, q_{v2f}, q_{v3}, v_{q3f},$ \linebreak $q_{v4}\}$, and has the following transitions:

\begin{enumerate}
\item $\la q_0, *, *, stay, q_{v3}\ra$
\item $\{ \la q_{v1}, fval(fq_P, =, 1.0) ~\rm{and}~ fval(fq_L, =, j + 1.0) ~\rm{and}~ 
                      (fval(e_{robot}, \leq, 1,5)$ and $fgrd(q_{robot},$ $>, North)),
               *, goNorth, q_{v1f}\ra, \\
          \la q_{v1}, fval(fq_P, =, 1.0) ~\rm{and}~ fval(fq_L, =, k + 1.0) ~\rm{and}~ 
                      (fval(e_{robot}, \leq, 1,5)$ and $fgrd(q_{robot},$ $>, North)),
               *, goNorth, q_{v1}\ra, \\
          \la q_{v1f}, fval(fq_L, < |V| + 2.0) ~\rm{and}~ 
                       (fval(e_{robot}, \leq, 1,5)$ and $fgrd(q_{robot}, >,$ \linebreak $ North)),$ $
               *, goNorth, q_{v1f}\ra, \\
          \la q_{v1}, fval(fq_P, =, 0.0) ~\rm{and}~ fval(fq_L, < |V| + 2.0) ~ \rm{and} ~
                      (fval(e_{robot}, \leq, 1,5)$ and $fgrd(q_{robot},$  $>, North)),
               *, goNorth, q_{v1}\ra \\
          ~ | ~ v_j \in N_C(v_i) ~\rm{and}~ v_k \not\in N_C(v_i)\}$
\item $\{ \la q_{v1f}, fval(fq_E, =, 1.0) ~\rm{and}~ fval(fq_L, =, |V| + 2,0) ~\rm{and} ~ 
                       (fval(e_{robot}, \leq, 1,5)$ and $fgrd(q_{robot},$ $ >, East)),
              *, goEast, q_{v2f}\ra, \\
          \la q_{v2f}, fval(fq_E, =, 1.0) ~\rm{and}~ fval(fq_L, =, |V| + 2,0)  ~\rm{and} ~
                       (fval(e_{robot}, \leq, 1,5)$ and $fgrd(q_{robot},$ $ >, East)),
              *, goEast, q_{v2f}\ra, \\
          \la q_{v1}, fval(fq_E, =, 1.0) ~\rm{and}~ fval(fq_L, =, |V| + 2,0) ~\rm{and} ~
                      (fval(e_{robot}, \leq, 1,5)$ and $fgrd(q_{robot},$ $ >, East)),
              *, goEast, q_{v2}\ra, \\
          \la q_{v2}, fval(fq_E, =, 1.0) ~\rm{and}~ fval(fq_L, =, |V| + 2,0)  ~\rm{and} ~
                      (fval(e_{robot}, \leq, 1,5)$ and $fgrd(q_{robot},$ $ >, East)),
              *, goEast, q_{v2}\ra\}$
\item $\{ \la q_{v2f}, fval(fq_S, =, 1.0) ~\rm{and}~ fval(fq_L, =, |V| + 2,0) ~\rm{and} ~
                       (fval(e_{robot}, \leq, 1,5)$ and $fgrd(q_{robot},$ $ >, South)),
           *, goSouth, q_{v3f}\ra, \\
          \la q_{v2}, fval(fq_S, =, 1.0) ~\rm{and}~ fval(fq_L, =, |V| + 2,0)  ~\rm{and} ~
                      (fval(e_{robot}, \leq, 1,5)$ and $fgrd(q_{robot},$ $ >, South)),
           *, goSouth, q_{v3}\ra\}$
\item $\{ \la q_{v3f}, fval(fq_L, <, |V| + 2,0) ~\rm{and}~ fval(fq_L, >, 1.0) ~\rm{and ~ not}~
                       fval(fq_L, =, i + 1.0) ~\rm{and} ~
                       (fval(e_{robot}, $ $ \leq, 1,5)$ and $fgrd(q_{robot}, >, South)),
            *, goSouth, q_{v3f}\ra, \\
          \la q_{v3f}, fval(fq_L, <, |V| + 2,0) ~\rm{and}~ fval(fq_L, >, 1.0) ~\rm{and}~
                       fval(fq_L, =, i + 1.0)~\rm{and} ~\linebreak
                       (fval(e_{robot},$ $ \leq, 1,5)$ and $fgrd(q_{robot}, >, South)), 
            fmod(s_X, (-1,0)), \linebreak goSouth, q_{v3f}\ra, \\
          \la q_{v3}, fval(fq_L, <, |V| + 2,0) ~\rm{and}~ fval(fq_L, >, 1.0)~\rm{and} ~
                      (fval(e_{robot}, \leq, 1,5)$ and $fgrd(q_{robot},$ $ >, South)),
            *, goSouth, q_{v3f}\ra\}$
\item $\{ \la q_{v3f}, fval(fq_W, =, 1.0) ~\rm{and}~ fval(fq_L, =, 1,0) ~\rm{and} ~ 
                       (fval(e_{robot}, \leq, 1,5)$ and $fgrd(q_{robot},$ $ >, West)),
           *, goWest, q_{v4}\ra, \\
          \la q_{v3}, fval(fq_W, =, 1.0) ~\rm{and}~ fval(fq_L, =, 1,0)  ~\rm{and} ~
                      (fval(e_{robot}, \leq, 1,5)$ and \linebreak $fgrd(q_{robot},$ $ >, West)),
           *, goWest, q_{v4}\ra, \\
          \la q_{v4}, fval(fq_W, =, 1.0) ~\rm{and}~ fval(fq_L, =, 1,0)  ~\rm{and} ~
                      (fval(e_{robot}, \leq, 1,5)$ and \linebreak $fgrd(q_{robot},$ $ >, West)),
           *, goWest, q_{v4}\ra\}$
\item $\la q_{v4}, fval(fq_N, =, 1.0) ~\rm{and}~ fval(fq_L, =, 1,0) ~\rm{and} ~
                   (fval(e_{robot}, \leq, 1,5)$ and \linebreak $fgrd(q_{robot},$ $ >, West)),
           *, goNorth, q_{v1}\ra$
\end{enumerate}

\noindent
The transitions in (2) move $r_{vi}$ north along the first column and,
on detecting a vertex $v$ encoded in the candidate dominating set
in $E'$ such that $v \in N_C(v_i)$, set the state of $r_{vi}$ to $q_{v1f}$. 
The transitions in (3) and (4) ensure that this state-distinction is preserved until 
$r_{vi}$ is at position $E'_{3,i+1}$ in the third column of $E'$, at which point
the transitions in (5) places the point-field $s_X$ one square to the west in the
central scaffolding. All other transitions ensure repeated motion around the
outer movement track in a clockwise fashion, provided the correct 
motion-marker point-fields have been placed on this track (see the description of the
checker robot below).
}
\item[2)]{
{\bf Checker robot}: The checker robot $r_{chk}$ moves clockwise around the movement-track
to ensure on its first pass around that $E'$ has the proper initial form. On its second
pass around, it checks if all vertex neighborhood robots have fully filled in the 
central scaffolding; if so, it creates the requested structure $X$ at $p_X$. Robot $r_{chk}$
is initially positioned at $E'_{1,1}$, is based on state-set $Q = \{q_0, q_{ce10},
q_{ce11} \ldots, 1_{ce1k}, q_{ce2}, q_{ce3}, q_{ce4}, q_{cv1}, q_{cv2}, 
q_{cv3f}, q_{cv3nf}, q_{c4}\}$, and has the following transitions:

\begin{enumerate}
\item $\la q_0, fval(fq_N, =, 1.0) ~\rm{and}~ fval(fq_L, =, 1,0) ~\rm{and} ~
                (fval(e_{robot}, \leq, 1,5)$ and \linebreak $fgrd(q_{robot}, >,$ $ North)),
           *, goNorth, q_{ce10}\ra$
\item $\{ \la q_{ce1(i-1)}, fval(fq_P, =, 1.0) ~\rm{and}~ fval(fq_X, =, 0.0) ~\rm{and}~
                            (fval(e_{robot}, \leq, 1,5)$ and $fgrd(q_{robot},$ $ >, North)),
              *, goNorth, q_{ce1i}\ra,$ \newline
         $\la q_{ce1(i-1)}, fval(fq_P, =, 0.0) \rm{and}~ fval(fq_E, =, 0.0) 
                            ~\rm{and}~ fval(fq_X, =, 0.0) ~\rm{and}~ \linebreak
                           (fval(e_{robot}, \leq, 1,5)$ and $fgrd(q_{robot}, >, North)),
              *, goNorth, q_{ce1(i-1)}\ra$ $ ~ | ~ 1 \leq i \leq k\}$
\item $\la q_{ce1k}, fval(fq_E, =, 1.0) ~\rm{and}~ fval(fq_L, =, |V| + 2,0) ~\rm{and}~
                     (fval(e_{robot}, \leq, 1,5)$ and $fgrd(q_{robot},$ $ >, East)),
           *, goEast, q_{ce2}\ra$
\item $\la q_{ce2}, fval(fq_E, =, 1.0) ~\rm{and}~ fval(fq_L, =, |V| + 2,0) ~\rm{and}~
                    (fval(e_{robot}, \leq, 1,5)$ and $fgrd(q_{robot},$ $ >, East)),
           *, goEast, q_{ce2}\ra$
\item $\la q_{ce2}, fval(fq_S, =, 1.0) ~\rm{and}~ fval(fq_L, =, |V| + 2,0) ~\rm{and}~
                    (fval(e_{robot}, \leq, 1,5)$ and $fgrd(q_{robot},$ $ >, South)),
           *, goSouth, q_{ce3}\ra$
\item $\la q_{ce3}, fval(fq_L, <, |V| + 2,0) ~\rm{and}~ fval(fq_L, >, 1.0) ~\rm{and}~ 
                    fval(fq_X, =, 0.0)$ $ ~\rm{and}~\linebreak
                    (fval(e_{robot}, \leq, 1,5)$ and $fgrd(q_{robot}, >, South)),
           *, goSouth, q_{ce3}\ra$
\item $\la q_{ce3}, fval(fq_W, =, 1.0) ~\rm{and}~ fval(fq_L, =, 1,0) ~\rm{and}~
                    (fval(e_{robot}, \leq, 1,5)$ and $fgrd(q_{robot},$ $ >, West)),
           *, goWest, q_{ce4}\ra$
\item $\la q_{ce4}, fval(fq_W, =, 1.0) ~\rm{and}~ fval(fq_L, =, 1,0) ~\rm{and}~
                    (fval(e_{robot}, \leq, 1,5)$ and $fgrd(q_{robot},$ $ >, West)),
           *, goWest, q_{ce4}\ra$
\item $\la q_{ce4}, fval(fq_N, =, 1.0) ~\rm{and}~ fval(fq_L, =, 1,0) ~\rm{and}~
                    (fval(e_{robot}, \leq, 1,5)$ and \linebreak  $fgrd(q_{robot},$ $ >, North)),
           *, goNorth, q_{cv1}\ra$

\item $\{ \la q_{cv1}, fval(fq_P, =, 1.0) ~\rm{and}~
                       (fval(e_{robot}, \leq, 1,5)$ and $fgrd(q_{robot}, >, North)), \linebreak 
              *, goNorth, q_{cv1}\ra, \\
          \la q_{cv1}, fval(fq_P, =, 0.0) ~\rm{and}~ fval(fq_E, =, 0.0) ~\rm{and}~
                       (fval(e_{robot}, \leq, 1,5)$ and \linebreak $fgrd(q_{robot}, >,$  $ North)),
              *, goNorth, q_{cv1}\ra\}$
\item $\la q_{cv1}, fval(fq_E, =, 1.0) ~\rm{and}~ fval(fq_L, =, |V| + 2,0) ~\rm{and}~
                    (fval(e_{robot}, \leq, 1,5)$ and $fgrd(q_{robot},$ $ >, East)),
           *, goEast, q_{cv2}\ra$
\item $\la q_{cv2}, fval(fq_E, =, 1.0) ~\rm{and}~ fval(fq_L, =, |V| + 2,0) ~\rm{and}~
                    (fval(e_{robot}, \leq, 1,5)$ and $fgrd(q_{robot},$ $ >, East)),
           *, goEast, q_{cv2}\ra$
\item $\la q_{cv2}, fval(fq_S, =, 1.0) ~\rm{and}~ fval(fq_L, =, |V| + 2,0) ~\rm{and}~
                    (fval(e_{robot}, \leq, 1,5)$ and $fgrd(q_{robot},$ $ >, South)),
           *, goSouth, q_{cv3f}\ra$
\item $\la q_{cv3f}, fval(fq_L, <, |V| + 2,0) ~\rm{and}~ fval(fq_L, >, 2.0) ~\rm{and}~ 
                     fval(fq_X, =, 0.5) ~\rm{and}~ \\
                     (fval(e_{robot}, \leq, 1,5)$ and $fgrd(q_{robot}, >, South)),
           *, goSouth, q_{cv3f}\ra$
\item $\la q_{cv3f}, fval(fq_L, =, 2.0) ~\rm{and}~ fval(fq_X, =, 0.5) ~\rm{and}~
                     (fval(e_{robot}, \leq, 1,5)$ and $fgrd(q_{robot},$ $ >, South)),
           fmod(s_X, (0.0)), goSouth, q_{cv3f}\ra$
\item $\la q_{cv3f}, fval(fq_L, <, |V| + 2,0) ~\rm{and}~ fval(fq_L, >, 2.0) ~\rm{and}~ 
                     fval(fq_X, =, 0.0) ~\rm{and}~$ 
                    $(fval(e_{robot}, \leq, 1,5)$ and $fgrd(q_{robot}, >, South)),
           *, goSouth, q_{cv3nf}\ra$
\item $\la q_{cv3nf}, fval(fq_L, >, 1.0) ~\rm{and}~
                      (fval(e_{robot}, \leq, 1,5)$ and $fgrd(q_{robot}, >, South)), \linebreak
           *, goSouth,$ $ q_{cv3nf}\ra$
\item $\la q_{cv3f}, fval(fq_W, =, 1.0) ~\rm{and}~ fval(fq_L, =, 1,0) ~\rm{and}~
                     (fval(e_{robot}, \leq, 1,5)$ and $fgrd(q_{robot},$ $ >, West)),
           *, goWest, q_{cv4}\ra$
\item $\la q_{cv3nf}, fval(fq_W, =, 1.0) ~\rm{and}~ fval(fq_L, =, 1,0) ~\rm{and}~
                      (fval(e_{robot}, \leq, 1,5)$ and $fgrd(q_{robot},$ $ >, West)),
           *, goWest, q_{cv4}\ra$
\item $\la q_{cv4}, fval(fq_W, =, 1.0) ~\rm{and}~ fval(fq_L, =, 1,0) ~\rm{and}~
                    (fval(e_{robot}, \leq, 1,5)$ and $fgrd(q_{robot},$ $ >, West)),
           *, goWest, q_{cv4}\ra$
\item $\la q_{cv4}, fval(fq_N, =, 1.0)) ~\rm{and}~ fval(fq_L, =, 1,0) ~\rm{and}~
                    (fval(e_{robot}, \leq, 1,5)$ and $fgrd(q_{robot},$ $ >, North)),
           *, goNorth, q_{cv1}\ra$
\end{enumerate}

\noindent
The transitions in (1)--(9) ensure that the environment $E$ initially has the proper 
form, i.e.,

\begin{itemize}
\item Square $E'_{1,1}$ contains a point-field of type $s_N$;
\item Squares $E'_{1,i}$, $2 \leq i \leq |V| + 1$, encode a candidate dominating set
       consisting of $k$ distinct vertices in $G$ (indicated by point-fields
       of type $s_P$ at the appropriate positions);
\item Squares $E'_{1,|V|+2}$ and $E'_{2,|V|+2}$ contain point-fields of type $s_W$;
\item Square $E'_{3,|V|+2}$ contains a point-fields of type $s_S$;
\item Squares $E'_{2,1}$ and $E'_{3,1}$ contain point-fields of type $s_W$; 
\item The north edge of $E'$ has an edge-field $s_L$; and
\item There are no point-fields of type $e_X$ in the middle $|V|$ squares of the second or third column of $E'$.
\end{itemize}

\noindent
With respect to the final point, there may be other types of point-fields in these squares,
but they are ignored by the vertex neighborhood and checker robots.
The transitions in (14) check if the central 
scaffolding has been fully filled in with point-fields of type $e_X$ by the vertex 
neighbourhood robots, i.e., the encoded dominating set is an actual dominating set of size 
$k$ in $G$; if this is so, the transition (15) constructs $X$ at $p_X$.
All other transitions ensure repeated motion around the
outer movement track in a clockwise fashion, provided the correct 
motion-marker point-fields have been placed on this track.
}
\end{description}

\noindent
Note that all of the transitions described above use the robot collision-avoidance 
formula $(fval(e_{robot}, \leq, 1,5) ~\rm{and}~ fgrd(q_{robot}, >, dir))$ derived in Section
\ref{SectResTeamDesLS} that allows moves in direction $dir$ along the movement-track.
This instance of EnvDesSF$^{fast}_D$ can be constructed
in time polynomial in the size of the given instance of {\sc Dominating set}.

Let us now prove that this reduction is correct, in that the answer to the given
instance of {\sc Dominating set} is ``Yes'' if and only if the answer for the
constructed instance of EnvDesSF$^{fast}_D$ is ``Yes''. 

\begin{itemize}
\item{
Suppose there is 
dominating set of size at most $k$ in the given graph $G$. Let the set $V' \subseteq V$
be the vertices in this dominating set; if $|V'| < k$, add an arbitrary set of
$k - |V'|$ vertices in $G$ to $V'$. Construct an environment $E'$ as described above.
In such an environment, observe that (1) the vertex neighbourhood robots in $T$ started at 
the specified positions in $p_I$ will, on their first pass around the
movement-track, fully fill in the squares in the central scaffolding with point-fields
of type $s_X$ and (2) the checker robot, on its second pass around the movement-track,
will construct $X$ at $p_X$.  At any point in the operation of $T$, at most one transition 
is enabled in any robot, which means that the operation of $T$ in $E'$ is deterministic in 
the sense described in Section \ref{SectForm}. With respect to $(c_1,c_2)$-completability, 
observe that in the worst case, each of robots in $T$ needs to do at most
two passes around the movement-track. As there are $2|V| + 6$ squares in the movement track
and at least one robot in $T$ moves forward in every timestep, this means that
$T$ requires at most $(|V| + 1)(2|V| + 6)$ timesteps to construct $X$ at $p_X$.
As in turn $(|V| + 1)(2|V| + 6) < (3|V| + 6)^2 = |E|^2 < c_1(|E| + |Q|)^{c_2}$ when
$c_1 = 1$ and $c_2 = 2$, this means that the task of constructing
$X$ at $p_X$ is $(c_1,c_2)$-completable when $c_1 = 1$ and $c_2 = 2$.
}
\item{
Conversely, suppose there is an environment $E'$ based on
$G'$ and $S$ such that robot team $T$ started at $p_I$ constructs $X$ at $p_X$. In order
for $X$ to be constructed at $p_X$, the checker robot the checker robot must have been
able to complete its second pass of $E'$ and all of the vertex neighborhood robots in
$T$ must have fully filled in the squares in the central scaffolding in $E'$ with 
point-fields of type $s_X$. However, by the structures of the transition-sets of the
vertex neighborhood robots, the latter can only happen if, for each vertex robot $e_v$ in 
$T$, there is a vertex $v'$ in the dominating set encoded in $E'$ such that $v' \in N_C(v)$ 
in $G$.\footnote{
Note that here, as in the reduction in the proof of Result D.SF.1, a vertex $v$ may appear 
multiple times in the dominating set encoded in $E'$. Once again, this is not a problem
--- the vertices in the encoded dominating set need not be distinct as we are are only 
interested in the presence of a dominating set of size {\em at most} $k$ in $G$.
}
Thus, $T$ started at $p_I$ can construct $X$ at $p_X$ in $E'$ only if a dominating set of 
size at most $k$ in $G$ is encoded in the middle $|V|$ squares of the first
column in $E'$.
}
\end{itemize}

\noindent
This proves the correctness of the reduction. The result then follows from the 
$W[2]$-hardness of {\sc Dominating set} \cite{DF99}, the reduction above, the fact that
$W[1] \subseteq W[2]$, and Lemma \ref{LemAppProp2}. To complete the proof, note that in
the constructed instance of EnvDesSF$^{fast}_D$, $|X| = 1$, $d = 4$,
$|S| = 8$, $|f| = 12$, and $|Q| = k + 10$.
\end{proof}

\begin{description}
\item[{\bf Result D.SF.7}] \cite[Result B.7]{WV21}: 
$\la |E|, |S|\ra$-EnvDesSF$^{fast}$ is fp-tractable.
\end{description}
\begin{proof}
Follows from the algorithm in the proof of Result N in \cite{WV18_SI}.
\end{proof}

\subsection{Proofs for Team / Environment Co-design by Library Selection}

\vspace*{0.1in}

\begin{description}
\item[{\bf Result E.SF.1}:] If TeamEnvDesSF$^{fast}$ is polynomial-time exact solvable then \linebreak $P = NP$.
\end{description}
\begin{proof}
Follows from the reduction in the proof of Result D.SF.1, modified such that
$L$ consists of the single robot in $T$.
\end{proof}

\begin{description}
\item[{\bf Result E.SF.2}:] 
If $\la |T|, h, |f|, |L|, |E|, |S_E|, |q_E|, |X|\ra$-TeamEnvDesSF$^{fast}$ is \linebreak fp-tractable 
then $FPT = W[1]$.
\end{description}
\begin{proof}
Follows from the reduction in the proof of Result D.SF.2, modified such that
$L$ consists of the single robot in $T$.
\end{proof}

\begin{description}
\item[{\bf Result E.SF.3}:] 
If $\la |T|, h, |Q|, d, |L|, |E|, |S_E|, |q_E|, |X|\ra$-TeamEnvDesSF$^{fast}$ is \linebreak fp-tractable
then $FPT = W[1]$.
\end{description}
\begin{proof}
Follows from the reduction in the proof of Result D.SF.3, modified such that
$L$ consists of the single robot in $T$.
\end{proof}

\begin{description}
\item[{\bf Result E.SF.4}:] 
If $\la |T|, h, |f|, |L|, |S|, |S_E|, |q_E|, |X|\ra$-TeamEnvDesSF$^{fast}$ is \linebreak fp-tractable
then $FPT = W[1]$.
\end{description}
\begin{proof}
Follows from the reduction in the proof of Result D.SF.4, modified such that
$L$ consists of the single robot in $T$.
\end{proof}

\begin{description}
\item[{\bf Result E.SF.5}:] 
If $\la |T|, h, d, |L|, |S|, |S_E|, |q_E|, |X|\ra$-TeamEnvDesSF$^{fast}$ is \linebreak fp-tractable
then $FPT = W[1]$.
\end{description}
\begin{proof}
Follows from the reduction in the proof of Result D.SF.5, modified such that
$L$ consists of the single robot in $T$.
\end{proof}

\begin{description}
\item[{\bf Result E.SF.6}:] 
If $\la |Q|, |f|, d, |S|, |X|\ra$-TeamEnvDesSF$^{fast}$ is fp-tractable
then \linebreak $FPT = W[1]$.
\end{description}
\begin{proof}
Follows from the reduction in the proof of Result D.SF.6, modified such that $L$ consists of
the $|V| + 1$ robots in $T$. Observe that given its transition structure,
the checker robot must be selected from $L$ and initially positioned 
in $E'_{1,1}$ in order to both progress around the movement-track and (if the central
scaffolding is filled in with point-fields of $s_X$) create $X$ at $p_X$. Moreover, in
order for all squares in the central scaffolding to be filled in, all $|V|$ vertex
neighbourhood robots from $L$ must be selected from $L$ and initially positioned
in the middle $|V|$ squares of the eastmost column in $E'$. The order of
the vertex neighborhood robots in this initial region is immaterial, as all of these
roots will have a chance to fill in their respective squares in the central scaffolding
before the second pass of the checker robot around the movement-track. The proof
of correctness of the reduction follows modulo these observations, as does the result.
\end{proof}

\begin{description}
\item[{\bf Result E.SF.7}:] 
$\la |L|, |E|, |S|\ra$-TeamEnvDesSF$^{fast}$ is fp-tractable.
\end{description}
\begin{proof}
Follows from the algorithm in the proof of Theorem 7 in \cite{TW19}.
\end{proof}

\end{document}